\documentclass[12pt,letterpaper]{JHEP3}

\usepackage{amsmath,amssymb,amsfonts,xspace}
\usepackage{epsfig}

\def\lfig#1#2#3#4{
\begin{figure}
\centerline{\hfill \includegraphics[height=#3]{#2}\hfill}
\caption{#1 \label{#4}}
\end{figure}
}

\def\Im{\,{\rm Im}\, }
\def\Re{\,{\rm Re}\, }

\def\rangl{\right\rangle   }
\def\langl{\left\langle  }
\def\({\left(}
\def\){\right)}
\def\[{\left[}
\def\]{\right]}
\def\hf{{1\over 2}}

\renewcommand{\d}{\mathrm{d}}
\newcommand{\de}{\mathrm{d}}
\newcommand{\De}{\mathrm{D}}
\newcommand{\I}{\mathrm{i}}
\newcommand{\e}{\mathrm{e}}
\newcommand{\cL}{\mathcal{L}}

\def\vrh{\varrho}

\newcommand{\p}{\partial}

\newcommand{\half}{\frac{1}{2}}

\newcommand{\cV}{\mathcal{V}}

\newcommand{\cC}{\mathcal{C}}
\newcommand{\cS}{\mathcal{S}}

\newcommand{\cK}{\mathcal{K}}
\newcommand{\cM}{\mathcal{M}}
\newcommand{\cN}{\mathcal{N}}

\newcommand{\CX}{\mathcal{X}}
\newcommand{\hCX}{\hat{\mathcal{X}}}
\newcommand{\cR}{\mathcal{R}}

\newcommand{\vb}{\bar{v}}

\DeclareSymbolFont{AMSa}{U}{msa}{m}{n}
\DeclareSymbolFont{AMSb}{U}{msb}{m}{n}
\DeclareMathSymbol{\fieldR}{\mathalpha}{AMSb}{"52}

\newcommand{\N}{{\mathcal N}}
\newcommand{\kahler}{{K\"ahler}\xspace}
\newcommand{\hk}{{hyperk\"ahler}\xspace}
\newcommand{\qk}{{quaternionic-K\"ahler}\xspace}

\newcommand{\cZ}{\mathcal{Z}}
\newcommand{\cO}{\mathcal{O}}
\newcommand{\cU}{\mathcal{U}}
\newcommand{\hU}{\hat{\mathcal{U}}}
\newcommand{\cA}{\mathcal{A}}

\newcommand{\tC}{{\tilde C}}

\newcommand{\pa}{\partial}
\newcommand{\nn}{\nonumber}

\newcommand{\eps}{\epsilon}
\newcommand{\IR}{\mathbb{R}}
\newcommand{\IC}{\mathbb{C}}
\newcommand{\IZ}{\mathbb{Z}}

\newcommand{\tzeta}{\tilde\zeta}

\newcommand{\txi}{\tilde\xi}

\newcommand{\CP}{\IC P^1}

\def\bea{\begin{eqnarray}}
\def\eea{\end{eqnarray}}
\def\be{\begin{equation}}
\def\ee{\end{equation}}
\def\ba{\begin{align}}
\def\ea{\end{align}}
\def\bse{\begin{subequations}}
\def\ese{\end{subequations}}

\def\bF{\bar F}

\def\bZ{\bar Z}
\def\bC{\bar C}

\def\bcZ{\bar \cZ}

\def\ba{\bar a}

\def\bi{\bar \imath}
\def\bj{\bar \jmath}

\def\bw{\bar w}

\def\bz{\bar z}
\def\bu{\bar u}
\def\bv{\bar v}

\def\bpi{\bar \pi}

\newcommand{\CL}{{\cal{L}}}
\newcommand{\CA}{{\cal{A}}}
\newcommand{\CB}{{\cal{B}}}

\def\vr{{\vec r}}

\def\etaf{\eta^{\flat}}

\def\vf{v^{\flat}}
\def\bvf{\bv^{\flat}}
\def\uf{u^{\flat}}

\def\xf{x^{\flat}}
\def\rf{r^{\flat}}
\def\brf{{\breve r}^{\flat}}
\def\vrf{\vec r^{\flat}}

\def\vrhf{\vrh_{\flat}}

\def\etap{\eta_+}
\def\etam{\eta_-}
\def\etapm{\eta_\pm}

\def\ze{\zeta}

\def\ztp{\zeta_+}
\def\ztm{\zeta_-}
\def\brztp{\breve\zeta_+}
\def\brztm{\breve\zeta_-}
\def\ztpm{\zeta_\pm}

\def\epsm{\epsilon_-}
\def\epsp{\epsilon_+}

\def\epst{\epsilon_3}

\def\nupk#1{\hat\nu_{#1}}

\def\bnupk#1{\bar {\hat\nu}_{#1}}

\def\Hp{H_{\scriptscriptstyle{\smash{(1)}}}}

\def\HpI#1{H_{{\scriptscriptstyle\smash{(1)}}#1}}

\def\mui#1{\mu^{[#1]}}
\def\txii#1{{\tilde\xi}^{[#1]}}

\def\ai#1{{\alpha}^{[#1]}}

\def\bxii#1{{\breve\xi}_{[#1]}}
\def\btxii#1{\lefteqn{\breve{\tilde\xi}}\vphantom{\tilde\xi}\hphantom{\tilde\xi}^{[#1]}}

\def\nui#1{\nu_{[#1]}}
\def\xii#1{\xi_{[#1]}}
\def\pmui#1{\mu^{'[#1]}}

\def\tmui#1{\tilde\mu^{[#1]}}
\def\tnui#1{\tilde\nu_{[#1]}}
\def\muzi#1{\breve\mu^{[#1]}}
\def\mupi#1{\hat\mu^{[#1]}}

\def\nupi#1{\hat\nu_{[#1]}}

\def\cij#1{c^{[#1]}}
\def\ci#1{c^{[#1]}}
\def\Sij#1{S^{[#1]}}

\def\hSij#1{\hat S^{[#1]}}
\def\Gi#1{G^{[#1]}}

\def\Hij#1{H^{[#1]}}
\def\pHij#1{H^{'[#1]}}
\def\tHij#1{\tilde H^{[#1]}}
\def\hHij#1{\hat H^{[#1]}}
\def\Hpij#1{\Hij{#1}_{\scriptscriptstyle{\smash{(1)}}}}
\def\tHpij#1{\tHij{#1}_{\scriptscriptstyle{\smash{(1)}}}}
\def\hHpij#1{\hHij{#1}_{\scriptscriptstyle{\smash{(1)}}}}

\def\nuor{\nui{0}}
\def\muor{\mui{0}}

\def\nupp{\hat\nu_+}
\def\nuppm{\hat\nu_\pm}
\def\nuppz{\hat\nu_{\zeta,+}}
\def\nuppmz{\hat\nu_{\zeta,\pm}}
\def\nuppzz{\hat\nu_{\zeta\zeta,+}}
\def\nuppmzz{\hat\nu_{\zeta\zeta,\pm}}

\def\nupm{\hat\nu_-}
\def\nupmz{\hat\nu_{\zeta,-}}

\def\Rf{r^{\flat}}
\def\hw{\hat w}

\def\ui#1{^{[#1]}}
\def\di#1{_{[#1]}}

\def\zxi{\xi_{(0)}}
\def\bbz{\breve \varpi}

\def\varpi{{\bf z}}

\title{\vspace*{-4mm} Linear perturbations of quaternionic metrics}

\preprint{LPTA/08-054,
LPTENS-08/35, IPhT-T08/115\\
ITP-UU-08-42, SPIN-08-34}

\author{Sergei Alexandrov$^1$, Boris Pioline$^{2}$,
Frank Saueressig$^3$, Stefan Vandoren$^4$\\

$^1$ {\it Laboratoire de Physique Th\'eorique \&
Astroparticules\footnote{Unit\'e mixte de recherche du CNRS UMR 5207}, \\
Universit\'e Montpellier II, 34095 Montpellier Cedex 05, France}\\

\vspace*{-3mm}

$^2$ -- {\it Laboratoire de Physique Th\'eorique et Hautes
Energies\footnote{Unit\'e mixte de recherche du CNRS UMR 7589}, \\
Universit\'e Pierre et Marie Curie,
4 place Jussieu, 75252 Paris cedex 05, France} \\
-- {\it Laboratoire de Physique Th\'eorique de l'Ecole Normale
Sup\'erieure\footnote{Unit\'e mixte
de recherche du CNRS UMR 8549},\\
24 rue Lhomond, 75231 Paris cedex 05, France}\\

\vspace*{-3mm}

$^3$ {\it Institut de Physique Th\'eorique\footnote{Unit\'e de recherche
associ\'ee au CNRS URA 2306},
 CEA,  F-91191 Gif-sur-Yvette, France}\\

\vspace*{-3mm}

$^4$ {\it   Institute for Theoretical Physics and
           Spinoza Institute,
           Utrecht University
           Leuvenlaan 4,
           3508 TD Utrecht
           The Netherlands.}\\

\vspace*{-3mm} {\tt e-mail: \email{alexandrov@lpta.univ-montp2.fr},
\email{pioline@lpthe.jussieu.fr}, \email{frank.saueressig@cea.fr},
\email{S.J.G.Vandoren@uu.nl}} \vspace*{-3mm}}

\abstract{
We extend the twistor methods developed in our earlier work  on linear deformations of
hyperk\"ahler  manifolds \cite{Alexandrov:2008ds} to the case of quaternionic-K\"ahler
manifolds. Via Swann's construction, deformations of a $4d$-dimensional
quaternionic-K\"ahler manifold $\cM$ are in one-to-one correspondence
with deformations of its $4d+4$-dimensional hyperk\"ahler cone $\cS$.
The latter can be encoded in variations of the complex symplectomorphisms
which relate different locally flat patches of the twistor space $\cZ_\cS$,
with a suitable homogeneity condition that ensures that the hyperk\"ahler cone
property is preserved. Equivalently, we show that  the deformations of $\cM$ can be encoded in
variations of the complex contact transformations which relate different locally flat patches
of the twistor space $\cZ_\cM$ of $\cM$, by-passing the Swann bundle and its twistor space.
We specialize these general results to the case
of quaternionic-K\"ahler metrics with $d+1$ commuting isometries, obtainable by
the Legendre transform method,  and linear deformations thereof.  We illustrate
our methods for the hypermultiplet moduli space in string theory compactifications
at tree- and one-loop level.
}

\begin{document}

\section{Introduction}

Quaternionic-\kahler (QK) manifolds play an important role in string
and supergravity theories, primarily because the hypermultiplet
moduli spaces appearing in string theory backgrounds with 8
supercharges fall into this class. In this work, we study general aspects of
QK manifolds and of their twistor spaces, and provide a general
formalism for describing linear perturbations of $4d$-dimensional QK
manifolds with $d+1$ commuting  isometries. For this purpose we build on our previous
study of linear deformations of \hk (HK) manifolds obtainable by the
Legendre transform method \cite{Alexandrov:2008ds}.

A key fact for the present study is the (local) one-to-one
correspondence between $4d$-dimensional QK manifolds $\cM$ and
$4d+4$-dimensional ``\hk cones'' (HKC) $\cS$,
i.e. $4d+4$-dimensional HK manifolds with a
homothetic Killing vector and an isometric $SU(2)$
action rotating the three complex structures (see Figure 1 for orientation).
In particular, Swann's construction produces $\cS$ as a $\IC^2/\IZ^2$ bundle over $\cM$,
twisted by the $SU(2)$ part of the spin connection on $\cM$
 \cite{MR1096180}. The converse relation goes under
the name of ``superconformal quotient'' in the physics literature
\cite{deWit:1999fp,deWit:2001dj}. Moreover, any isometry of $\cM$
can be lifted to a tri-holomorphic isometry of $\cS$, see e.g.
\cite{Bergshoeff:2004nf,deWit:2001bk}. Therefore, the formalism of
\cite{Alexandrov:2008ds} is directly applicable to the Swann bundle
$\cS$, with a suitable restriction to ensure the \hk cone (or ``superconformal invariance'') property.

For this purpose, one introduces the twistor space  $\cZ_\cS=\cS\times
\CP$ of the HK manifold $\cS$, an open covering $\hU_i$ of $\cZ_\cS$
projecting to open disks $\cU_i$ on $\CP$,
and a local Darboux coordinate system
$(\nui{i}^I, \mui{i}_I)$ ($I=\flat,0,\dots, d-1$) for the $\cO(2)$-twisted  complex
symplectic structure $\Omega\ui{i}=\de \mui{i}_I \wedge \de \nui{i}^I$ on $\hU_i$.
Since\footnote{Here $f_{ij}$ are the transition functions of the
$\cO(1)$ bundle on $\CP$ with coordinate $\zeta$.}
$\Omega\ui{i} =f_{ij}^2\, \Omega\ui{j} \mod \de\zeta$
on the overlap $\hU_i\cap \hU_j$, the coordinate systems
$(\nui{i}^I, \mui{i}_I)$ and $(\nui{j}^I, \mui{j}_I)$  must be related by
a symplectomorphism on $\hU_i\cap \hU_j$; the latter can be
parametrized in the usual way by a generating function
$\Sij{ij}(\nui{i}^I, \mui{j}_I,\zeta)$. The set of all $\Sij{ij}$, subject
to consistency relations, reality conditions and gauge equivalence,
encodes the complex symplectic structure on $\cZ_\cS$,
and therefore the HK metric on $\cS$.

As we show in Section \ref{sec-condsup} below, superconformal invariance
restricts $\Sij{ij}$ to be a function
of $f_{ij}^{-2n} \nui{i}^I$ and $\mui{i}_I$ only, with no further $\zeta$
dependence, and to be homogeneous of degree one when  $(\nui{i}^I,
\mui{i}_I)$ are rescaled with weight $(n,1-n)$, respectively\footnote{See
\cite{Ketov:2000hw,Kuzenko:2007qy,Kuzenko:2008ep} for other discussions
of superconformal invariance in projective superspace and
\cite{deWit:2006gn,deWit:2007qz} for an analysis in components.}. The integer
$n$ characterizes the transformation rules of the local coordinates
under both dilations and $SU(2)$ rotations. For $n=1$, the relevant
case for QK manifolds with isometries, the $\cO(0)$ sections
$\mui{i}_I$ may acquire anomalous scaling dimensions, and the
homogeneity condition may be relaxed into a ``quasi-homogeneity''
property, as explained further in Section \ref{sec-condsup}.

Deformations of $\cS$ correspond to variations $\Hpij{ij}(\nui{i}^I,
\mui{i}_I,\zeta)$ of the generating functions $\Sij{ij}$, subject to the
consistency, reality and quasi-homogeneity conditions and gauge equivalence.
When $\cS$ is obtainable by the Legendre transform method, which is the
case when $\cM$ admits $d+1$ commuting isometries, the
deformed twistor lines and \hk potential are easily computed
to first order in the perturbation. The deformed QK metric may
in principle be obtained by the standard superconformal quotient
procedure. In Appendix C, we construct a natural set of coordinates
on the deformed QK manifold, but stop short of writing the deformed
metric explicitly, as the expressions would be too cumbersome.

\lfig{Summary of various coordinate systems on the QK space $\cM$, its twistor space $\cZ_\cM$,
its Swann bundle $\cS$ and the twistor space of the Swann bundle $\cZ_\cS$.}{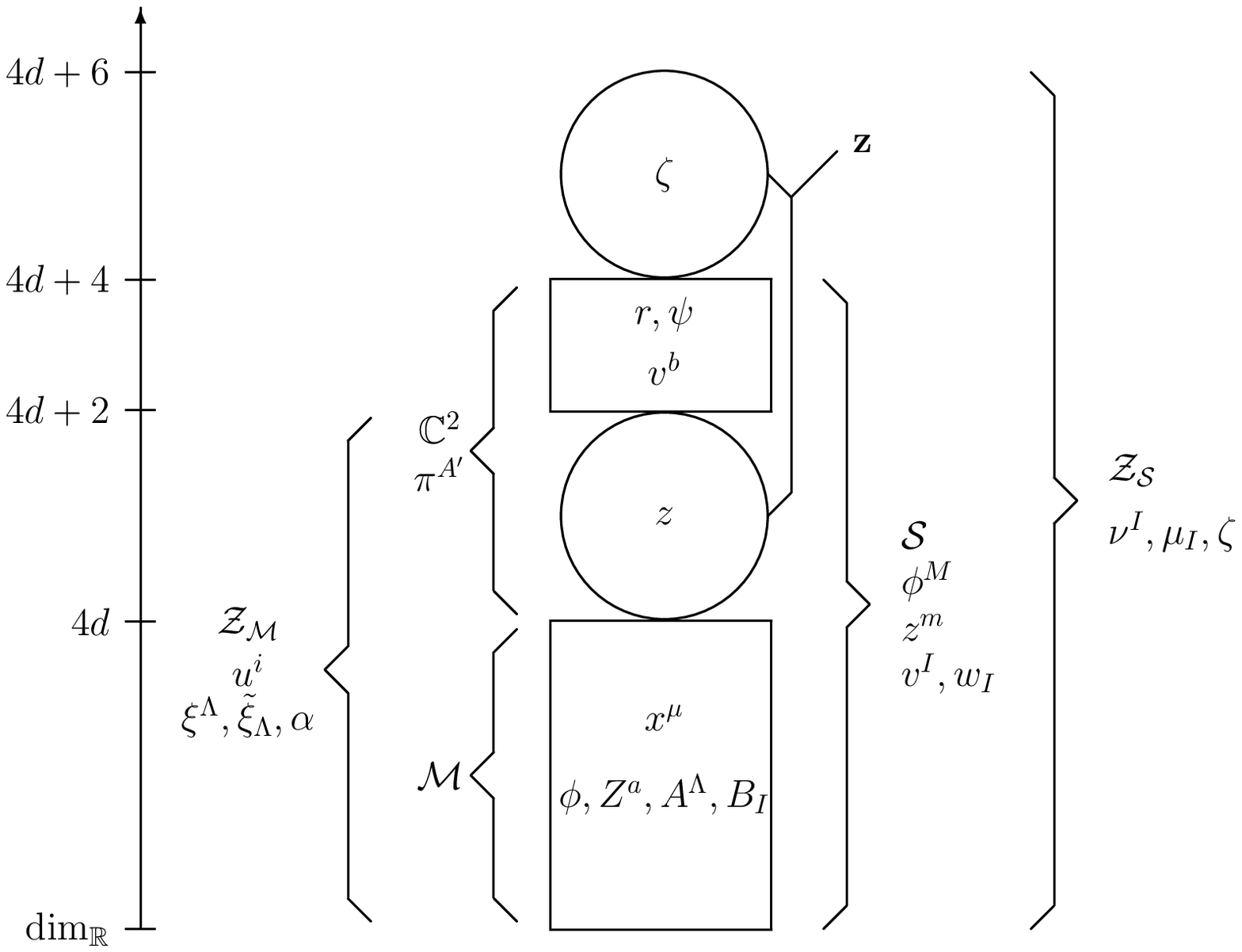}{11.5cm}{twifig}

While the strategy outlined above is conceptually straightforward, it is rather unpractical.
As we explain in detail in Section \ref{seccontact}, one may by-pass the twistor
space of the Swann bundle $\cZ_\cS$, and work directly with the
twistor space $\cZ_\cM$ of the QK manifold $\cM$, as emphasized in
particular by Salamon and Lebrun \cite{MR664330,MR1327157,lebrun1988rtq,lebrun1994srp}.
While $\cZ_\cS$ carries a complex
$\cO(2)$-valued symplectic structure and a HK metric degenerate
along the $\CP$ fiber, $\cZ_\cM$ carries a complex $\cO(2)$-valued
contact structure and a non degenerate \kahler-Einstein
metric\footnote{Moreover, in contrast to $\cZ_\cS$, the projection
from $\cZ_\cM$ to $\CP$ is not holomorphic.} \cite{MR664330}.
Conversely, Fano contact manifolds
with a  \kahler-Einstein metric are twistor spaces of QK
manifolds \cite{MR1327157}\footnote{In fact, our local analysis seems to
support Lebrun's conjecture \cite{MR1327157} that every Fano contact manifold is a twistor space.}.
Similarly to $\cZ_\cS$, the complex contact
structure $\CX$ on $\cZ_\cM$ admits local Darboux coordinates
$\xii{i}^\Lambda$, $\txii{i}_\Lambda$, $\ai{i}$ ($\Lambda=0,\dots, d-1$)
such that locally, the contact one-form takes the canonical form
$\CX\ui{i}=\de\ai{i}+ \xii{i}^\Lambda \de\txii{i}_\Lambda$.
These Darboux coordinates on $\cZ_\cM$ are essentially the
projectivization\footnote{The equivalence between contact structures and
homogeneous symplectic structures is a standard trick in contact
geometry, see e.g.  \cite{MR1327157} and references therein.}
of the Darboux coordinates $\nui{i}^I, \mui{i}_I$ on  $\cZ_\cS$.
More precisely, we show that the
projectivized complex Darboux coordinates depend on the coordinates $(\zeta,\pi^1,\pi^2)$
on the $\CP\times (\IC^2/\IZ_2)$ fiber over a given point on $\cM$ only through the
ratio $\varpi$ defined in \eqref{varpidef} below. Together with the
projectivization, this provides the desired reduction
from  $\cZ_\cS$ to  $\cZ_\cM$. The homogeneous generating functions
$\Sij{ij}(\nui{i}^I, \mui{i}_I,\zeta)$ of complex symplectomorphisms on
$\cZ_\cS$ yield generating functions
$\hSij{ij}(\xii{i}^\Lambda,\txii{i}_\Lambda,\ai{i})$ of complex
contact transformations on $\cZ$. Their deformations can be encoded in
functions $\hHpij{ij}$ of the same variables, subject to consistency relations,
reality conditions, and gauge equivalence.
This recovers Lebrun's assertion that the QK deformations of $\cM$ are classified by
the \v{C}ech cohomology group $H^1(\cZ_\cM, \cO(2))$ \cite{MR1327157}.
The deformed QK metric on $\cM$ can then
be extracted in a systematic way from the knowledge of the ``contact
twistor lines" (referred to as the "twistor map" in
\cite{Neitzke:2007ke}), i.e. the complex coordinates
$\xii{i}^\Lambda, \txii{i}_\Lambda, \ai{i}$ on $\cZ_\cM$ expressed as functions of
the coordinates $\varpi$ on $\CP$ and $x^\mu$ on the base
$\cM$.

We end this introduction with an important remark. In string theory
or supergravity, only QK manifolds with negative scalar curvature
appear as a consequence of supersymmetry \cite{Bagger:1983tt}. Such
QK spaces are generically non-compact. The linear deformation theory
set up in this paper is local and applies to both compact and
non-compact manifolds. Possible obstructions to extend and integrate
infinitesimal deformations into finite global deformations, however,
depend strongly on the (non-)compactness.  For instance, it is known
that complete QK manifolds with positive scalar curvature admit no deformations,
see e.g. \cite{lebrun1994srp,lebrun1988rtq}.
In contrast, the hypermultiplet moduli spaces
arising from string theory compactifications are in general deformed by quantum corrections,
as explained e.g. in the introduction of \cite{Alexandrov:2008ds}
and to be discussed  further in \cite{apsv3}.

This paper is organized as follows. $\bullet$ In Section 2, we review general
aspects of QK manifolds, their twistor spaces,  HKC and twistor spaces
thereof, and study the consequences of superconformal
invariance on the symplectomorphisms used in the patchwork construction
of the complex symplectic structure. In particular, in Section \ref{seccontact},
we explain in detail how  the homogeneous
complex symplectic structure on $\cZ_\cS$ reduces to a complex contact
structure on $\cZ_\cM$, thus allowing to by-pass the Swann bundle and its twistor space.
$\bullet$ In Section 3 we specify to the case when the $4d$-dimensional QK space
has $d+1$ commuting isometries, i.e. when its Swann bundle is obtainable
by the Legendre transform construction. We find the corresponding restriction on the
symplectomorphisms, perform the superconformal quotient and obtain
the contact twistor lines. $\bullet$ In
Section 4, we illustrate these methods on the example of the hypermultiplet
moduli space in type II string theory, both at tree and
one-loop level, and in the process strengthen the case for the absence of perturbative
corrections beyond one-loop.  $\bullet$ Section 5 studies deformations of QK manifolds
with $d+1$ commuting isometries. We determine the allowed linear perturbations
which preserve superconformal invariance,  and find the deformed
twistor lines and contact twistor lines. These results will be applied to the
hypermultiplet moduli space of type II string theories in \cite{apsv3}. $\bullet$
In Appendix A, we spell out the $SU(2)$ action on the various multiplets at the
infinitesimal level. In Appendix B, we briefly discuss an alternative description
of the hypermultiplet moduli space using a different choice of contour, and show
that it is related to the one in Section 4 by a local symplectomorphism. In Appendix C
we generalize the superconformal quotient of Section 3.2 to the perturbed case,
and provide an independent check on the results of Section 5.2.

\section{Quaternionic-K\"ahler geometry and twistors}
\label{sec_caseisom}

In this section, we review the relation between quaternionic-K\"ahler (QK) manifolds
and hyperk\"ahler cones (HKC). This relation is one-to-one up to coverings
(Theorem (5.9) in \cite{MR1096180}), and can be established ``bottom up'', by
constructing the Swann bundle $\cS$ over the QK manifold ${\cal M}$,
or ``top down'', by performing the superconformal quotient of  $\cS$.
These two constructions are  summarized in Sections
2.1 and 2.2 following  \cite{deWit:2001dj,Neitzke:2007ke}.  In Section 2.3, we
recall the patchwork construction of the twistor space $\cZ_\cS$ of the HK space $\cS$
developed in our previous work \cite{Alexandrov:2008ds}. In Section 2.4  we derive the
restrictions on the transition functions imposed by superconformal invariance.
In Section 2.5, we study the reduction of the homogeneous
complex symplectic structure on $\cZ_\cS$
to a complex contact structure on $\cZ_\cM$.
The reader will find it helpful to refer to  Figure \ref{twifig} for the various coordinate
systems involved in these constructions.

\subsection{Bottom-up: from QK to HKC}
\label{Sect:2.1}

A quaternionic-K\"ahler manifold $\cM$ is a $4d$-dimensional
manifold with Riemannian metric $g_\cM$ and Levi-Civita connection
$\nabla$ whose holonomy group is contained in $USp(d)\times SU(2)$
\cite{MR664330}. $\cM$ admits a triplet of almost complex Hermitian
structures $\vec J$ (defined up to $SU(2)$ rotations) satisfying the algebra of the unit imaginary
quaternions. The quaternionic two-forms
$\vec\omega_\cM(X,Y)=g_\cM(\vec J X,Y)$ are covariantly closed with
respect to the $SU(2)$ part  $\vec p$ of the Levi-Civita connection,
and are proportional with a fixed coefficient $\nu$ to the curvature
of $\vec p$,
\be \label{ompp} {\rm d}\vec\omega_\cM + \vec p \times
\vec\omega_\cM = 0\ ,\qquad {\rm d}\vec p+ \half \vec p \times \vec
p = \frac12 \nu\, \vec \omega_\cM\, ,
\ee
where we use the notation
$(\vec a \times \vec b)_i=\epsilon_{ijk}a_j\wedge b_k.$ As a
consequence, the metric on $\cM$ is Einstein, with constant Ricci
scalar curvature $R=4d(d+2)\nu$. HK manifolds are degenerate limits
of QK manifolds, where $\nu=0$. We are mainly concerned in this work
with the case of negative curvature, $\nu<0$.

The Swann bundle $\cS$ associated to $\cM$ is the total space of a
$\IC^2$ bundle (more precisely $\IC^2/\IZ_2$ with the zero section
deleted) over $\cM$. It is a hyperk\"ahler manifold of dimension
$4(d+1)$  with an $SU(2)$ isometric action which rotates the complex
structures into each other, and a homothetic Killing vector. The
homothetic Killing vector ensures that the hyperk\"ahler manifold is
actually a cone, and the $SU(2)$ isometries guarantee that this is a
cone over a three-Sasakian space with $S^3$ fibres over the
quaternionic base $\cM$ \cite{Boyer:1998sf}. In physics
terminology, these properties follow from $\cN=2$ superconformal
invariance of the associated sigma model \cite{deWit:2001dj}. We denote by $\pi^{A'}$
the complex coordinates on the $\IC^2/\IZ^2$ fiber, ${\bar\pi}_{A'}\equiv
(\pi^{A'})^*$ their complex conjugate, and use the antisymmetric
tensor $\epsilon_{A'B'}$ to raise and lower the indices.\footnote{We
use conventions in which $\epsilon_{12}=1=-\epsilon_{21}$ and ${\bar
\pi}_{A'}={\bar \pi}^{B'}\epsilon_{B'A'}$.} The HK metric on $\cS$
is given by
\begin{equation}
\label{S-metric}
\de s^2_{\cal S}=|D\pi^{A'}|^2+\frac{\nu}{4}\, r^2 \de s^2_{{\cal M}}\, ,
\end{equation}
where $ \de s^2_{{\cal M}}$ is the QK metric on ${\cal M}$,
$r^2\equiv |\pi^1|^2+|\pi^2|^2=\pi^{A'}{\bar \pi}_{A'}$
is the squared norm on the fiber, and
\begin{equation}
D\pi^{A'}\equiv\de \pi^{A'}+p^{A'}{}_{B'}\,\pi^{B'} \, ,
\end{equation}
is the covariant differential of $\pi^{A'}$. The isometric $SU(2)$ action
on $\cS$ is given by the  infinitesimal transformations
\be\label{su2pi}
\delta \pi^{A'} = \frac{\I}{2} \eps_3  \pi^{A'}  + \eps_+   \bpi_{A'}\ ,\quad
 \delta \bpi_{A'} = -\frac{\I}{2} \eps_3  \bpi_{A'}  + \eps_-  \pi_{A'}\, .
\ee
In particular, the norm $r^2$ is $SU(2)$ invariant. The homothetic
Killing vector $r\partial_r=\pi^{A'}\pa_{\pi^{A'}}+\bpi_{A'}\pa_{\bpi_{A'}}$
corresponds to dilations of the fiber.  With respect to the complex
structure $J^3$ where $D\pi^{A'}$ are $(1,0)$ forms, the
\kahler form is
\begin{equation} \label{s-kahler-form}
\omega_\cS^3 = \I\left( \De \pi^{A'} \wedge \De \bar \pi_{A'}
+ \frac{\nu}{2}\,\pi_{A'}\bar \pi_{B'} \,\omega^{A' B'} _{\cM}\right)\, ,
\end{equation}
while the holomorphic symplectic form $\omega^+_\cS=-\half(\omega_\cS^1-\I \omega_\cS^2)$
is given by
\begin{equation}
\label{holsym}
\omega^+_\cS = \De \pi^{A'} \wedge \De \pi_{A'} + \frac{\nu}{2}\,\pi_{A'}
\pi_{B'} \omega^{A'B'}_\cM = \d\,\left( \pi_{A'} \De \pi^{A'} \right)\, .
\end{equation}
This construction directly defines the HKC, or Swann bundle ${\cal S}$,
given a QK manifold ${\cal M}$, see \cite{Neitzke:2007ke} for more details.

For many purposes, it is useful to decompose the construction above in two steps,
by first introducing the twistor space ${\cal Z}_{{\cal M}}$ \cite{MR664330}, a $\CP$ bundle over $\cM$,
and then obtaining the Swann bundle $\cS$ as a $\IC^\times$ bundle
over ${\cal Z}_{{\cal M}}$. The twistor space ${\cal Z}_{{\cal M}}$ over $\cM$ should not be
confused with the twistor space ${\cal Z}_{{\cal S}}$ of $\cS$ itself, to be introduced
in Section 2.3 below. ${\cal Z}_{\cal M}$  is a complex manifold with a canonical
K\"ahler-Einstein metric and a complex contact structure\footnote{Recall
that a complex contact form on a complex manifold of complex dimension $2d+1$
is a  holomorphic one-form $\hCX$, defined globally, such that $\hCX \wedge (\de \hCX)^d$
is a nowhere vanishing
holomorphic top form. A contact structure corresponds to the case where $\hCX$
is a local one-form defined up to multiplication by a nowhere
vanishing smooth function.}.
Introducing a complex
coordinate $z$ on $\CP$, the line element is given by
\begin{equation}\label{Z-metric}
\de s^2_{{\cal Z}_{{\cal M}}}=
\frac{|\de z+{\cal P}|^2}{(1+z{\bar z})^2}+\frac{\nu}{4}\de s_{\cal M}^2\, ,
\end{equation}
while the \kahler form on $\cZ_{\cal M}$ is given by
\be
\omega_{\cZ_{\cal M}}
= \I \left( \frac{(\de z+{\cal P}) \wedge (\de \bar z+\bar {\cal P})}{(1+z\bar z)^2}
+ \frac{\nu}{2}
\frac{(1-z\bz) \omega_\cM^3 - 2 \I z \omega^+_\cM + 2 \I \bar{z} \omega^-_\cM}
{1+z\bz} \right)\, ,
\ee
where $\omega^{\pm}_\cM=-\half(\omega^1_\cM\mp\I\omega^2_\cM)$,
$\omega^+_\cM=(\omega^-_\cM)^*$.
In these expressions, ${\cal P}$ stands for the ``projectivized connection'',
defined from the $SU(2)$ connection $p^{A'}{}_{B'}$ as
\begin{equation}
{\cal P}=p_+ -\I p_3\, z + p_-\, z^2\, ,
\end{equation}
where $p_+ \equiv p^1{}_2$, $p_3\equiv \I(p^1{}_1-p^2{}_2)$,
$p_-\equiv - p^2{}_1$, with $p_3$ real and
$(p_-)^*=p_+$.

The complex contact structure on $\cZ_{\cM}$ is induced from the Liouville
form $\CX$ on $\cS$,
\be
\label{cxcont}
 \CX \equiv  \pi_{A'} \De \pi^{A'} = \frac12 (\pi^2)^2\,
({\rm d}z+{\cal P}) \, ,
\ee
and, as apparent from the overall factor of $(\pi^2)^2$, is a
section\footnote{More precisely, $\CX$ is defined on $\cS$.
In \eqref{defhX} we define a contact one-form $\hCX$ proportional to $\CX$,
which does live on $\cZ_\cM$.}  of the $\cO(2)$ line bundle on $\CP$.
From the complex contact structure one may easily extract the
$SU(2)$ connection $\vec{p}$, and therefore the triplet of
quaternionic two-forms $\vec\omega_\cM$ via eq.\ \eqref{ompp}. Thus,
the knowledge of the complex structure and contact structure
on $\cZ_\cM$ is sufficient to reconstruct the \qk metric.

To construct the Swann bundle $\cS$ we introduce two more real coordinates $r$ and $\psi$
parametrizing the fiber of a $\IC^\times$ bundle over ${\cal Z}_{{\cal M}}$, with
metric\footnote{We follow the conventions
of \cite{Neitzke:2007ke}, but with a slightly different notation. E.g. the
coordinate $\psi$ here is denoted by $\phi$ in \cite{Neitzke:2007ke}. Also, in
\cite{Neitzke:2007ke}, the $SU(2)$ index in $\pi^{A'}$ was not lowered after
complex conjugation.}
\be
\label{dshkcr}
\de s^2_{\cS}=\de r^2 + r^2 \left[
(D\psi)^2 + \de s^2_{{\cal Z}_{\cM}} \right]\, ,
\ee
where
\begin{equation}
D\psi =\de \psi + \frac{\I}{2(1+z\bar z)} \left[ (z \de \bar z -
\bar z \de z) -\I (1-z\bar z)p_3 +2 z \, p_- -2 {\bar z}\, p_+
\right] \, .
\end{equation}
The metrics \eqref{S-metric} and \eqref{dshkcr} are identical, provided
the coordinates $r,\psi,z,\bz$ are related to $\pi^{A'},\bpi_{A'}$ via
\begin{equation}
r^2= |\pi^1|^2+|\pi^2|^2\ ,\quad e^{i\psi}=\sqrt{\pi^2/{\bar\pi}_2}\, ,\quad
z=\frac{\pi^1}{\pi^2}\ ,\quad \bz=\frac{\bpi_1}{\bpi_2}\, .
\end{equation}
or conversely
\begin{equation}
\label{hkcpi}
\begin{pmatrix}\pi^1\\ \pi^2 \end{pmatrix}=
\frac{r\, e^{\I\psi}}{\sqrt {1+z\bz}} \,
\begin{pmatrix} z \\ 1 \end{pmatrix}\, .
\end{equation}
For more details, we again refer the reader to \cite{Neitzke:2007ke}.

\subsection{Top down: from HKC to QK}

The characterizing property of an HKC is that there exists a function
$\chi$ on $\cS$, known as the hyperk\"ahler potential, such that the metric,
in local (real) coordinates $\phi^M, M=1,...,4(d+1)$,
satisfies~\cite{MR1096180,deWit:2001dj}
\begin{equation}\label{hkpot}
g_{MN}=D_M\partial_N \chi(\phi)\, .
\end{equation}
For any Hermitian complex structure, in adapted complex coordinates
$z^m, m=1,...,2(d+1)$,  \eqref{hkpot} implies that
\begin{equation}\label{K-pot}
g_{m{\bar n}}=\partial_m\partial_{\bar n}\chi(z,\bar z)\, ,
\qquad
D_m\partial_n \chi(z,\bar z)=0\, .
\end{equation}
In particular, $\chi$ provides a K\"ahler potential in any complex structure.
The dilation and $SU(2)$ symmetries are generated by the vector fields
\begin{equation}\label{su2KV}
\chi^M=g^{MN}\chi_N\, , \qquad {\vec k}^M={\vec J}^M{}_N\chi^N\, ,
\end{equation}
where $\chi_M\equiv\partial_M\chi$, $g^{MN}$ is the inverse HKC metric, and
${\vec J}$ is
a triplet of complex structures. The $SU(2)$ Killing vector fields are not
tri-holomorphic but rotate the complex structures into each other. It follows
from \eqref{hkpot} that the four vector fields $\chi^M$ and ${\vec k}^N$
satisfy
\begin{equation}
\label{holo-quaternion}
D_M\chi^N=\delta_M^N\ ,\qquad D_M{\vec k}^N={\vec J}^N{}_M\, .
\end{equation}
In particular, $\chi^m \pa_{z^m}$ is holomorphic. One can also express
the hyperk\"ahler potential in terms of the metric and the homothetic Killing vector
fields,
\begin{equation}\label{eqn-hkpot}
\chi = \frac{1}{2}\chi^Mg_{MN}\chi^N = \frac{1}{2}\chi^M\partial_M \chi\, ,
\end{equation}
consistent with \eqref{hkpot}. It is easy to check that this form of the hyperk\"ahler
potential is $SU(2)$ invariant. In the coordinates that appear in
the construction of the Swann bundle, the homothetic Killing vector is
generated by the vector field $\chi^M\partial_M = r\partial_r$, and so
\eqref{eqn-hkpot} yields
\be
\label{chir2}
\chi = r^2 = \pi^{A'} \bpi_{A'}\, .
\ee

One can descend from the HKC $\cS$ to the twistor space $\cZ_\cM$ by
performing a $U(1)$ K\"ahler quotient. For any choice of complex structure
${\vec n}\cdot {\vec J}$ with ${\vec n}$ a unit vector,  ${\vec n}\cdot {\vec k}$ is a
holomorphic Killing vector. The K\"ahler quotient of $\cZ_\cM$ with respect to
 ${\vec n}\cdot {\vec k}$ provides a K\"ahler manifold of real dimension $4d+2$,
 independent of the choice of $\vec n$, which is just
the twistor space ${\cal Z}_{{\cal M}}$. By Frobenius' theorem, one may choose a
set of independent complex coordinates $\lambda, u^i, i=1,...,2d+1$ adapted to the action of the
holomorphic vector field $\chi^M$,
\begin{equation}\label{Frob}
\chi^m(z)\partial_m=\partial_\lambda\vert_{u^i}\, .
\end{equation}
The K\"ahler potential on ${\cal Z}_{{\cal M}}$ is then determined from the
hyperk\"ahler potential $\chi$ by means of
\begin{equation}
\label{chikz}
\chi(\lambda,{\bar \lambda},u,\bar u)=e^{\lambda+{\bar \lambda}+
K_{{\cal Z}_\cM}(u,\bar u)}\, .
\end{equation}
Defining the $\cO(2)$-twisted holomorphic contact
form on $\cZ_\cM$
\be
\label{defhX}
\hat \CX \equiv e^{-2 \lambda} \CX =e^{\bar\lambda- \lambda+2\I\psi+K_{\cZ_\cM}}
\frac{  \de z+ \mathcal{P}}{2(1+z\bz)} \, ,
\ee
one may rewrite the metric on the fiber as  the modulus square of the contact form
\cite{deWit:1999fp},
\be
\label{modX2}
 \frac{|\de z+{\cal P}|^2}{(1+z{\bar z})^2} =4\, e^{-2K_{{\cal Z}_\cM}} |\hat \CX|^2\, .
\ee
Note that $\psi$ is not an independent coordinate, but rather will be determined
in terms of $\lambda,z,x^\mu$ in Eq. \eqref{psieq} below, in such a way that
$\bar\lambda- \lambda+2\I\psi$ is a function on $\cZ_\cM$ only.

The QK metric on $\cM$ can be computed from the holomorphic contact
form $\CX$  as indicated below \eqref{cxcont}, or by decomposing the
metric on the twistor space as in \eqref{Z-metric}, see
\cite{deWit:2001dj,Bergshoeff:2004nf} for more details. To express
the metric on $\cM$ in closed form, one needs to express the complex
coordinates $z^m$ on $\cS$ (or $u^i$ on $\cZ_\cM$) in terms of $4d$
independent real coordinates, corresponding to $\IR^+ \times SU(2)$
invariant combinations of $\phi^M$, and coordinates on the $\IC^2$
fiber $z,\bz,\lambda,\bar\lambda$. As we shall see shortly, this
problem is a QK analog of the problem of ``parametrizing the twistor
lines'' in HK geometry.

\subsection{Patchwork construction of twistor spaces of HK manifolds - a summary}

As explained e.g. in \cite{MR506229,Hitchin:1986ea, hitchin1987sde, Ivanov:1995cy,
Alexandrov:2008ds, Lindstrom:2008gs}, HK geometry is equivalent
to complex symplectic geometry on the twistor space, compatible
with the real structure. This, of course, also applies to the HKC metric
on the Swann bundle $\cS$, with suitable restrictions on the complex
symplectic structure to ensure the HK cone property.
In this subsection, we briefly review the twistorial description
of general HK manifolds $\cS$ following \cite{Alexandrov:2008ds},
before studying the implications of superconformal invariance in
Section \ref{sec-condsup}.

In contrast to the \qk case described in Section \ref{Sect:2.1}, the
twistor space $\cZ_\cS$ over a $4d+4$-dimensional HK
manifold\footnote{For obvious reasons, we deviate from the notations
of \cite{Alexandrov:2008ds} which considered $4d$ dimensional HK manifolds $\cM$.}
$\cS$ is a trivial product ${\cal Z}_{\cal S}={\cal S}\times \CP$.
Its structure was developed from a physics viewpoint in \cite{Hitchin:1986ea,Ivanov:1995cy},
and its relation to
projective superspace was recently further analysed in \cite{Lindstrom:2008gs}.

We denote by $\zeta$ a
complex coordinate on the projective line $\CP$ around the north pole $\zeta=0$.
$\cZ_\cS$ carries an integrable complex structure given by
\be\label{J-twistor}
J(\zeta,\bar\zeta)=\frac{1-\zeta\bar\zeta}{1+\zeta\bar\zeta}\, J^3
+\frac{\zeta+\bar\zeta}{1+\zeta\bar\zeta}\, J^2
+ \I \frac{\zeta-\bar\zeta}{1+\zeta\bar\zeta}\, J^1
\ee
on the base $\cS$ (where $J_i$ are the three complex structures on $\cS$)
and the standard complex structure on $\CP$. Moreover, in this complex structure,
$\cZ_\cS$ carries a holomorphic two-form (more accurately, a section of $\Lambda^2 T_F^*(2)$,
see  \cite{Hitchin:1986ea}) and a K\"ahler form given locally by
\be
\label{Omega}
\Omega(\zeta) = \omega^+_\cS -\I \zeta \omega^3_\cS + \zeta^2\ \omega^-_\cS \, ,
\ee
and
\be
\omega(\zeta,\bar\zeta)=\frac{1}{1+\zeta\bar\zeta}
\left[(1-\zeta\bar\zeta) \omega^3_\cS - 2\I\zeta \omega^+_\cS + 2\I \bar\zeta \omega^-_\cS  \right]\, ,
\label{omegzz}
\ee
where $\omega^{\pm}_\cS=-\half(\omega^1_\cS\mp\I\omega^2_\cS)$.
Note that, in contrast to the \qk case,
both of these forms are degenerate along the $\CP$ fiber direction $\de\zeta$.
The K\"ahler form $\omega$ coincides with $\omega^3_\cS$
at the north pole $\zeta=0$, and with $-\omega^3_\cS$ at the south pole $\zeta=\infty$.

The holomorphic two-form $\Omega$ however, while coinciding with
$\omega^+_\cS$ at the north pole, diverges with a second order pole at $\zeta=\infty$.
As explained in \cite{Alexandrov:2008ds}, it is useful to introduce a set of patches $\hU_i$,
$i=1,\dots,N$ on $\cZ_\cS$, which project to open disks\footnote{In principle, one should
introduce a local coordinate $\zeta\ui{i}$ on each connected disk; to avoid cluttering
we shall use a single coordinate $\zeta$ to parametrize all patches at once, with
each connected disk $\cU_i$ being centered at $\zeta=\zeta_i$.}
$\cU_i$ on $\CP$, and a local section
$\Omega\ui{i}$ which is regular on each patch. In order for the holomorphic
section $\Omega$ to be well defined, one must require that,  on the overlap
$\hU_i \cap \hU_j$,
\be
\label{omij}
\Omega^{[i]}= f_{ij} ^2(\zeta) \, \Omega^{[j]}  \,\quad \mod\, \de\zeta\, .
\ee
The factor $f_{ij}(\zeta)$ corresponds to the transition function of the $\cO(1)$ bundle
on $\CP$, and was discussed in detail in \cite{Alexandrov:2008ds}. In particular,
we recall that
\be
f_{ij} f_{jk} = f_{ik}\, , \quad f_{ii}=1\, ,\quad \overline{\tau(f_{ij}^2)} = f_{\bi \bj}^2\, ,
\ee
where $\tau$ is the antipodal map $[\tau(\nu)](\zeta)\equiv \nu(-1/\bar\zeta)$,
and $\bi$ labels the patch $\cU_{\bi}$ opposite to the patch $\cU_i$
under the involution $\tau$. Defining $\Omega\ui{0}=\Omega$ and using
$f_{0\infty}=\zeta$ one finds that
\be
\label{Ominfty}
\Omega\ui{\infty}\equiv \zeta^{-2}\, \Omega\ui{0}
= \omega^-_\cS -\I  \omega^3_\cS \zeta^{-1}
+  \omega^+_\cS \zeta^{-2}
\ee
is regular at the south pole $\zeta=\infty$.

Now, we may choose the covering $\hU_i$ such that, on each patch, the
holomorphic section $\Omega\ui{i}$ takes the Darboux form
\be
\label{darboux}
\Omega^{[i]}=\de\mui{i}_I\wedge \de \nui{i}^I \, ,
\ee
where $(\nui{i}^I,\mui{i}_I,\zeta)$ is a local
complex coordinate system on $\cZ_\cS$, regular throughout
the patch $\hU_i$  (here $I$ runs over $d+1$ values,
which we shall denote $\flat,0,\dots,d-1$).
Eq. \eqref{omij} implies that on the overlap
of two patches, $(\nui{i}^I,\mui{i}_I)$ and $(\nui{j}^I,\mui{j}_I)$
must be related by a complex ($\cO(2)$-twisted)
symplectomorphism. This is conveniently encoded by a
generating function $\Sij{ij}$ of the initial "position"
and final "momentum" coordinates, such that
\be
\nui{j}^I = \p_{\mui{j}_I}\Sij{ij}(\nui{i},\mui{j},\zeta) \, , \qquad
\mui{i}_I =f_{ij}^2\,\p_{\nui{i}^I}\Sij{ij}(\nui{i},\mui{j},\zeta) \, .
\label{cantr}
\ee
To check \eqref{omij}, one may use the identity
\begin{equation}
\label{diff-S}
{\rm d}S^{[ij]}=\nu^I_{[j]}{\rm d}\mu^{[j]}_I+f_{ij}^{-2}\mu_I^{[i]}{\rm d}
\nu^I_{[i]}\qquad {\mbox {mod d$\zeta$}}\, .
\end{equation}
The transition functions $S^{[ij]}$ are restricted by consistency
conditions which ensure that the symplectomorphisms compose properly
(see  \cite{Alexandrov:2008ds} for more details). As a result, the holomorphic
symplectic structure on the twistor space $\cZ_\cS$ is
entirely specified by $N-1$ freely chosen functions
$\Sij{0i}(\nui{0},\mui{i},\zeta)$. In order to ensure the reality
of the resulting metric, it is also necessary to require that the
sections $\nui{i}^I$, $\mui{i}_I$ transform under the real structure as
\be
\overline{\tau\bigl(\nui{i}^I\bigr)}=-\nui{\bi}^I\, ,
\qquad
\overline{\tau\bigl(\mui{i}_I\bigr)}=-\mui{\bi}_I \, .
\label{realcon}
\ee
The condition \eqref{realcon} requires that
the functions $\Sij{i\bi}$ are related by complex conjugation to their Legendre
transform \cite{Alexandrov:2008ds}.

For a suitably generic choice
of such transition functions, it is a general property of
HK manifolds that the space of solutions of \eqref{cantr}
has dimension $4d+4$, i.e. all $\nui{i}^I$, $\mui{i}_I$
can be expressed as infinite Taylor series around  $\zeta=\zeta_i$
whose coefficients are all functions of $4d+4$ parameters. The
moduli space of solutions is isomorphic to the HK base $\cS$,
and  the map $\zeta \mapsto (\nui{i}^I, \mui{i}_I)$ defines the
``twistor lines'', i.e. realizes the $\CP$ fiber over any point in $\cS$
as a rational curve  in $\cZ_\cS$.

Having found the twistor lines, the geometry of $\cS$ can be
computed by Taylor expanding the holomorphic section $\Omega$
around any point $\zeta\in \CP$. When $\cS$ is a HKC, as we discuss
further in the next section, all points of $\CP$ are equivalent, and we
can therefore expand around $\zeta=0$. Since $\Omega$ is a global
section of $\cO(2)$, the Taylor expansion stops at quadratic order,
\be
\Omega^{[0]} =  \de w_I \wedge \d v^I -\I\, \omega^3_\cS \ze
+\de\bw_I \wedge \de\bv^I \zeta^2\, .
\ee
where $v^I,w_I$ are the complex coordinates in the complex structure
$J^3=J(0,0)$,
\be
v^I =\nui{0}^I(\zeta=0)\, ,\quad
w_I=\mui{0}_I (\zeta=0)\, ,
\ee
Knowing the complex coordinates and the \kahler form $\omega^3_\cS$, it is
straightforward to obtain the metric and a \kahler potential. When $\cS$ is a HKC,
as we will discuss in the next section
it is always possible to choose the \kahler potential such that it is invariant
under $SU(2)$, and therefore equal to the \hk potential $\chi$.

 \subsection{Conditions for superconformal invariance \label{sec-condsup}}

We now discuss the implications of superconformal invariance for the general
construction of the twistor space $\cZ_\cS$ of a HK manifold $\cS$. We recall
from Section 2.2 that superconformal invariance requires the existence of
a homothetic Killing vector and an $SU(2)$ group of Killing vectors
that rotates the complex structures and commutes with the dilations.


As follows from the first equation in \eqref{holo-quaternion}, the
dilations rescale the hyperk\"ahler cone
metric and leave the complex structures invariant.
We normalize the action of the dilations such that the metric has weight 2,
\begin{equation}
g' = \Lambda^2 \, g \, ,\qquad  
{\vec J}'={\vec J}\, .
\end{equation}
This implies that all the two-forms $\vec{\omega}_\cS$ on the Swann bundle $\cS$
scale with weight two. The action of the dilations can be extended to the twistor
space $\cZ_\cS$ by assigning a scaling weight zero to $\zeta$. In this
way, the holomorphic two-form $\Omega$ from \eqref{Omega} transforms
uniformly throughout the $\zeta$ plane,
\begin{equation}\label{scaling-Om}
{\Omega'}\ui{i} = \Lambda^2 \,\Omega\ui{i} \, .
\end{equation}
The local Darboux coordinates $\nui{i}$ and $\mui{i}$ must transform in such a way
that \eqref{scaling-Om} is obeyed, so we postulate\footnote{One may also consider giving
a different scaling weight $n_I$ for each conjugate pair $(\nu^I,\mu_I)$.  The generalization
of the following discussion is immediate.}
\begin{equation}\label{dil-twistor}
{\nu'}^{I}_{[i]}=\Lambda^{2n} \nu^I_{[i]}\, ,
\qquad
{\mu'}_I^{[i]}=\Lambda^{(2-2n)}\mu_I^{[i]}\, ,
\end{equation}
for some constant $n$. This is a symmetry of the gluing conditions
\eqref{cantr} provided the generating functions  are homogenous functions of degree
one when $\nu$ and $\mu$ are scaled with degree $n$ and $1-n$ respectively,
\begin{equation}\label{scaling-S}
S^{[ij]}(\Lambda ^{2n} \nu_{[i]}, \Lambda^{(2-2n)}\mu^{[j]},\zeta)=\Lambda^2\,
S^{[ij]}(\nu_{[i]},\mu^{[j]},\zeta)\, .
\end{equation}

We now turn to the $SU(2)$ action. In order for the
complex structure $J(\zeta,\bar\zeta)$ given in \eqref{J-twistor}
to be invariant, one should compensate the rotation of $\vec J$ by
a rotation on the $\CP$ fiber. Thus the fiber coordinate $\zeta$ must transform as
\be
\label{su2ze}
\zeta' = \frac{ \alpha \zeta + \beta}{-\bar\beta \zeta + \bar\alpha}\, ,
\quad
\quad \alpha\bar\alpha+\beta\bar\beta = 1\, .
\ee
Under this transformation, $\Omega$ should transform as a $\cO(2)$ section,
\be
\label{su2om}
{\Omega'}\ui{0}(\zeta)=
\left( -\bar\beta \zeta + \bar\alpha \right)^{2}\,
\Omega\ui{0}\left( \zeta' \right) \, .
\ee
Here, we have written the action in the patch $\cU_0$ around the north pole of $\CP$,
parametrized by the local coordinate $\zeta=\zeta\ui{0}$. The action in the patch $\cU_i$
can be obtained by replacing $\zeta\to\zeta\ui{i}$, $\Omega\ui{0}\to\Omega\ui{i}$. If
we continue to use $\zeta$ as a coordinate in $\cU_i$, then from \eqref{omij} the
transformation of $\Omega\ui{i}$ becomes
\begin{equation}
\label{su2omi}
{\Omega'}\ui{i}(\zeta)=\[\frac{f_{i0}(\zeta)}{f_{i0}(\zeta')}\]^{2}
\(-\bar\beta\zeta+\bar\alpha\)^{2}  \Omega\ui{i}
\left( \zeta' \right) \, .
\end{equation}
In order to ensure that $\Omega$ transforms as \eqref{su2omi} in every patch, we postulate that
the local Darboux coordinates $\nui{i}^I$, $\mui{i}_I$ transform
locally as $\cO(2n)$ and $\cO(2-2n)$ sections

\be
\label{su2diff}
\begin{split}
\nu_{[i]}^{'I}(\zeta) =&\[\frac{f_{i0}(\zeta)}{f_{i0}(\zeta')} \(-\bar\beta\zeta+\bar\alpha\) \]^{2n}
\nu_{[i]}^{I}(\zeta')\, ,\\
\mu^{'[i]}_{I}(\zeta) =&\[\frac{f_{i0}(\zeta)}{f_{i0}(\zeta')} \(-\bar\beta\zeta+\bar\alpha\)\]^{2-2n}
\mu^{[i]}_{I}(\zeta')\, .
\end{split}
\ee
This is a symmetry of the gluing equations \eqref{cantr} provided
\begin{equation}
\label{transf-S}
S^{[ij]}\(\nui{i}^{'}(\zeta),\mu^{'[j]}(\zeta),\zeta\)=
\[\frac{f_{j0}(\zeta)}{f_{j0}(\zeta')}\]^2
(-\bar\beta\zeta+\bar\alpha)^2 \, S^{[ij]}\Big(\nui{i}(\zeta'),\mu^{[j]}(\zeta'),\zeta'\Big)\, .
\end{equation}
Using the homogeneity property \eqref{scaling-S}, this translates into
\begin{equation}
S^{[ij]}\[\frac{f^{2n}_{ij}(\zeta)}{f^{2n}_{ij}(\zeta')}\nu_{[i]}(\zeta'),\mu^{[j]}(\zeta'),\zeta\]
=S^{[ij]}\[\nu_{[i]}(\zeta'),\mu^{[j]}(\zeta'),\zeta'\]\, .
\end{equation}
This equation fixes the $\zeta$ dependence to be of the form
\be
\label{o2o2s}
\Sij{ij}(\nui{i},\mui{j},\zeta)=\hSij{ij}\(f_{ij}^{-2n}\, \nui{i},\mui{j}\)\, .
\ee
In particular, note that the special case where $\nui{i}^I$ and $\mui{i}_I$ are
global sections of  $\cO(2n)$ and $\cO(2-2n)$,
\be
\label{o2o2sg}
\Sij{ij}(\nui{i},\mui{j},\zeta)=f_{ij}^{-2n}\nui{i}^I\mui{j}_I \, ,
\ee
solves the conditions of superconformal invariance. In addition, as in \cite{Alexandrov:2008ds},
one must impose the reality conditions
\be\label{Sreal}
\overline{ \tau\(\Sij{ij}(\nui{i}, \mui{j}, \zeta\ui{i}) \)} =
\Sij{\bi \bj}(\nui{\bi}, \mui{\bj}, \zeta\ui{\bi}) \, .
\ee

Thus, we conclude that superconformal invariance is guaranteed provided
the generating functions $\Sij{ij}(\nui{i},\mui{j},\zeta)$ are functions of $f_{ij}^{-2n} \nui{i}$
and $\mui{j}$, without explicit dependence on $\zeta$, homogeneous of degree 1 when
their first and second arguments
are scaled with weight $n$ and $1-n$, respectively,
and satisfying the reality condition \eqref{Sreal}.

\subsubsection*{Anomalous $\cO(0)$ multiplets}

In fact, the above conditions are sufficient but not strictly speaking necessary. Indeed,
we have assumed that the Darboux coordinates are adapted to the superconformal
action, in the sense that dilations and $SU(2)$ act canonically as in \eqref{dil-twistor} and
\eqref{su2diff}, respectively. Clearly, a local gauge transformation depending on
$\zeta$ only would not affect the existence of an isometric $SU(2)$ action, but would just make
it look more complicated. More importantly, when $n=1$ (or equivalently $n=0$, after exchanging
$\mu_I$ with $\nu^I$), it is possible that $\mu_I$ transforms anomalously under dilations,
namely
\begin{equation}
\label{mu-anom0}
\mu_I^{'[i]}=\mu_I^{[i]}-c_I^{[i]} \log \Lambda^2\, ,
\end{equation}
for some constants $c_I^{[i]}$, which we shall refer to as ``anomalous dimensions".
This anomalous transformation may be
generated from the standard transformation \eqref{dil-twistor} with $n=0$
by a local symplectomorphism generated by
\be
T\ui{i}= \tmui{i}_I \nui{i}^I - c\ui{i}_I \nui{i}^I \log\nui{i}^\flat\, ,
\ee
where $\nui{i}^\flat$ is any one of the $\nui{i}^I$.
This however need not be a regular gauge transformation in the patch $\cU_i$, and so the geometry
will in general depend non-trivially on $c_I^{[i]}$. After this local symplectomorphism,
the generating functions $S\ui{ij}$ are now of the form\footnote{The $\hSij{ij}$
appearing in this equation differs from the one in \eqref{o2o2s}, the relation
between the two being transcendental in general.}
\be
\label{o2o2sq}
S\ui{ij} = f_{ij}^{-2}\, \left[ \hSij{ij} \left( \nui{i},\mui{j}_I
+ c_I^{[j]} \log( f_{ij}^{-2} \nui{i}^\flat ) \right)
- c_I^{[i]} \nui{i}^I  \log \nui{i}^\flat  \right] \, ,
\ee
where $\hSij{ij}$ is a homogeneous function of degree one in its first argument.
In particular,  $S\ui{ij}$ satisfy a ``quasi-homogeneity condition''
\begin{equation}
S^{[ij]}\left(\Lambda^2 \nu^I_{[i]}, \mu_I^{[j]} - c_I^{[j]}\log \Lambda^2,\zeta\right)
=\Lambda^2 \left[ S^{[ij]} \(\nu^I_{[i]},\mu_I^{[j]},\zeta\) - f_{ij}^{-2}
c_I^{[i]}\nu_{[i]}^I  \,\log \Lambda^2\right] \, .
\end{equation}
Such generating functions are consistent with $SU(2)$ invariance and dilations
provided $\mu_I^{[i]}$ transforms in the same way as $- c_I^{[i]} \log \nui{i}^\flat$,
namely as \eqref{mu-anom0} under dilations and
\be
\label{mu-anom}
\mu^{'[i]}_{I}(\zeta) =\mu^{[i]}_{I}(\zeta')
- 2 c_I^{[i]} \log  \[\frac{f_{i0}(\zeta)}{f_{i0}(\zeta')} \(-\bar\beta\zeta+\bar\alpha\)\]
\ee
under rotations.
Anomalous transformations play an important role, e.g., in describing the one-loop
correction to the hypermultiplet metric in Section \ref{sec-oneloop}.

Note that the constants $\ci{i}_I$ are not arbitrary. Firstly, they must
satisfy the reality conditions $(\ci{i}_I)^*=-\ci{\bar \imath}_I$.
Besides, they are also subject to additional consistency constraints, which follow from
from the requirement that the open contours around the logarithmic branch cuts in $\zeta$ plane
(as discussed in \cite{Alexandrov:2008ds}) combine consistently into closed contours.
This requires in particular that the anomalous dimensions associated with the patches
containing the zeros of $\nui{i}^\flat$ are real.
In this paper we assume that $\nui{i}^\flat$ has always two first order zeros $\ztpm$
in the patches $\cU_\pm$ related by the antipodal map, and therefore
demand that $\ci{+}_I=-\ci{-}_I$ are real constants.
For a similar reason, we impose the same condition on $\ci{0}_I=-\ci{\infty}_I$
(see footnote \ref{foot_realcond}).

For later reference, we give the action of the symplectomorphism generated by
\eqref{o2o2sq},
\be
\label{symph}
\begin{split}
\nui{j}^I =& f_{ij}^{-2}\, \pa_{\mui{j}_I} \hSij{ij}\, ,
\\
\mui{i}_I =&  \pa_{\nui{i}^I} \hSij{ij} - \ci{i}_I \log\nui{i}^\flat + \delta_I^\flat
\left(  \frac{1}{\nui{i}^\flat}  \ci{j}_J \pa_{\mui{j}_J} \hSij{ij}
- \frac{\nui{i}^J}{\nui{i}^\flat} \ci{i}_J
\right)\, .
\end{split}
\ee

\subsubsection*{From $\CP_\zeta \times \IC^2_{\pi}$ to $\CP_\varpi$}

We close this discussion of $SU(2)$ transformations with an important observation,
which will be instrumental for understanding
the relation between the twistor spaces $\cZ_\cS$ and $\cZ_\cM$.
Notice that the isometric $SU(2)$ action on $\cS$ corresponds to an $SU(2)$ action on
the fiber coordinates $\pi^{A'}, \bpi_{A'}$  \eqref{su2pi}, at a fixed position on the
QK base $\cM$. Thus, any local $\cO(2n)$ section $\nui{i}$, viewed as a function
of $(\zeta, \pi^{A'}, \bpi_{A'})$ and $x^\mu$, satisfies differential equations
\be
\begin{split}
\left( \pa_\zeta+\bpi_1\pa_{\pi^2}  - \bpi_2\pa_{\pi^1}  \right) (f_{i0}^{-2n} \nui{i}) &= 0 \, , \\
\left( 2\zeta\pa_\zeta-2n+ \pi^1\pa_{\pi^1} +\pi^2\pa_{\pi^2}
-\bpi_1\pa_{\bpi_1}-\bpi_2\pa_{\bpi_2}   \right) (f_{i0}^{-2n} \nui{i}) &= 0 \, , \\
\left( \zeta^2 \pa_\zeta - 2n \zeta +  \pi^1\pa_{\bpi_2} - \pi^2\pa_{\bpi_1} \right)
(f_{i0}^{-2n} \nui{i}) &= 0 \, .
\end{split}
\ee
It follows that there exists a function $\tnui{i}( \varpi, x^\mu)$ of the coordinates $x^\mu$
on $\cM$ and of the ratio
\be
\varpi \equiv  \frac{\bpi_2 \zeta +\pi^1}{-\bpi_1 \zeta +\pi^2} \, ,
\label{varpidef}
\ee
such that
\be
\nui{i}(\zeta, \pi^{A'}, \bpi_{A'},x^\mu) =
f_{i0}^{2n} (\pi^2-\zeta\bpi_1)^{2n}\, \tnui{i}( \varpi, x^\mu)\, .
\label{varpinu}
\ee
For anomalous $\cO(0)$ sections, the same argument guarantees the existence of
a function $\tmui{i}( \varpi, x^\mu)$ such that
\be
\mui{i}(\zeta, \pi^{A'}, \bpi_{A'},x^\mu) = \tmui{i}( \varpi, x^\mu) - \ci{i}_I \log
\left( f_{i0}^2 (\pi^2-\zeta\bpi_1)^2 \right)\,  .
\label{varpimu}
\ee
Moreover, under the action of the antipodal map,
\be
\overline{\tau(\varpi)} = -1/\varpi\, ,\qquad
\overline{\tau(\tnui{i})} = - \tnui{\bar \imath}/ \varpi^{2n}\, ,\qquad
\overline{\tau(\tmui{i})} = - \tmui{\bar \imath} -2 \ci{\bar \imath} \log\varpi\, .
\ee

The coordinate $\varpi$ can be viewed as a coordinate on the $\CP$ fiber of
the twistor space $\cZ_\cM$. After an appropriate $SU(2)$ rotation on the $\IC^2/\IZ_2$ fiber,
we can always assume that the zero and the pole of \eqref{varpidef} occur at the zeros $\ztpm$ of
the singled-out section $\nu^\flat$,
\be
\label{zztpm}
\varpi = -\frac{1}{\bz}\, \frac{\zeta-\ztp}{\zeta-\ztm}\, ,
\qquad \zeta_+=-\frac{\pi^1}{\bpi_2}\ , \qquad \zeta_-=\frac{\pi^2}{\bpi_1} \, ,\qquad
z=\frac{\pi^1}{\pi^2}\ .
\ee
In particular, the points $(0,\ztp,\ztm,\infty)$ in the $\zeta$ plane are mapped to
$(z, 0, \infty, -1/\bar z)$ in the $\varpi$ plane, respectively.
Since $\nui{i}$ is assumed to be  regular at $\zeta=\zeta_i$,  $\tnui{i}( \varpi, x^\mu)$
is regular at the point $\varpi_i\equiv \varpi(\zeta_i)$, except
when $i=-$ where the factors $(\pi^2-\zeta\bpi_1)$ introduce extra singularities at $\varpi=\infty$.

In the next subsection, we elaborate on these observations
and relate the symplectic and contact structures on the twistor spaces  $\cZ_\cS$
and $\cZ_\cM$.

\subsection{Homogeneous symplectic vs. contact geometry\label{seccontact}}

Having understood the constraints of superconformal invariance on the transition functions
$\Sij{ij}$, we now explain how the homogeneous symplectic structure on $\cZ_\cS$
descends to a contact structure on $\cZ_\cM$. For definiteness, and since this is the
case of most physical interest, we restrict to twistor spaces with $n=1$ from this section
onward.

\subsubsection*{From homogeneous symplectic to contact}

Let us return to \eqref{diff-S}: the term proportional to $\de\zeta$,
usually unspecified in HK geometry, can be computed explicitly in the case of HKC
manifolds. Indeed,  by
differentiating the factors $f_{ij}$ appearing explicitly in \eqref{o2o2sq},
and integrating by parts, one obtains
\begin{equation}
\label{diff-S3}
\begin{split}
{\rm d}\left(S^{[ij]} -f_{ij}^{-2} \mu_I^{[i]} \nu^I_{[i]} \right)
&=\nu^I_{[j]}{\rm d}\mu^{[j]}_I-f_{ij}^{-2}  \nu^I_{[i]} {\rm d} \mu_I^{[i]} \\
&+ \left(  \hSij{ij} + \ci{j}_I \pa_{\mui{j}_I} \hSij{ij} - c_I^{[i]} \nui{i}^I  \log \nui{i}^\flat
-  \mu_I^{[i]} \nu^I_{[i]} \right) \, \de (f_{ij}^{-2})\, .
\end{split}
\end{equation}
Re-expressing $ \mu_I^{[i]}$ using \eqref{symph} and using the homogeneity property of $\hSij{ij}$,
one concludes that
\be
\label{glue1}
\CX\di{i} = f_{ij}^{2} \, \CX\di{j}\, ,
\ee
where $\CX\di{i}$ is the $\cO(2)$-valued complex Liouville form
\be
\label{contact1}
\CX\di{i} = \nu^I_{[i]} {\rm d} \mu_I^{[i]} + \ci{i}_I \de \nu^I_{[i]}\, ,
\ee
satisfying the reality condition $\overline{\tau(\CX\di{i})}=\CX\di{\bar i}$.

Let us now introduce the dilation-invariant $\cO(0)$ sections\footnote{Our notations
are related to the ones in \cite{Neitzke:2007ke} via
$\xi^{\Lambda,{\rm NPV}}=\xii{0}^\Lambda, \,\txi_\Lambda^{\rm NPV}=-2\I \txii{0}_\Lambda, \,
\alpha^{\rm NPV}=4\I \txii{0}_\flat+ 2\I \xii{0}^\Lambda \txii{0}_\Lambda$, where the quantities
on the r.h.s. are evaluated at $\zeta=0, \varpi=z$.}
\be
\label{defxixit}
\xii{i}^I \equiv \frac{\nui{i}^I}{\nui{i}^\flat} \, ,\qquad
\txii{i}_I \equiv \mui{i}_I + \ci{i}_I \log \nui{i}^\flat\, ,
\ee
where we have singled out one coordinate\footnote{Eq. \eqref{defxixit} is
singular at the zeros of $\nui{i}^\flat$, and one should in principle single out
a second coordinate $\nui{i}^0$ to cover these patches. Rather than  doing
so, we allow for poles and logarithmic cuts in
$\xii{i}^\Lambda$ and $\txii{i}_I$, as in \eqref{behap} below, in effect trivializing
the $\cO(2)$ bundle over $\CP_\varpi$.}
$\nui{i}^\flat$, and denoted by
$\nui{i}^\Lambda$ the remaining $d$ coordinates. In this trivialization,
the Liouville form $\CX\ui{i}$  leads to a contact form $\hat\CX\ui{i}$,
\be
\label{con1fo}
\CX\ui{i} = \nui{i}^\flat \hat\CX\ui{i} \, ,\qquad
\hat\CX\ui{i}  \equiv  \de \txii{i}_\flat
+  \xii{i}^\Lambda \, \de \txii{i}_\Lambda + \ci{i}_\Lambda \de\xii{i}^\Lambda \, .
\ee
The term linear in $\ci{i}_I$
may be reabsorbed by defining
\be
\label{defaa}
\ai{i}\equiv \txii{i}_\flat + \ci{i}_I \xii{i}^I\, ,
\qquad
\hat\CX\ui{i}  =  \de \ai{i}
+  \xii{i}^\Lambda \, \de \txii{i}_\Lambda \, .
\ee
The gluing condition \eqref{glue1} becomes
\be
\label{glue2}
\hat \CX\ui{i} =  \hat f_{ij}^{2} \, \hat\CX\ui{j} \, ,\qquad \hat f_{ij}^{2}
\equiv f_{ij}^{2} \, \nui{j}^\flat / \nui{i}^\flat \, .
\ee
while $\hat\CX$ satisfies the reality condition $\overline{\tau(\hat\CX\ui{i})}=-\hat\CX\ui{\bar i}$.

According to the remark at the end of the previous section, $\xii{i}^\Lambda,
\txii{i}_I$ and $\hat f_{ij}^{2}$ are all functions of the $\CP$ coordinate $\varpi$
defined in \eqref{varpidef} and  of the coordinates $x^\mu$ on $\cM$,
\be
\xii{i}^\Lambda = \frac{\tnui{i}^\Lambda(x^\mu,\varpi)}{\tnui{i}^\flat(x^\mu,\varpi)}\,  ,\quad
\txii{i}_I = \tmui{i}_I (x^\mu,\varpi) + \ci{i}_I \log \tnui{i}^\flat (x^\mu,\varpi)\, ,\quad
\hat f_{ij}^{2}  = \frac{\tnui{j}^\flat(x^\mu,\varpi)}{\tnui{i}^\flat(x^\mu,\varpi)}\ .
\label{transtnu}
\ee
Thus, the sections $(\xii{i}^\Lambda,\txii{i}_I)$
provide local complex Darboux coordinates for the
complex contact structure $\hat\CX$ on $\cZ_\cM$. They satisfy the reality conditions
\be
\label{rexixit}
\overline{\tau(\xii{i}^\Lambda)} = \xii{\bar \imath}^\Lambda\, ,\qquad
\overline{\tau(\txii{i}_I)} = -\txii{\bi}_I  + \I \pi c_I^{[\bi]}\, .
\ee
On the overlap of two patches, the Darboux coordinates are related
by contact transformations following directly from \eqref{symph},
\bea\label{xi:trafo}
\xi\di{j}^\Lambda &=&  \hat f_{ij}^{-2} \, \pa_{\txii{j}_\Lambda} \hSij{ij}\, ,\qquad
 \txii{i}_\Lambda = \pa_{\xii{i}^\Lambda} \hSij{ij}\, ,\\
  \txii{i}_\flat &=& \hSij{ij} - \xii{i}^\Lambda \pa_{\xii{i}^\Lambda} \hSij{ij}
 + \ci{j}_I \pa_{\txii{j}_I} \hSij{ij} - \ci{i}_I \,\xii{i}^I\, ,\nn
\eea
where $\hSij{ij}$ is a general function of $\xii{i}^\Lambda$
and $\txii{j}_I+ \ci{j}_I \log \hat f_{ij}^{-2}$,
related to the original quasi-homogeneous generating function $\Sij{ij}$ via
\be
\Sij{ij}(\nui{i}^I,\mui{j}_I,\zeta)=f_{ij}^{-2}\[ \nui{i}^\flat \,\hSij{ij}\left( \xii{i}^\Lambda,
\txii{j}_I + \ci{j}_I \log( \hat f_{ij}^{-2} ) \right)
- \ci{i}_I \xii{i}^I  \nui{i}^\flat  \log \nui{i}^\flat \]\, ,
\ee
and it is understood that $\xii{i}^\flat=1$.
In particular, note that the transition functions $\hat f_{ij}^2$ are holomorphic
functions on $\cZ_\cM$ given by
\be
\label{fhflat}
\hat f_{ij}^{2} =
\pa_{\txii{j}_\flat} \hSij{ij}\, ,
\ee
and are equal to one if and only if $\nu^\flat$ is a global $\cO(2)$ section.

\subsubsection*{Recovering the metric from the contact twistor lines}

The functions $\xii{i}^\Lambda( \varpi, x^\mu)$ and $\txii{i}_I( \varpi, x^\mu)$ specify the
twistor fiber over each point in $\cM$, and are the analogs of the twistor lines
$\nui{i}^I(\zeta), \mui{i}_I(\zeta)$ on $\cS$. The knowledge of these
``contact twistor lines'' allows to reconstruct the K\"ahler-Einstein metric on $\cZ_\cM$
and the \qk metric on $\cM$,
in the following manner.


First,  identifying $\CX\ui{0}=\CX$ in \eqref{cxcont}
and using \eqref{varpinu}, the holomorphic contact
form in any patch $\cU_i$ may be written as
\be
\hat\CX\ui{i} =\frac{\CX\ui{0}}{f_{0i}^2\nui{i}^\flat}
=\frac12 \left(\frac{\pi^2}{\pi^2-\zeta \bpi_1}\right)^2
\frac{\de z+\mathcal{P}}{\tnui{i}^\flat(x^\mu,\varpi)}\, ,
\label{cXnu}
\ee
Since $\hat\CX\ui{i}$ depends on the fiber coordinates $\pi^{A'}, \bpi_{A'},\zeta$ only
through the combination $\varpi$, we may set $\zeta=0,\ \varpi=z$ in this expression,
and obtain
\be
\frac{\de \varpi}{\varpi} +\frac{p_+}{\varpi} - \I\, p_3\,  + p_- \varpi = \frac12
 e^{-\Phi_{[i]}}\,
\hat\CX\ui{i}
\label{cXnu0}
\ee
where we define the ``contact potential",
\be
\label{phipmnu}
e^{-\Phi_{[i]}(x^\mu,\varpi)}  \equiv 4 \, \tnui{i}^{\flat}(x^\mu,\varpi) / \varpi \ ,\qquad
\overline{\tau(\Phi_{[i]})} = \Phi_{[\bi]}\ .
\ee
Applying \eqref{modX2}, we conclude that the \kahler potential
on $\cZ_\cM$ is given by
\be
\label{Knuflat}
K_{{\cal Z}_\cM} = \log\frac{4(1+z \bz)}{|z|}
+ \Re\Phi_{[i]} (x^\mu,z)
+ \log | \hat f_{0i} |^2\, .
\ee
Since $\hat f_{0i}$ is a holomorphic function, the last term in
\eqref{Knuflat} can be absorbed by a \kahler transformation, leading to
a \kahler potential valid in the patch $\cU_i$. In order to derive the
metric on $\cZ$, one could therefore express $z, \bz$ and $\Phi_{[i]}$ in
terms of the complex coordinates $(\xii{i}^\Lambda, \txii{i}_\Lambda, \ai{i})$
in $\hat\cU_i$. For the purpose of computing the QK metric on $\cM$,
this step is unnecessary and it suffices to study the contact twistor lines,
as we show below. For later reference, we record the \hk potential
which follows from \eqref{chikz} using $v^\flat=e^{2\lambda}$,
\be
\label{chinuflat}
\chi = 4|\vf\hat f_{0i}^2|\,\frac{ 1+z \bz}{|z | }\,
e^{\Re[\Phi_{[i]} (x^\mu,z) ]}\, .
\ee
By comparing \eqref{defhX} with \eqref{cXnu0} for $i=0$, we can also relate
the coordinate $\psi$ in \eqref{hkcpi} to the coordinates $\vf, z, x^\mu$ on $\cS$,
\be
\label{psieq}
e^{2\I\psi}= \sqrt{\frac{\vf\bz}{\bvf z}} \,e^{\,\I \Im[ \Phi_{[0]}(x^\mu,z)]}\ .
\ee


We now restrict our attention to the patches $\hat\cU_+$ and $\hat\cU_-$
around $\varpi=0$ and $\varpi=\infty$, respectively,
corresponding to $\zeta=\zeta_+$ and $\zeta=\zeta_-$.
Using \eqref{defxixit}, \eqref{varpinu}, \eqref{varpimu} and
$f_{0+}\sim\zeta-\zeta_-$, we find that the
contact twistor lines behave near $\varpi=0$ as
\be
\label{behap}
\begin{split}
\xii{+}^\Lambda &= \xii{+}^{\Lambda,-1} \varpi^{-1} + \xii{+}^{\Lambda,0}
+ \xii{+}^{\Lambda,1}  \varpi + \cO(\varpi^2)\, ,  \qquad\\
\txii{+}_\Lambda &=  \ci{+}_\Lambda \log\varpi + \txi^{[+]}_{\Lambda,0}
+ \txi^{[+]}_{\Lambda,1} \varpi+  \cO(\varpi^2)\, , \\
\ai{+} &= \ci{+}_\flat  \log\varpi + \ci{+}_\Lambda \xii{+}^{\Lambda,-1} \varpi^{-1}
+ \alpha^{[+]}_0+ \alpha^{[+]}_1 \varpi+  \cO(\varpi^2)\, .
\end{split}
\ee
Similarly, near $\varpi=\infty$,
\be
\label{beham}
\begin{split}
\xii{-}^\Lambda &= \xii{-}^{\Lambda,-1} \varpi + \xii{-}^{\Lambda,0}
+ \xii{-}^{\Lambda,1}  \varpi^{-1} + \cO(\varpi^{-2})\, ,  \qquad\\
\txii{-}_\Lambda &= - \ci{-}_\Lambda \log\varpi + \txi^{[-]}_{\Lambda,0}
+ \txi^{[-]}_{\Lambda,1} \varpi^{-1}+  \cO(\varpi^{-2})\, , \\
\ai{-} &=  - \ci{-}_\flat  \log\varpi + \ci{-}_\Lambda \xii{-}^{\Lambda,-1} \varpi
+ \alpha^{[-]}_0+ \alpha^{[-]}_1 \varpi ^{-1} +  \cO(\varpi^{-2})\, ,
\end{split}
\ee
where the Laurent coefficients are related to those at $\varpi=0$ by the reality
conditions \eqref{rexixit}. It is also useful to specify the Laurent expansion of
the contact potentials,
\be
\label{launuf}
\begin{split}
\Phi\di{+}=& \phi\di{+}^0 +  \phi\di{+}^1 \varpi +\cO( \varpi^2) \, ,
\\
\Phi\di{-}=&  \phi\di{-}^0 +  \phi\di{-}^1 \varpi^{-1} + \cO( \varpi^{-2}) \, ,
\end{split}
\ee
related by the antipodal map $\Phi\di{-}=\overline{\tau(\Phi\di{+})}.$

For generic choices of contact transformations, we
expect\footnote{In the case where $\cM$
admits $d+1$ commuting isometries, or for perturbations thereof, this will be demonstrated
in Sections 3 and 5 below.} that similarly to the HK case \cite{Hitchin:1986ea}, the moduli space of
solutions to the gluing conditions \eqref{xi:trafo} and reality conditions \eqref{rexixit} is
of real dimension $4d+1$, and can
be parametrized by  the lowest Laurent coefficients $\xii{+}^{\Lambda,-1}=-(\xii{-}^{\Lambda,-1})^*$,
$\txi^{[+]}_{\Lambda,0}=-(\txi^{[-]}_{\Lambda,0})^*$ and
the real coefficient $\I(\alpha^{[+]}_0+\alpha^{[-]}_0)$.
This parameter space admits a $U(1)$ action induced by phase rotations of $\varpi$, which
can be quotiented out to produce the QK manifold $\cM$ itself.

Expanding the contact form \eqref{defaa} for $i=\pm$ at $\varpi=0,\infty$
and identifying the coefficients of $\varpi^n$ on either sides of \eqref{cXnu0}
allows to extract the $SU(2)$ connection,
\be
\begin{split}\label{connection}
p_\pm &=\frac12\, e^{-\phi\di{\pm}^0}
\left( \xi^{\Lambda,-1}_{[\pm]}  \de\txi^{[\pm]}_{\Lambda,0}
+ \ci{\pm}_\Lambda \de\xi^{\Lambda,-1}_{[\pm]}
\right)\, ,
\\
p_3 &=\frac{\I}{2}\, e^{-\phi\di{+}^0} \left( \de\alpha^{[+]}_0 +
\xi^{\Lambda,0}_{[+]}  \de\txi^{[+]}_{\Lambda,0} +
\xi^{\Lambda,-1}_{[+]}  \de\txi^{[+]}_{\Lambda,1}  \right)
- \I \phi\di{+}^1 p_+\, ,
\end{split}
\ee
and to express the Laurent coefficients of the contact potentials in terms of the
Laurent coefficients of the contact twistor lines,
\be
\label{contpot}
\begin{split}
e^{\phi\di{\pm}^0} =&  \pm\hf  \left( \xi^{\Lambda,-1}_{[\pm]} \txi^{[\pm]}_{\Lambda,1}
+ \ci{\pm}_\Lambda \xii{\pm}^{\Lambda,0} +   \ci{\pm}_\flat \right)\, ,\\
\phi\di{\pm}^1 = & \pm\hf \,e^{-\phi\di{\pm}^0}
\left( \alpha^{[\pm]}_1 + 2 \xi^{\Lambda,-1}_{[\pm]}  \txi^{[\pm]}_{\Lambda,2}
+  \xi^{\Lambda,0}_{[\pm]}  \txi^{[\pm]}_{\Lambda,1} + \ci{\pm}_\Lambda \xi^{\Lambda,1}_{[\pm]}  \right)\ .
\end{split}
\ee
Via \eqref{ompp}, one  obtains the triplet of quaternionic forms $\vec\omega_\cM$,
in particular
\be
\label{om3p}
\omega_{\cM,3} =\frac{2}{\nu} \left( \de p_3 +2\I \, p_+ \wedge p_- \right) \, .
\ee
As anticipated above, the $U(1)$ action induced by phase rotations
of $\varpi$ shifts $p_3$ by a total derivative and acts on $p_\pm$ in opposite ways,
so lies in the kernel of $\omega_3$.

In order to obtain the metric from $\omega_{\cM,3}$, it is still necessary to specify
the almost complex structure $J_3$. This is achieved
by expanding the holomorphic one-forms $\de\xii{+}^\Lambda$ and $\de\txii{+}_I$
around $\varpi=0$, and projecting them along the base $\cM$:
\be
\label{dxi}
\begin{split}
\de\xii{+}^\Lambda &= \xi_{[+]}^{\Lambda,-1}  p_+\varpi^{-2} + \cV^\Lambda \varpi^{-1}
+ \cO(\varpi^0) \quad \mod D\varpi\, ,
\\
\de\txii{+}_I &=-\ci{+}_I  p_+ \varpi^{-1} +
\tilde{\cV}_I + \cO(\varpi^1) \quad \mod D\varpi\, ,
\end{split}
\ee
where $D\varpi= d\varpi + p_+ - \I p_3 \varpi + p_- \varpi^2$ and
\be
\label{xi10}
\cV^\Lambda\equiv  (\de - \I p_3) \xi_{[+]}^{\Lambda,-1}\, ,
\qquad
\quad \tilde{\cV}_I \equiv  \de\txi^{[+]}_{I,0} - \txi^{[+]}_{I,1}  p_+  + \I \ci{+}_I p_3 \, .
\ee
$(1,0)$ forms with respect to the almost complex structure~$J_3$ on $\cM$ can then be obtained
by forming linear combinations of $\de\xii{+}^\Lambda$ and $\de\txii{+}_I$ which are regular
at $\varpi=0$, and setting $\varpi=0$ in the corresponding expressions. Thus,
singling out the index 0 out of $\Lambda$, a basis of
$(1,0)$ forms on $\cM$ is given by
\be
\label{defPi}
\Pi^a =\xi_{[+]}^{0,-1} \cV^a - \xi_{[+]}^{a,-1} \cV^0\, ,
\qquad
\tilde\Pi_I = \xi_{[+]}^{0,-1}  \tilde{\cV}_I + \ci{+}_I \cV^0\, ,
\ee
where $a$ runs from 1 to $d-1$. Note that the (1,0) form
$p_+$ is not linearly independent from those, as it satisfies
\be
\xi^{0,-1}_{[+]} \left( 1 - \xi^{\Lambda,-1}_{[+]}  \txi^{[+]}_{\Lambda,1} \right)
p_+ = \frac12 \, e^{-\phi\di{+}^0} \left( \xi_{[+]}^{a,-1}  \tilde\Pi_a +
\ci{+}_\Lambda \Pi^\Lambda\right)\, .
\ee
Having determined $J_3$ in this way,
the QK metric then follows from $\omega_3(X,Y)=g_\cM(J_3 X,Y)$.
Of course, the $SU(2)$ connection and almost complex structure  can equivalently be
obtained by expanding near $\varpi=\infty$.

Before closing this section, let us note that the above discussion simplifies considerably
in the special case where $\nu^\flat$ is a global $\cO(2)$ section:
in this case, the transition functions $\hat f_{ij}^2$ become equal to one,
and the contact potentials $\Phi_{i}(x^\mu,\varpi)$ become independent of $\varpi$,
defining a real function on $\cM$.


\section{Quaternionic geometry with commuting isometries}
\label{Sect.2}

In this section, we study aspects of the twistor space $\cZ_\cS$ of a HKC $\cS$
(of real dimension $4d+4$) with $d+1$ commuting tri-holomorphic isometries.
As explained in the introduction, this situation arises when $\cS$ is the Swann
bundle of a QK manifold $\cM$ with $d+1$ commuting isometries.

\subsection{Tri-holomorphic isometries and superconformal invariance}

As explained in \cite{Alexandrov:2008ds}, the moment maps associated to the $d+1$ commuting
tri-holomorphic isometries provide  $d+1$ global $\cO(2)$ sections,
which can be taken as the ``position'' coordinates  $\nu^I$ ($I=\flat,0,\dots,d-1$)
for the holomorphic section $\Omega$. The fact that $\nu^I$ are global
$\cO(2)$ sections restricts the form of the transition functions $\Sij{ij}$ to
 \be
\Sij{ij}(\nui{i},\mui{j},\zeta)=f_{ij}^{-2}\,\nui{i}^I\mui{j}_I
-\tHij{ij}(\nui{i},\zeta) \, ,
\label{trans_2n}
\ee
in such a way that, on the overlap of two patches,
\be
\nui{j}^I = f_{ij}^{-2}\,\nui{i}^I \ ,\quad
\mui{i}_I =\mui{j}_I -f_{ij}^2\,\p_{\nui{i}^I}\tHij{ij}(\nui{i},\zeta) \, .
\label{oncantr1}
\ee
The condition of superconformal invariance \eqref{o2o2s}
further restricts $\tHij{ij}(\nui{i},\zeta)$ to be of the form
\be
\label{confth}
\tHij{ij}(\nui{i},\zeta) = f_{ij}^{-2} \,\hHij{ij}(\nui{i})\, ,
\ee
where $\hHij{ij}(\nui{i})$ is a homogeneous function of degree one
in $\nui{i}$\footnote{R. Ionas and A. Neitzke have independently
shown that the condition that the generalized prepotential is a
section of $\cO(2)$ implies that $\cS$ is HKC \cite{neitzkeprivate}.}.
Following \cite{Alexandrov:2008ds}, we like to express $\tHij{ij}$ in terms of
the standard $\cO(2)$ multiplet
\be
\eta^I(\zeta) \equiv \zeta^{-1}\nui{0}^I(\zeta) = \frac{v^I}{\zeta} + x^I - \bv^I \zeta \, .
\label{defeta}
\ee
Thus, we define (cf. Eq. (3.7) in \cite{Alexandrov:2008ds})
\be
\label{HHt}
\Hij{ij}(\eta^I,\zeta) \equiv \zeta^{-1} f_{0j}^2 \tHij{ij}( \zeta f_{0i}^{-2} \eta, \zeta)\, .
\ee
Using \eqref{confth}, this reduces to
\be
\Hij{ij}(\eta^I,\zeta) = \hHij{ij}(\eta^I) \equiv \Hij{ij}(\eta^I)\, .
\ee
In terms of $\Hij{ij}(\eta)$, the gluing conditions \eqref{oncantr1} simply become
\be
\mui{i}_I=\mui{j}_I- \,\p_{\eta^I}\Hij{ij}(\eta)\, .
\label{eq2n}
\ee

The consistency conditions on $H^{[ij]}(\eta,\zeta)$ were analyzed in \cite{Alexandrov:2008ds},
and just need to be restricted to the superconformal case. Thus, we require that
\be
\Hij{ji}=-\Hij{ij}, \qquad \Hij{ik}+\Hij{kj}=\Hij{ij}\, ,
\label{consist_condH}
\ee
subject to the equivalence relation
\be
\Hij{ij}\mapsto \Hij{ij}+\Gi{i}-\Gi{j}\, ,
\label{gaugeHG}
\ee
and reality conditions
\be
\overline{\tau(\Hij{ij})} = -\Hij{\bi\bj}\, ,
\label{realH}
\ee
 where all quantities are $\zeta$-independent functions of $\eta^I$, homogeneous of degree one.
As in \cite{Alexandrov:2008ds}, we shall abuse notation and define $\hHij{ij}$ away from 
the overlap $\hU_i \cap \hU_j$
(in particular when the two patches do not intersect) using analytic continuation and the 
second equation in \eqref{consist_condH} to interpolate from $\hU_i$ to $\hU_j$.

\subsubsection*{Anomalous $\cO(0)$ multiplets}

As discussed in the previous section, it is possible to relax the homogeneity condition
\eqref{o2o2s} into the ``quasi-homogeneity'' condition \eqref{o2o2sq}. In this case,
$\tHij{ij}$ is restricted to be of the form
\be
\tHij{ij}(\nui{i},\zeta) =  f_{ij}^{-2} \left[ \hHij{ij}(\nui{i}) + \ci{i}_I \nui{i}^I \log \nui{i}^\flat
- \ci{j}_I  \nui{i}^I  \log\left( f_{ij}^{-2}  \nui{i} ^\flat\right) \right] \, ,
\ee
where $\hHij{ij}(\nui{i})$ is again homogeneous of degree one in its argument. Defining
$\Hij{ij}$ as in \eqref{HHt}, we find
\be
\Hij{ij}(\eta,\zeta) = \hHij{ij}(\eta) + \ci{i}_I \eta^I \log\left( \zeta f_{0i}^{-2} \eta^\flat\right)
- \ci{j}_I \eta^I \log\left( \zeta f_{0j}^{-2} \eta^\flat\right) \, .
\ee
The explicit dependence on $\zeta$ may be removed by
a local symplectomorphism
\be
\label{quasG}
\Gi{i}=-\ci{i}_I \eta^I \log\left( \zeta f_{0i}^{-2}\right)+\ci{0}_I \eta^I  \log\zeta \, ,
\ee
where the second, $i$ independent term was added to ensure regularity
in the patches $i=0$ and $i=\infty$.\footnote{This is the place where the additional
reality condition $\ci{0}_I=-\ci{\infty}_I$ becomes necessary. \label{foot_realcond}} After this
gauge transformation, we find
\be
\label{quasH}
\Hij{ij}(\eta,\zeta) = \hHij{ij}(\eta) + \ci{ij}_I \eta^I \log \eta^\flat \equiv \Hij{ij}(\eta)\, ,
\ee
where $\ci{ij}_I\equiv\ci{i}_I-\ci{j}_I$,
while the momentum coordinate $\mui{i}$ is replaced by
\be
\mui{i}_{T;I} = \mui{i}_I +\ci{i}_I \, \log\left( \zeta f_{0i}^{-2}\right)-\ci{0}_I \,  \log\zeta \, .
\ee
Note that \eqref{quasH} is no longer a homogeneous function of $\eta^I$, but rather satisfies
the quasi-homogeneity condition
\be\label{def:cij}
\left( \eta^I \pa_{\eta^I} - 1 \right) \Hij{ij} =  \ci{ij}_I \eta^I \, .
\ee

\subsubsection*{Complex contact structure}

As in Section \ref{seccontact}, using the (quasi)-homogeneity property of the transition
functions $\Hij{ij}$, one may reduce the complex symplectic structure on $\cZ_\cS$ to
a complex contact structure on $\cZ_\cM$. One should only be careful that due to the gauge
transformation \eqref{quasG}, the anomalous $\cO(0)$ sections satisfy
\be
\mui{i}_{T;I}(\zeta, \pi^{A'}, \bpi_{A'},x^\mu) = \tmui{i}_{T;I}( \varpi, x^\mu) - \ci{i}_I \log
\left[ (\pi^2-\zeta\bpi_1)^2 \right] - \ci{0i}_I \log \zeta\, ,
\label{varpimua}
\ee
rather than \eqref{varpimu}, while
$\txii{i}_I(\varpi,x^\mu)$ defined in \eqref{defxixit} becomes
\be
\label{truemu}
\txii{i}_I\equiv  \mui{i}_{T;I}(\zeta) + \ci{i}_I \log\eta^\flat +  \ci{0}_I \log\zeta\, .
\ee
Using the fact that $\hat f_{ij}=1$ when $\nu^\flat$ is a globally defined
$\cO(2)$ section,
the transition function \eqref{trans_2n} with $\Hij{ij}$ as in \eqref{quasH} then leads to
\be
\hSij{ij}(\xii{i}^\Lambda,\txii{j}_I) = \txii{j}_\flat
 + \xii{i}^\Lambda \,  \txii{j}_\Lambda
- \hHij{ij}( \xii{i}^\Lambda)\, .
\ee

The section $\xi^\Lambda\equiv\xii{j}^\Lambda$ is globally well defined, and therefore takes
the form
\be
\label{xiY}
\xi^\Lambda  = \varpi^{-1} Y^\Lambda_+ + A^\Lambda - \varpi\, Y^\Lambda_-\, ,
\ee
where $A^\Lambda$ is real and $(Y^\Lambda_+)^*=Y^\Lambda_-$. The vector
$(2\Re(Y^\Lambda_+),2\Im(Y^\Lambda_+),A^\Lambda)$ is in fact the generalized
moment map for the translational isometry along $A^\Lambda$, as defined in
\cite{MR872143}. The relation
between $A^\Lambda, Y^\Lambda_+$ and $x^I,v^I$ will be discussed in Section \ref{sec33}
below.

On the other hand, the sections $\txii{i}_I$ are defined only in the patch $\cU_i$, and
are related on the overlap of two patches by the complex contact transformation
\be
\label{contactmu}
 \begin{split}
 \txii{i}_\Lambda &=  \txii{j}_\Lambda - \pa_{\xi^\Lambda} \hHij{ij}(\xi^\Lambda)\, ,\\
  \txii{i}_\flat &=  \txii{j}_\flat - \hHij{ij}(\xi^\Lambda)
  +\xi^\Lambda\pa_{\xi^\Lambda} \hHij{ij}(\xi^\Lambda)-\ci{ij}_I \xi^I\, .
\end{split}
\ee
It should be noted that the term proportional to $\ci{ij}_\Lambda$ in this expression
disappears when $\txii{i}_\flat$ is traded for $\ai{i}$ as in \eqref{defaa},
\be
\ai{i}=  \ai{j} - \hHij{ij}(\xi^\Lambda)
  +\xi^\Lambda\pa_{\xi^\Lambda} \hHij{ij}(\xi^\Lambda)  \, .
\ee

\subsubsection*{Lagrangian, \hk potential and twistor lines}

As explained in \cite{Alexandrov:2008ds}, the transition functions $\Hij{ij}(\eta)$
determine the holomorphic symplectic structure of $\cZ_\cS$, and therefore a HK
metric on $\cS$. The latter  can be computed
from the  ``Lagrangian'', a function of the components $v^I, x^I, \bv^I$ of
$\eta^I$ defined by the contour integral
\be
\label{tlag}
\mathcal{L}=\sum_j
\oint_{C_j} \frac{\de\zeta}{2\pi \I\,\zeta}\, \Hij{0j}(\eta(\zeta))\, ,
\ee
where the contours $C_j$ encircle the centered disks $\cU_j$ in the complex
$\zeta$-plane. Note that due to the consistency conditions \eqref{consist_condH}, the index
$0$ on the right-hand side of this expression may be substituted with any
other value without changing the result.  A \kahler potential for the HK metric
on $\cS$ is then obtained by Legendre transformation with
respect to $x^I$ \cite{Hitchin:1986ea},
\be
\label{Legtrafo}
\chi(v^I,w_I,\bv^I,\bw_I) = \cL - x^I \p_{x^I} \cL \, ,\quad
\p_{x^I} \cL = w_I + \bw_I\, .
\ee
As shown in \cite{Alexandrov:2008ds}, the ``momentum" coordinates $\mu_I^{[i]}(\zeta)$
are given by a single expression valid for all patches $i$,
\be
\mu_{T;I}^{[i]} (\zeta) =
\frac{\I}{2}\, \vrh_I +\sum_{j} \oint_{C_j} \frac{\de\zeta'}{2\pi\I\,\zeta'}\,
\frac{\zeta+\zeta'}{2 (\zeta'-\zeta)}\,
 \p_{\eta^I}H^{[0j]}(\eta(\zeta'))\, ,
\label{contintrep}
\ee
provided $\zeta$ lies in the open disk $\cU_i$. In particular, it is manifestly regular in
$\cU_i$. The coordinates
$\vrh_I\equiv - \I (w_I - \bw_I)$ correspond to overall additive constants unconstrained
by \eqref{eq2n}, and are adapted to the tri-holomorphic isometries $\pa_{\vrh_I}$ of $\cS$.

We now discuss the homogeneity and $SU(2)$ transformation properties
of $\cL$ and $\chi$. Taking into
account the quasi-homogeneous property \eqref{def:cij}, we readily find the
scaling relation
\be
\label{Lscal}
\cL\left( \Lambda^2 v^I, \Lambda^2 \bv^I, \Lambda^2 x^I \right) =
\Lambda^2 \left[ \cL( v^I, \bv^I, x^I ) -2\cij{0}_I x^I \log \Lambda^2 \right] \, .
\ee
On the other hand, the \hk potential
satisfies
\be
\chi\left( \Lambda^2 v^I, \Lambda^2 \bv^I, w_I - \ci{0}_I \log \Lambda^2,
\bw_I - \ci{0}_I \log \Lambda^2\right) = \Lambda^2  \, \chi( v^I, \bv^I, w_I, \bw_I)\, .
\ee
The $SU(2)$ action on $v^I$ and $w_I$ can be obtained from \eqref{su2tay} and
\eqref{su2defci}, in Appendix A, leading to
\be
\begin{split}
\delta v^I  &= \I \eps_3 v^I + \eps_+ x^I \, ,  \quad
\delta \bv^I = - \I \eps_3 \bv^I + \eps_- x^I\, ,  \quad  \\
\delta w_I &= \eps_+ \cL_{v^I} - \I \, \eps_3\, \ci{0}_I\, ,\quad
\delta \bw_I = \eps_- \cL_{\bv^I} + \I \,  \eps_3\, \ci{0}_I\, ,\quad
 \end{split}
\ee
while the real combinations $x^I$ and $\vrh_I$ transform as
\be
\label{su2xrho}
\delta x^I = -2 \left( \eps_+ \bv^I + \eps_- v^I \right)\, ,\qquad
\delta \varrho_I =
- \I \left( \eps_+ \cL_{v^I} - \eps_- \cL_{\bv^I} \right) -2 \eps_3 \ci{0}_I\, .
\ee
Note that in the quasi-homogeneous case, $w_I$ has anomalous
transformations under dilations and $U(1)$ transformations \cite{Alexandrov:2007ec},
compared to the  transformations found in \cite{deWit:2001dj}.
The anomalous terms can be removed by defining $\hw_I\equiv w_I+\ci{0}_I \log \vf$.

It is instructive to check explicitly that $\chi$ is $SU(2)$ invariant. Keeping only  the homogeneous
term in \eqref{quasH},  one may rewrite
\bea
\label{Legtrafo2}
\hat\chi &=& \sum_j \oint_{C_j} \frac{\de\zeta}{2\pi\I\,\zeta}\, \left[ \hHij{0j}(\eta^I)
- x^I \pa_{\eta^I} \hHij{0j}(\eta^I) \right]\\
&=&\sum_j\oint_{C_j}\frac{\de\zeta}{2\pi\I\, \zeta}\[
\left(\frac{\vf}{\zeta}-\bvf\zeta\right)\frac{\hHij{0j}}{\etaf}
+\left(\frac{v^\Lambda \xf-x^\Lambda \vf}{\zeta}
+(x^\Lambda \bvf-\bv^\Lambda \xf)\zeta\right)\frac{\pa_{\eta^\Lambda}\hHij{0j}}{\etaf} \]\nn\, .
\eea
Integrating the round bracket in the first term by parts, a short computation establishes
\be
\label{chisu2}
\hat\chi=(\rf)^2 \sum_j\oint_{C_j}\frac{\de\zeta}{2\pi\I\zeta}
\frac{\eta^\Lambda \pa_{\eta^\Lambda}\, {\hHij{0j}}}{(\etaf)^2}
- (\vec r^\flat \cdot \vec r^\Lambda)\sum_j\oint_{C_j}\frac{\de\zeta}{2\pi\I\zeta}
\frac{\pa_{\eta^\Lambda}\,  \hHij{0j}}{\etaf}\, ,
\ee
where
\be
\vec r^I \cdot \vec r^J = x^I x^J + 2 v^I \bv^J + 2 \bv^I v^J
\ee
is the inner product of the 3-vectors $\vr^I=\(2\, \Re(v^I),2 \,\Im(v^I),x^I\)$
associated to the $\cO(2)$ multiplets $\eta^I$.
Each term is the product of a $SU(2)$ invariant quantity times a contour integral
of an $\cO(-2)$ section, and so (according to a general argument discussed
at the end of Appendix A)
is  $SU(2)$ invariant. In the quasi-homogeneous
case, the same line of argument combined with the contour deformations
discussed in Section 3.4 of \cite{Alexandrov:2008ds} leads to
\be
\label{chihchi}
\chi = \hat\chi + \ci{+-}_I \,  \frac{\vec r^\flat \cdot \vec r^I}{\rf} \, ,
\ee
where $\ci{+-}_I$ denotes the quasi-homogeneity coefficient relating the
patches around the two roots of $\etaf(\zeta)$.
Thus $\chi$ is $SU(2)$ invariant, and therefore
equal to the \hk potential on $\cS$. This concludes the proof that transition functions
of the form \eqref{quasH} indeed lead to a HKC metric on $\cS$.

\subsection{Superconformal quotient}
\label{subsec_QKmap}

In this subsection and the following one, we perform the superconformal quotient explicitly
for a general HKC $\cS$ described by the formalism of the previous subsection.
We start by constructing a convenient
set of coordinates $x^\mu, \pi^{A'},\bpi_{A'}$ on $\cM \times \IC^2/\IZ^2$,  in terms of the
complex coordinates $\nu^I(\zeta),\mu_I(\zeta)$ on the Swann bundle $\cS$ in
an arbitrary complex structure $\zeta$.  In the next subsection, we find the reciprocal
change of variables, and determine the twistor lines.

The real coordinates $x^\mu$ on $\cM$ are characterized by their invariance under the scaling and
isometric $SU(2)$ actions on $\cZ_\cS$. Instead, the coordinates $\pi^{A'},\bpi_{A'}$
should transform  as a pair of doublets under $SU(2)$, and have a squared norm
dictated by the \hk potential, $\pi^{A'}\bpi_{A'}=\chi$. These constraints do not
determine the coordinates $x^\mu, \pi^{A'},\bpi_{A'}$ uniquely. In the case of
QK spaces obtained by the classical and quantum corrected $c$-map, studied in
\cite{Neitzke:2007ke,Alexandrov:2007ec} , it was convenient to choose coordinates
$x^\mu$ adapted to the action of a $2d+1$ dimensional Heisenberg  group of isometries.
In the general $\cO(2)$ case, the only isometries are the $d+1$ abelian shift symmetries,
and there is no such "canonical" choice. Our construction below is tailored to
reproduce the results \cite{Neitzke:2007ke,Alexandrov:2007ec} for $c$-map spaces,
as we illustrate later in Section \ref{secthyper}. It also follows
from considerations in contact geometry, as discussed in greater generality
in Section \ref{pertqline}.

We start by singling out two multiplets $\etaf, \eta^0$ out of the $d+1$ $\cO(2)$ multiplets
$\eta^I$, and denote by $\eta^a$, $a=1,\dots, d-1$,  the remaining ones.
The zeros of $\nui{0}^\flat=\zeta \etaf$ are now
\be
\label{zepmrf}
\ztpm=\frac{\xf\mp\rf}{2\bvf}, \qquad \rf=\sqrt{(\xf)^2+4\vf\bvf}\, .
\ee
As explained below \eqref{su2G}, $SU(2)$-invariant quantities can be constructed by
contour--integrating $\cO(-2)$ sections on $\cS$. The simplest example is
\be
\oint_{C_+} \frac{\de\zeta}{2\pi\I\, \zeta}\,\frac{1}{\etaf}=\frac{1}{\rf}\, ,
\ee
which recovers the $SU(2)$ invariant $\rf$, homogeneous of degree one
under dilations. Other convenient $SU(2)$ and dilation invariants
are given by
\be
\label{intQKmap}
\begin{split}
A^I &\equiv \rf
\oint_{C_+} \frac{\de\zeta}{2\pi\I\, \zeta}\,\frac{\eta^I}{(\etaf)^2}
=\frac{\(\vrf\cdot \vr^I\)}{(\rf)^2}\, ,
 \quad\\
Z^\Lambda \equiv
\rf \oint_{C_+} \frac{\de\zeta}{2\pi\I\, \zeta}\,& \frac{\eta^\Lambda}{\etaf\eta^0}
=\frac{\etap^\Lambda}{\etap^0}\, ,
\quad
\bZ^\Lambda \equiv - \rf \oint_{C_-} \frac{\de\zeta}{2\pi\I\, \zeta}\,\frac{\eta^\Lambda}{\etaf\eta^0}
=\frac{\etam^\Lambda}{\etam^0}\, ,
\end{split}
\ee
and, when $\mui{i}$ is a non-anomalous $\cO(0)$ local section,
\be
\label{intQKmap2}
B_I\equiv
-\I \, \rf \oint_{C_+} \frac{\de\zeta}{2\pi\I\, \zeta}\,\frac{\mui{+}_{I}}{\etaf}
+\I \, \rf \oint_{C_-} \frac{\de\zeta}{2\pi\I\, \zeta}\,\frac{\mui{-}_{I}}{\etaf}
=-\I\(\mu^{+}_I+\mu^{-}_I \)  \, .
\ee
Here $C_\pm$ denote the contours containing
$\ztpm$, $\mui{\pm}_{I}$  are the multiplets which are regular in the patch
containing $\ztpm$,
$\etapm^{\Lambda}=\eta^{\Lambda}(\ztpm)$, and
$\mu_I^\pm=\mui{\pm}_I(\ztpm)$.
Moreover, an additional invariant can be constructed out
of the \hk potential itself,
\be
\label{coor}
e^{\phi}\equiv \frac{\chi}{4\rf}\, ,
\ee
As we shall see the $4d$ variables $x^\mu=\{\phi, Z^a, \bZ^a, A^\Lambda, B_I\}$ provide a
convenient coordinate system on~$\cM$. In particular,
the coordinates $B_I$ correspond to the directions along the $d+1$ isometries.

In the quasi-homogeneous case, \eqref{intQKmap2},
is no longer $SU(2)$-invariant. One may imagine replacing $\mui{+}_I$ by $\txii{+}_I$,
which is non-anomalous, however this quantity is singular in the patch $\cU_+$, as
apparent from \eqref{behap}. The logarithmic singularity can be cancelled without
affecting the $SU(2)$ transformation properties by adding
$\ci{+}_I \log (\eta^0/\etaf)$, which leads us to define\footnote{This definition
arises naturally from the general procedure explained in Section \ref{pertqline}.}
\be
\begin{split}
\label{coorBqh}
B_I\equiv &
-\I\, \rf \oint_{C_+} \frac{\de\zeta}{2\pi\I\, \zeta\etaf}\,
\left( \mui{+}_{T;I}+\ci{+}_I \log \eta^0 +\ci{0}_I \log\zeta \right)
- ( + \leftrightarrow -)
\, .
\end{split}
\ee

It is important to note that there exists another manifestly $SU(2)$ and dilation invariant
quantity,
\be
\label{defcR}
\cR\equiv \frac{|\vec r^\flat \times \vec r^0|}{2(\rf)^2}=\frac{|\vf\etap^0|}{(\rf)^2}\, ,
\ee
where $\times$ denotes the inner product of vectors in $\IR^3$.
As we shall see shortly, it is $\cR$ rather than $\phi$ which appears most naturally
in the general formulae \eqref{gentwixi} for the twistor lines. Note that $\cR$ vanishes
when the zeros of $\etaf$ and $\eta^0$ collide.

As far as the coordinates on the fiber $\pi^{A'},\bpi_{A'}$ are concerned, the reasoning
below \eqref{su2G} shows that  $SU(2)$ doublets can be constructed by contour-integrating
$\cO(-3)$ sections. Thus, it is natural to consider\footnote{The contour integrals given below
suffer from ambiguities in the choice of square root branches. This is inherent to the fact that the
fiber of the Swann bundle is $\IC^2/\IZ_2$. We choose the branch cuts in such a way that the
reality conditions $\bpi_{A'}=(\pi^{A'})^*$ is obeyed, and $(\pi^{A'},\eps^{A'B'}\bpi_{B'})$
transform as $SU(2)$ doublets.}
\be
\begin{split}
\pi^1 &= C \oint_{\cC_+} \frac{\de\zeta}{2 \pi \I} \frac{1}{\eta^\flat \sqrt{\zeta\eta^0} }
= \frac{C}{\rf }\,\sqrt{\frac{\zeta_+}{\eta^0_+}} \, ,
\label{pieq}\\
\pi^2 &= \bar C \oint_{\cC_-} \frac{\de\zeta}{2 \pi \I} \frac{1}{ \eta^\flat\sqrt{-\zeta\eta^0} }
= -\frac{ \bar C}{\rf}\,  \sqrt{-\frac{\zeta_-}{\eta^0_-}} \, , \\
\bar{\pi}_1 &= -\bar C \oint_{\cC_-} \frac{\de\zeta}{2 \pi \I} \frac{1}{\zeta \eta^\flat\sqrt{-\zeta\eta^0} }
=  \frac{\bar C}{\rf \sqrt{-\zeta_-\eta^0_-}} \, , \\
\bar{\pi}_2 &= -C  \oint_{\cC_+} \frac{\de\zeta}{2 \pi \I} \frac{1}{\zeta  \eta^\flat\sqrt{\zeta\eta^0}}
= -\frac{C}{\rf \sqrt{\zeta_+\eta^0_+} } \, .
\end{split}
\ee
where the proportionality constant $C$ can be chosen to be real, and adjusted such
that \eqref{chir2} is obeyed. This gives the $SU(2)$ invariant
\be
C=2\rf e^{\phi/2} \sqrt{|\vf \etap^0|}\, .
\label{valueC}
\ee
Eq. \eqref{hkcpi} may now be rewritten as in  \cite{Neitzke:2007ke},
\be
\label{hkc-coord}
\begin{pmatrix} \pi^1 \\ \pi^2 \end{pmatrix} = 2\,\e^{\phi/2} \sqrt{v^\flat}
\begin{pmatrix} z^\half \\ z^{-\half} \end{pmatrix} \, ,\qquad
z = \ztp\sqrt{\frac{\bvf \bar\eta_+^0}{\vf\etap^0}}\ ,\qquad
e^{4\I\psi}= \frac{\vf}{\bvf}\, \frac{\bz}{z}\, .
\ee
The following further relations are also often useful:
\be
\zeta_- = \frac{1}{|z|}\sqrt{\frac{\vf}{\bvf}}\, ,\quad
\zeta_+ = -|z|\sqrt{\frac{\vf}{\bvf}}\, ,\quad
\rf = \frac{|\vf|}{|z|}(1+ z\bz) \, ,\quad
\xf = \frac{|\vf|}{|z|}(1-z\bz)\, .
\label{relzzz}
\ee
Finally, we note that the reduced $\cO(2)$ global
section $\tnui{i}^\flat$ defined in \eqref{varpinu}
takes the simple form
\be
\label{nuom}
 \tnui{i}^\flat  = \frac{1}{4} \,  \, \e^{-\phi} \, \varpi \, .
\ee
Comparing with \eqref{phipmnu}, we see that the contact potential $\Phi\di{+}$
is real, and coincides with the invariant $\phi$ defined in \eqref{coor}. This is
also apparent from \eqref{chinuflat}, using the third equation in \eqref{relzzz} and
the fact that $\hat f_{ij}=1$ in $\cO(2)$ geometries.


\subsection{Contact twistor lines \label{sec33}}

We now consider the converse problem, of determining the complex coordinates
$\nu^I,\mu_I$ on  $\cS$ in terms of the coordinates $x^\mu$ on the QK base $\cM$
and of the coordinates $\pi^{A'}$ on the $\IC^2$ fiber. This is known
in mathematics as ``parametrizing the twistor lines''. In view of the discussion in
Section \ref{seccontact}, this is equivalent to expressing the ``contact
twistor lines" $(\xi^\Lambda,\txii{i}_I)$ in terms of the coordinates $(\varpi,x^\mu)$
on $\cZ_\cM$.

Let us consider first  $\xi^\Lambda(\varpi,x^\mu)$. As explained in \eqref{behap}, $\xi^\Lambda$
(equal to $\xii{i}^\Lambda$ for any $i$) admits a single pole at $\varpi=0$ and $\varpi=\infty$.
The coefficients of the Laurent expansion of $\xi^\Lambda(\varpi,x^\mu)$ around $\varpi=0$
can be extracted from the contour integral
\be
\label{xiLk}
\xi^{\Lambda,k} =\oint_0 \frac{d\varpi}{2\pi\I \varpi^{k+1}}\, \xi^\Lambda(\varpi,x^\mu)
= \rf \oint_{C_+} \frac{d\zeta}{2\pi\I\zeta}\, \frac{\eta^\Lambda}{(\etaf)^2}\,
\varpi^{-k}\, ,
\ee
where we used
\be
\label{jacopi}
\frac{\de\varpi}{2 \pi \I \varpi} = \rf \frac{\de\zeta}{2 \pi \I \zeta\etaf}\, .
\ee
Eq. \eqref{xiLk} vanishes for $k\leq -2$ and $k\geq 2$, as can be seen by  deforming
the contour around $\zeta_+$ to a contour around $\zeta_-$. For $k=0$,
one immediately recovers the first quantity  in \eqref{intQKmap}. For $k=-1$,
one may decompose
\be
\xi^{\Lambda,-1} = \rf \oint_{C_+} \frac{d\zeta}{2\pi\I\zeta} \,\frac{\eta^\Lambda}{\etaf \eta^0}
\left[ \frac{\varpi\, \eta^0}{\etaf} \right]  = \cR\, Z^\Lambda\, ,
\ee
where $Z^0\equiv 1$, and we used the fact that the term in brackets is regular at $\zeta=\ztp$ and equal
at that point to the invariant $\cR$ defined in \eqref{defcR}.
For $k=1$, a similar argument leads to $\xi^{\Lambda,1}=-\cR \bZ^\Lambda$.
As a side product, we obtain a contour integral representation for the quantity \eqref{defcR}
\be
\label{defcrcont}
\cR = \rf \, \oint_{C_+} \frac{\de\zeta}{2\pi\I\, \zeta}\,\frac{\eta^0}{(\etaf)^2} \, \varpi =
 \rf \, \oint_{C_-} \frac{\de\zeta}{2\pi\I\, \zeta}\,\frac{\eta^0}{(\etaf)^2} \, \varpi^{-1}\, .
 \ee
Thus, we conclude that the contact twistor line is parametrized by
\be
\label{gentwixi}
\xi^\Lambda(\varpi,x^\mu)
= A^\Lambda + \cR \left( \varpi^{-1} Z^{\Lambda}-\varpi \bZ^{\Lambda}   \right)\, ,
\ee
so that $Y^\Lambda_+ = \cR\, Z^{\Lambda}$ in \eqref{xiY}.
Setting $\zeta=0,\ \varpi=z$ in this expression allows to express the complex coordinate
$\xi^\Lambda$
on $\cZ_\cM$ in terms of the coordinates on the base and on the $\CP$ fiber.

For $\txii{i}_I$ defined in \eqref{defxixit}, and assuming $\ci{i}_I=0$ for simplicity, one may
eliminate $\vrh_I$ in \eqref{contintrep} in favor of $B_I$, and use the identity
\be
\frac{\de\zeta'}{2\pi\I \zeta'} \left[
\frac{\zeta+\zeta'}{\zeta'-\zeta}
-\half \frac{\zeta_++\zeta'}{\zeta'-\zeta_+}
-\half \frac{\zeta_-+\zeta'}{\zeta'-\zeta_-} \right] =
\frac{d\varpi'}{2\pi\I\, \varpi'}\,
\frac{\varpi+\varpi'}{\varpi'-\varpi}\, ,
\ee
where $\varpi'$ is obtained from \eqref{jacopi} by replacing $\zeta\to\zeta',\varpi\to\varpi'$.
This gives
\be
\txii{0}_I(\varpi, x^\mu)=\frac{\I}{2}\, B_I+\hf\sum_j\oint_{\tilde C_j}\frac{d\varpi'}{2\pi\I\, \varpi'}\,
\frac{\varpi'+\varpi}{\varpi'-\varpi}\, \Hij{0j}_I(\xi(\varpi'))\, ,
\label{eqmu}
\ee
where $\tilde C_i$ is the image of the contour $C_i$ in the $\varpi'$ plane, and
$\Hij{0j}_I(\xi)\equiv \pa_{\eta^I}\Hij{0j}(\eta^I)$.
In the quasi-homogeneous case, similar manipulations (explained in
greater generality in Section \ref{pertqline})
lead to
\bea
 \label{eqmuqh}
\txii{0}_\Lambda &=& \frac{\I}{2}\, B_\Lambda+\hf\sum_j\oint_{\tilde C_j}\frac{d\varpi'}{2\pi\I\, \varpi'}\,
\frac{\varpi'+\varpi}{\varpi'-\varpi}\, \pa_{\xi^\Lambda}  \hHij{0j}(\xi (\varpi'))
+\hf\,\cij{+-}_\Lambda \log \varpi\, ,\\
\txii{0}_\flat &=& \frac{\I}{2}\, B_\flat+\hf\sum_j\oint_{\tilde C_j}\frac{d\varpi'}{2\pi\I\, \varpi'}\,
\frac{\varpi'+\varpi}{\varpi'-\varpi}\, \left[ \hat H - \xi^\Lambda  \pa_{\xi^\Lambda} \hat H+
c_I \xi^I \right]^{[0j]}
+\hf\,\cij{+-}_\flat \log \varpi\, .\nn
\eea
Again, by substituting $\zeta=0,\ \varpi=z$ in these expressions, one may obtain the complex
coordinate $\txi_I$ on $\cZ_\cM$ in terms of the fiber coordinate $z$ and the coordinates on $\cM$.
As in the case of \eqref{contintrep}, the r.h.s. of \eqref{eqmuqh} gives the
contact twistor line $\txii{i}_I$ in any patch $\cU_i$, provided $\varpi$ is chosen to
lie in the corresponding patch (moreover, as in \eqref{contintrep}, one may replace the
superscript $[0j]$ with any $[kj]$ without changing the result). Indeed,
one may check that  the discontinuity
of the r.h.s. of \eqref{eqmuqh} across the contours $\tilde C_i$ precisely implements
the contact transformations given in \eqref{contactmu}.
Another important remark is that,
due to the fact that the argument $\xi^\Lambda$ of $\hHij{0j}$ has a pole at
$\varpi'=0$, the integrals appearing in  \eqref{eqmuqh} need not be regular at
$\varpi=0$: we shall see an example of this phenomenon in \eqref{1ltwistor} below.

For what concerns $\txii{+}_I$ however, the integrals are regular
and the first Laurent coefficients
needed to find the $SU(2)$ connection and the \qk metric are readily extracted:
\be
\label{xiplus0}
\begin{split}
\txi^{[+]}_{\Lambda,0} = &  \frac{\I}{2}\, B_\Lambda
+ \hf\sum_j\oint_{\tilde C_j}\frac{d\varpi}{2\pi\I\, \varpi}\,
 \pa_{\xi^\Lambda}  \hHij{0j}\, ,\quad
 \txi^{[+]}_{\Lambda,1} = \sum_j\oint_{\tilde C_j}\frac{d\varpi}{2\pi\I\, \varpi^2}\,
 \pa_{\xi^\Lambda}  \hHij{0j}\, , \\
&  \qquad \alpha^{[+]}_0 = \frac{\I}{2} \,B_\flat
+\hf\sum_j\oint_{\tilde C_j}\frac{d\varpi'}{2\pi\I\, \varpi'}
\( \hHij{0j} - \xi^\Lambda  \pa_{\xi^\Lambda} \hHij{0j}\) \, .
 \end{split}
\ee
From these expressions it is easy to find
the contact potentials using \eqref{contpot},
\be
\begin{split}
e^{\Phi_{[+]}} &= \frac12 \cR\,
 \sum_j\oint_{\tilde C_j}\frac{d\varpi}{2\pi\I\varpi^2}\,
 Z^{\Lambda} \pa_{\xi^\Lambda} \hat{H}^{[0j]}(\xi^\Lambda(\varpi))
+\half \,\ci{+}_I A^I\, ,\\
e^{\Phi_{[-]}} &= - \frac12 \cR\,
 \sum_j\oint_{\tilde C_j}\frac{d\varpi}{2\pi\I}\,
 \bZ^{\Lambda} \pa_{\xi^\Lambda} \hat{H}^{[0j]}(\xi^\Lambda(\varpi))
-\half \,\ci{-}_I A^I\, ,
\end{split}
\label{eqpot}
\ee
In particular, $e^{\Phi_{[\pm]}}$ are independent of the fiber coordinate $\varpi$ and
are equal to each other, since their difference is the integral of a total derivative.

On the other hand, changing variable from $\zeta'$ to $\varpi'$ in \eqref{chisu2}, we may rewrite
the \hk potential $\chi$, eq.\ \eqref{chihchi}, as a contour integral
\be
\chi=\rf \, \cR \, \sum_j\oint_{\tilde C_j}\frac{d\varpi}{2\pi\I\, \varpi}
\(\varpi^{-1} Z^{\Lambda}-\varpi \bZ^{\Lambda} \)\pa_{\xi^\Lambda} {\hat{H}^{[0j]}}
(\xi^\Lambda(\varpi))
+\rf\cij{+-}_I A^I  \, ,
\label{eqchi}
\ee
where we used the notation $A^\flat=1$. Eq. \eqref{eqchi} may be used to  express $\cR$
in terms of $e^\phi$ or vice-versa. Moreover, comparing  \eqref{eqchi}
and \eqref{eqpot}, one confirms that the invariant $\phi$ defined in \eqref{coor}
is indeed equal to the contact potential $\Phi_{[\pm]}$.

\section{The perturbative hypermultiplet moduli space}
\label{secthyper}
In order to illustrate the results in the previous section, we now discuss the
geometry of hypermultiplet moduli spaces in type II string theories compactified
on a Calabi-Yau three-fold, which is the main motivation for this study. In section
\ref{sec_firstflat} we focus on the tree-level geometry, deferring the inclusion of
the one-loop correction
to the next subsection.  Non-perturbative contributions will be considered in \cite{apsv3},
using the results on deformation in Section \ref{secdefhkc} below.

\subsection{Tree-level geometry}
\label{sec_firstflat}

At tree-level in the string perturbative expansion, the
hypermultiplet moduli space $\cM$ in type IIA (resp. IIB) string theory compactified
on a Calabi-Yau three-fold $Y$ (resp. $X$) is a QK space of quaternionic dimension
$d=h_{2,1}(Y)+1$ (resp. $d= h_{1,1}(X)+1$). It is obtained  by the $c$-map
construction from the vector multiplet moduli space $\cM_V$ in
type IIB (resp. IIA) theory  compactified on the same Calabi-Yau
manifold $Y$ (resp. $X$) \cite{Cecotti:1988qn,Ferrara:1989ik}.
$\cM_V$ is a  projective special K\"ahler manifold
of dimension $2d-2$, representing the moduli space of complex deformations of $Y$
(resp. complexified \kahler deformations of $X$),
described in the standard way \cite{deWit:1984pk,Cremmer:1984hj} by a
holomorphic prepotential $F(X^\Lambda)$ ($\Lambda=0,\dots,d-1$),
homogeneous of degree two.

As shown in \cite{Rocek:2005ij,Rocek:2006xb}, the Lagrangian describing the
Swann bundle of $\cM$ is given by
\be
\label{contcmap}
\cL = {\rm Im} \, \left[ \oint_{C_+} \frac{\de\zeta}{2 \pi\I\,\zeta}
\frac{F(\eta^\Lambda)}{\eta^\flat} \right] \, .
\ee
where $\eta^I$ ($I=\flat,0,\dots, d-1$) are $\cO(2)$ multiplets parameterized
as in \eqref{defeta}, and the contour $C_+$
encloses the root $\zeta_+$ of $\zeta \eta^\flat(\zeta)$ (given in eq.~\eqref{zepmrf})
counter-clockwise. This can be cast in our general framework \eqref{tlag} by introducing
four patches\footnote{Since $\Hij{0\infty}=0$, $\cU_0$ and $\cU_\infty$ are really one and the
same patch. We further assume that all singularities of $F(\eta)$ belong to
either $\cU_0$ or $\cU_\infty$, but not to $\cU_\pm$.}
on $\CP$, centered at $0,\infty,\zeta_+,\zeta_-$, with transition functions
\be\label{symp-cmap}
\Hij{0+}_{\rm tree}= - \frac{\I}{2} \frac{F(\eta^\Lambda)}{\etaf}\, ,
\qquad
\Hij{0-}_{\rm tree}= - \frac{\I}{2} \frac{\bar F(\eta^\Lambda)}{\etaf}\, ,
\qquad
\Hij{0\infty}_{\rm tree}=0\, .
\ee
The contour integral \eqref{contcmap} was evaluated in \cite{Neitzke:2007ke} (generalizing
a previous computation in \cite{Rocek:2005ij,Rocek:2006xb} restricted to
the locus $\vf=\bar{v}^\flat=0$), resulting in
\begin{equation}
\label{cont-cmap} \mathcal{L}(v,\bv,x)=
\frac{1}{2\I r^\flat}\left(F(\eta^\Lambda_+)- \bF(\eta^\Lambda_-)\right)\, .
\end{equation}
The \hk potential $\chi$ following by Legendre transform is given by~\cite{Neitzke:2007ke}
\be
\label{chihkc}
\chi= \frac{\vf\bvf}{(\rf)^3}\,K(\etap^\Lambda,\etam^\Lambda)\, ,
\ee
where
\be
K(Z,\bZ)\equiv \I\(\bZ^\Lambda F_\Lambda(Z)-Z^\Lambda \bF_\Lambda(\bZ)\)
\equiv e^{-\cK(Z,\bZ)}
\, .
\ee
The \hk potential $\chi$ may be further expressed in terms of the complex coordinates
$v^I, w_I$ and their complex conjugate by means of the
Hesse potential associated to the
special \kahler manifold $\cM_V$~\cite{Neitzke:2007ke}.

The momentum coordinates $\mu_I^{[0]}$ for this geometry can be
evaluated using \eqref{contintrep} (away from the locus where the zeros
of $\etaf$ collide with other singularities of $F(\eta^\Lambda)$):
\be
\begin{split}
\mui{0}_\Lambda = & \frac{\I}{2}\,\vrh_\Lambda -\tfrac{1}{4\I \rf}
\left[ \tfrac{\zeta + \ztp}{\zeta - \ztp} F_\Lambda(\etap)
- \tfrac{\zeta + \ztm}{\zeta - \ztm} \bar{F}_\Lambda(\etam)\right] \, ,
\\
\mui{0}_\flat = &\frac{\I}{2}\,\vrh_\flat
+ \tfrac{\zeta}{2\I(\rf)^2} \left[ \tfrac{\ztp\, F(\etap)}{(\zeta - \ztp)^2 }
+ \tfrac{\ztm \, \bar{F}(\etam)}{(\zeta- \ztm)^2}  \right]
+ \tfrac{\xf}{4\I (\rf)^3} \left[
\tfrac{\zeta + \ztp}{\zeta - \ztp} \, F(\etap)
- \tfrac{\zeta + \ztm}{\zeta - \ztm} \, \bar{F}(\etam) \right]
\\
& \,
- \tfrac{1}{4\I(\rf)^2} \left[
\tfrac{\zeta + \ztp}{\zeta - \ztp} \,
F_\Lambda(\etap)\left( \tfrac{v^\Lambda}{\ztp} + \ztp \vb^\Lambda \right)
+ \tfrac{\zeta + \ztm}{\zeta - \ztm} \,
\bar{F}_\Lambda(\etam) \left( \tfrac{v^\Lambda}{\ztm} + \ztm \vb^\Lambda \right)\right]   \, .
\end{split}
\label{mult_hyper}
\ee
Since $\Hij{0\infty}=0$, the momentum coordinates around the south pole
are given by $\mui{\infty}_I=\mui{0}_I$, and one may check that the
reality conditions \eqref{realcon} are indeed satisfied.

The multiplets $\mui{0}_\Lambda$ and $\mui{0}_\flat$ have a first order and second
order pole at $\zeta=\zeta_\pm$, respectively, while being regular elsewhere.
It is readily checked that the combinations
\be
\mu^{[+]}_\Lambda = \mui{0}_\Lambda -
\frac{\I}{2}\frac{F_{\Lambda}(\eta)}{\eta^\flat}\ ,\quad
\mu^{[+]}_\flat = \mui{0}_\flat +\frac{\I}{2}
\frac{F(\eta)}{(\eta^\flat)^2} \, ,
\ee
related to $\mui{0}_I$ by the symplectomorphism generated by
$\Hij{0+}_{\rm tree}$, are regular at $\zeta = \zeta_+$, while being singular
at $\zeta = \zeta_-$ and other possible singularities of $F(\eta)$.
Indeed,  evaluating $\mu^{[+]}_I$ at $\zeta = \zeta_+$ yields
\be\label{mufL}
\mu^{+}_\Lambda = \frac{\I}{2} \varrho_\Lambda -
\frac{\I \, x^\flat}{4 (\rf)^2} \left[F_\Lambda(\etap) - \bF_\Lambda(\etam) \right]
+ \frac{\I}{2 r^\flat} \, F_{\Lambda \Sigma}(\etap)
\left( \frac{v^\Sigma}{\ztp} + \bv^\Sigma \ztp \right)
\ee
and
\be\label{mufp}
\begin{split}
\mu^{+}_\flat = & \, \frac{\I}{2} \varrho_\flat
+ \I \frac{(x^\flat)^2 -
2 v^\flat \bv^\flat}{4 (\rf)^4} \left[ F(\etap) - \bF(\etam) \right]\\
& \, -\frac{\I \, x^\flat}{4 (r^\flat)^3} \left[
\left(\frac{v^\Lambda}{\ztp} + \ztp \bv^\Lambda \right) F_\Lambda(\etap)
+ \left(\frac{v^\Lambda}{\ztm} + \ztm \bv^\Lambda \right) \bF_\Lambda(\etam)
\right] \\
&
- \frac{\I \left( \bar{v}^\flat v^\Lambda
- \vf \bar{v}^\Lambda \right)}{2 (\rf)^3}F_\Lambda(\etap)
 + \frac{\I }{4 (r^\flat)^2} \,F_{\Lambda \Sigma}
\left( \frac{v^\Sigma}{\ztp} + \ztp \, \bv^\Sigma \right)
\left( \frac{v^\Lambda}{\ztp} + \ztp \, \bv^\Lambda \right) \, ,
\end{split}
\ee
Similarly, the combinations
\be
\mu^{[-]}_\Lambda = \mu^{[0]}_\Lambda
- \frac{\I}{2}\frac{\bar F_{\Lambda}(\eta)}{\eta^\flat} \, ,\quad
\mu^{[-]}_\flat = \mu^{[0]}_\flat
+\frac{\I}{2} \frac{\bar F(\eta)}{(\eta^\flat)^2} \, ,
\ee
related to $\mui{0}_I$ by the symplectomorphism generated by
$\Hij{0-}$, are regular at $\zeta = \zeta_-$, while being singular
at $\zeta = \zeta_+$ and other possible singularities of $F(\eta)$.

We note that the multiplets $\mu_I$ may be obtained independently by
making use of the special symmetry properties of the QK metrics in the
image of the $c$-map, namely the existence of an extended Heisenberg
group of tri-holomorphic isometries \cite{Cecotti:1988qn,Ferrara:1989ik}.
Upon lifting them to the Swann bundle,
these isometries are generated by the holomorphic Killing vector
fields \cite{Neitzke:2007ke}
\be
\begin{split}
K=-&\frac{\I}{4} \pa_{w_\flat}\ ,\quad P^\Lambda=\frac{\I}{2} \pa_{w_\Lambda}\, ,\quad
Q_\Lambda= w_\Lambda \pa_{w^\flat} - v^\flat \pa_{v^\Lambda}\, ,\\
& M=w_\flat \pa_{w^\flat}- v^\flat \pa_{v^\flat} +\frac12 w_\Lambda \pa_{w_\Lambda}
- \frac12 v^\Lambda \pa_{v^\Lambda}\,  .
\end{split}
\ee
The commuting isometries $K,P^\Lambda$ are manifest in the $\cO(2)$
projective superfield construction; their $\cO(2)$-valued moment maps
are just the $\cO(2)$ multiplets $\eta^\flat, \eta^\Lambda$. The
moment maps, $\lambda_\Lambda, \lambda_\flat$
associated to the remaining isometries $Q_\Lambda$ and $M$ provide $d+1$
additional global $\cO(2)$ sections
\bea
\lambda_\Lambda &=&
v^\flat w_\Lambda /\zeta + \left(w_\Lambda
\pa_{w^\flat}\chi- v^\flat \pa_{v^\Lambda}\chi \right)
+ \bv^\flat \bw_\Lambda\zeta  \, ,\nn
\\
\lambda_\flat &=& \left( v^\flat w_\flat
+ \frac12 v^\Lambda w_\Lambda \right){\zeta}^{-1} + \left( w_\flat \pa_{w^\flat}\chi-
v^\flat \pa_{v^\flat}\chi +\frac12 w_\Lambda \pa_{w_\Lambda}\chi-
\frac12 v^\Lambda \pa_{v^\Lambda}\chi
\right)\nn
\\
&& \hspace*{1cm} + \left( \bv^\flat \bw_\Lambda  + \frac12 \bv^\Lambda \bw_\Lambda \right) \zeta \, .
\eea
Matching the leading terms in the expansion around $\zeta=0$, one readily checks that the
momentum coordinates around the north pole are given in terms of the global $\cO(2)$
sections
\be
\mui{0}_\Lambda = \frac{\lambda_\Lambda}{\eta^\flat}\, ,
\qquad
\mui{0}_\flat = \frac{\lambda_\flat}{\eta^\flat} - \frac{\lambda_\Lambda
\eta^\Lambda}{2(\eta^\flat)^2} \, .
\ee

\subsection{One-loop correction\label{sec-oneloop}}

In type II theories compactified on a Calabi-Yau $Y$, the
metric on the hypermultiplet moduli space receives a one-loop correction,
proportional to the Euler number of $Y$ \cite{Antoniadis:1997eg}.
There is evidence that there are no perturbative corrections to the hypermultiplet
metric beyond one-loop~\cite{Robles-Llana:2006ez}\footnote{In the case of the universal
hypermultiplet, this was established rigorously in  \cite{Antoniadis:2003sw}. See the
end of Section \ref{sec-sq} for a strengthening of the
non-renormalization argument in \cite{Robles-Llana:2006ez}.}
As shown in~\cite{Robles-Llana:2006ez}, the one-loop correction
can be described in the projective superspace formalism by adding a term
\be
\label{Loneloop}
\cL_{\rm 1-loop} = 2 c \oint_\cC \frac{\de\zeta}{2 \pi\I\zeta} \, \eta^\flat \, \log \eta^\flat
= -4c\(\rf-\xf\log\frac{\xf+\rf}{2|\vf|}\) \, .
\ee
to the Lagrangian \eqref{contcmap}, where $c$ is a constant determined
in \cite{Robles-Llana:2006ez}, proportional to the
Euler character of the Calabi-Yau threefold.
Here the contour $\cC$ is a  figure-eight contour
around $\zeta_+$ and $\zeta_-$, and the branch cuts in $\log \eta^\flat$
are chosen to extend from $\zeta_+$ to $0$ and $\zeta_-$ to $\infty$
(see Section 3.4 in \cite{Alexandrov:2008ds} for a more detailed discussion).
Equivalently, the one-loop correction gives rise to additional contributions to the
transition functions  \eqref{symp-cmap}.
\be\label{1ltrans}
H^{[0+]}_{\rm 1-loop} = 2 c \, \eta^\flat \, \log \eta^\flat \, , \qquad
H^{[0-]} _{\rm 1-loop} = -2 c \, \eta^\flat \, \log \eta^\flat \, , \qquad
H^{[0\infty]}_{\rm 1-loop}  = 0 \, .
\ee
In particular, the transition functions are no longer homogeneous, but fall
in the ``quasi-homogeneous" class, with anomalous dimensions
\be
\ci{0}_\flat= \ci{\infty}_\flat=0, \qquad  \ci{\pm}_\flat=\mp 2c\ ,\qquad \ci{i}_\Lambda=0\ .
\ee
The one-loop contribution to the \hk potential is given by a simple correction
to the formula \eqref{chihkc} \cite{Alexandrov:2007ec}
\be
\chi =  \frac{\vf \bvf}{(\rf)^3} K(\eta_+,\eta_-) - 4 \, c \, \rf\, ,
\label{oneloopchi}
\ee
in agreement with the general result \eqref{chihchi}.

Let us now determine the twistor lines for the one-loop corrected hypermultiplet moduli space.
Starting from the general expression \eqref{contintrep}, the additional contribution
\eqref{1ltrans} to the transition functions gives rise to extra terms in~$\mu_{\flat;T}^{[i]}$,
\be
\begin{split}
\mu_{T;\,\flat}^{[0]} =& \mu_{\flat}^{[0]{\rm tree}}
+2c\log \left(|\ztp|\, \frac{\zeta-\ztm}{\zeta-\ztp}\right) \, , \\
\mu_{T;\,\flat}^{[+]} =& \mu_{\flat}^{[+]{\rm tree}}
+2c \log\left(-\frac{|\vf| (\zeta-\ztm)^2}{\zeta \ztm} \right) + 2c \, ,
\label{oneloopmu}
\end{split}
\ee
while the other momentum coordinates remain unaltered,
$\mu_\Lambda^{[i]} = \mu_{\Lambda}^{[i]{\rm tree}}$.
It is easy to check that the multiplet \eqref{oneloopmu} transforms
under $SU(2)$ transformations according to \eqref{su2defci}.

\subsection{Superconformal quotient\label{sec-sq}}

The superconformal quotient of the HKC defined by \eqref{contcmap} was studied in
\cite{Neitzke:2007ke,Alexandrov:2007ec}\footnote{Ref.
\cite{Alexandrov:2007ec} used a different contour prescription related to the
one used here by a local gauge transformation, as we explain in Appendix \ref{ap_hyper}.
As a result, the expressions for $\tilde{\zeta}_\Lambda$ and $\sigma$ acquired some additional terms.}.
The dilation and $SU(2)$ invariant coordinates used in these references
were given by
\be\label{TM:N}
\begin{split}
& e^{2 U} =   \frac{\chi}{4 r^\flat} \, , \; \; \zeta^\Lambda
= \frac{\vec{r}^\flat \cdot \vec{r}^\Lambda}{(r^\flat)^2} \, , \; \;
z^a = \frac{\eta^a_+}{\eta^0_+} \, , \; \;
\tilde{\zeta}_\Lambda =
- \I (w_\Lambda - \bar{w}_\Lambda) + \frac{\xf}{(\rf)^2} {\rm Re}\left[F_\Lambda(\eta_+) \right] \, ,
\\
& \sigma =   2 \I (w_\flat - \bar{w}_\flat) +
\I \left( \tfrac{v^\Lambda}{\vf} w_\Lambda
- \tfrac{ \bv^\Lambda}{\bv^\flat} \bar{w}_\Lambda \right)
- \tfrac{\xf}{(\rf)^2} {\rm Re} \left[ \eta^\Lambda_+ \tilde{\zeta}_\Lambda
- F_{\Lambda}(\eta_+) \zeta^\Lambda \right]
- 4 \I c \log \tfrac{\eta^0_+}{\eta^0_-}\,  .
\end{split}
\ee
While the $SU(2)$-invariance of $\tilde{\zeta}_\Lambda$ and $\sigma$
directly is rather tedious to check, it can be made manifest by casting
the resulting expressions in the form of a contour integral of an $\cO(-2)$ section,
e.g. when $c=0$,
\be
\tzeta_\Lambda= -2 \I \rf \oint_{C^+} \frac{\de\zeta}{2\pi\I \zeta} \frac{\mui{0}_\Lambda}{\eta^\flat}\, ,\quad
\sigma = 4\I \rf
\oint_{C^+} \frac{\de\zeta}{2\pi\I \zeta}
\left(  \frac{\mui{0}_\flat}{\eta^\flat} +  \frac{\eta^\Lambda \mui{0}_\Lambda}{ 2(\eta^\flat)^2} \right)\, .
\ee
In the presence of the one-loop correction, these expressions may be generalized by performing the same
replacement as in \eqref{coorBqh} and taking the real part.

We now relate the result \eqref{TM:N} to the general $SU(2)$ and dilation
invariant coordinates introduced in section \ref{subsec_QKmap}. Clearly,
$U = \phi/2, z^a = Z^a, \zeta^\Lambda = A^\Lambda$.
On the other hand, evaluating \eqref{coorBqh}  with the help of
 \eqref{mufL}, \eqref{mufp} and \eqref{oneloopmu}
leads to
\bea
B_{\Lambda}  &=&
\vrh_{\Lambda}+ \frac{\xf}{(\rf)^2}\, \Re
F_\Lambda(\etap)-A^\Sigma \Re
F_{\Lambda\Sigma}(\etap)\, ,
\\
B_\flat  &=&
\vrhf+\tfrac{1}{(\rf)^2}\, \Re F(\etap) -\tfrac{x^\Lambda+\xf
A^\Lambda}{2(\rf)^2}\, \Re F_{\Lambda}(\etap)
+\hf\, A^\Lambda A^\Sigma \Re F_{\Lambda\Sigma}(\etap)
+2\I c \log \tfrac{\eta_+^0}{\eta_-^0}\, .
\nn
\eea
Thus the coordinates $B_I$ differ from $\sigma, \tilde{\zeta}_\Lambda$ by $SU(2)$ invariant terms,
\be
\begin{split}
B_{\Lambda} &=\tzeta_\Lambda-A^\Sigma \Re F_{\Lambda\Sigma}(Z)\, , \qquad
B_\flat
=-\hf\,\sigma -\hf\, A^\Lambda B_\Lambda\, .
\end{split}
\label{reltwist}
\ee

The contact twistor lines can be found using the general formulae \eqref{gentwixi}, \eqref{eqmuqh},
\be\label{1ltwistor}
\begin{split}
\xi^\Lambda= &  A^\Lambda+ \cR\( \varpi^{-1}Z^\Lambda-\varpi\bZ^\Lambda\)\, , \\
\txii{0}_\Lambda = & \frac{\I}{2}\[
B_{\Lambda}+A^\Sigma \Re F_{\Lambda\Sigma}(Z)
+\cR\(\varpi^{-1}F_\Lambda(Z)-\varpi \bF_\Lambda(\bZ)\) \]\, ,
\\
\txii{0}_\flat =& \frac{\I}{2}\biggl[B_\flat-\hf\,A^\Lambda A^\Sigma\Re F_{\Lambda\Sigma}(Z)
+\cR^2\Re\(\bZ^\Lambda F_\Lambda(Z)\)
\biggr.
 \\
& \biggl.
-\cR A^\Lambda\(\varpi^{-1} F_\Lambda(Z)-\varpi \bF_\Lambda(\bZ)\)
-\cR^2\(\varpi^{-2} F(Z)+\varpi^2 \bF(\bZ)\)\biggr]
- 2 c  \log \varpi \, .
\end{split}
\ee
Finally, it remains  to express $\cR$, defined in \eqref{defcR} above,
in terms of the base coordinates \eqref{intQKmap} - \eqref{coorBqh}.
For this purpose, one may substitute
the one-loop corrected hyperk\"ahler potential \eqref{oneloopchi}
into the definition of $\phi$, eq.\ \eqref{coor},
and use the homogeneity property of $K(\cdot,\cdot)$ to obtain
\be
\cR
= 2 \,e^{\frac12 \, \cK(Z,\bar Z)} \, \sqrt{e^\phi+ c} \, .
\ee
Introducing
\be
W \equiv  F_\Lambda(Z)\ \zeta^\Lambda - Z^\Lambda \tzeta_\Lambda
\ee
and using \eqref{reltwist}, one may obtain the contact twistor lines
for the one-loop corrected hypermultiplet geometry
in the form found in \cite{Neitzke:2007ke},
\be
\label{gentwi}
\begin{array}{rcl}
\xi^\Lambda &=& \zeta^\Lambda + \cR
\left( \varpi^{-1} Z^{\Lambda} -\varpi \,\bar Z^{\Lambda}  \right) \, , \\
-2\I \txii{0}_\Lambda &=& \tzeta_\Lambda + \cR
\left( \varpi^{-1} F_\Lambda-\varpi \,\bar F_\Lambda \right) \, , \\
4\I  \txii{0}_\flat + 2 \I \txii{0}_\Lambda \xi^\Lambda &=& \sigma + \cR
\left(\varpi^{-1} W-\varpi \,\bar W \right) -8\I \,c  \log \varpi \, ,
\end{array}
\ee

We proceed to extract the one-loop corrected hypermultiplet metric from
the twistor data on $\cZ_\cM$. Following the procedure outlined
at the end of Section \ref{seccontact}, we first compute the Laurent coefficients
of $\txii{+}_I$ entering the $SU(2)$ connection \eqref{connection},
\be
\begin{split}
\txii{+}_{\Lambda,0} = &
\frac{\I}{2} \left( \tzeta_\Lambda - F_{\Lambda\Sigma} \,\zeta^\Sigma
 \right) \, ,\qquad
 \txii{+}_{\flat,0} =  -\frac{\I}{4} ( \sigma + \zeta^\Lambda \tzeta_\Lambda
 -  F_{\Lambda\Sigma} \,\zeta^\Lambda \zeta^\Sigma)
 - e^\phi  + c \, , \\
\txii{+}_{\Lambda,1} = &
\frac12\, \cR  N_{\Lambda \Sigma}
\bZ^\Sigma - \frac{\I}{4 \cR}\, F_{\Lambda \Sigma \Xi} \,\zeta^\Sigma \zeta^\Xi  \, ,
\\
\txii{+}_{\flat,1} = &
-\frac12 \, \cR N_{\Lambda \Sigma} \zeta^\Lambda \bZ^\Sigma
+ \frac{\I}{12 \cR}\, F_{\Lambda \Sigma \Xi}\,\zeta^\Lambda \zeta^\Sigma \zeta^\Xi\, .
\end{split}
\ee
where  $N_{\Lambda\Sigma} \equiv \I (F_{\Lambda\Sigma} - \bar F_{\Lambda\Sigma})$,
leading to the $SU(2)$ connection
\be
\begin{split}
p_+ &=
\frac{\I}{4}\, e^{-\phi} \,\cR\, Z^\Lambda
\( \de \tzeta_\Lambda - F_{\Lambda\Sigma} \de\zeta^\Sigma \)
= (p_-)^*\, ,
\\
p_3 &= \frac{1}{8} \, e^{-\phi}  \left( \de\sigma + \tzeta_\Lambda \de
\zeta^\Lambda - \zeta^\Lambda \de \tzeta_\Lambda + 4 (\e^\phi +c)
\cA_K \right)\, ,
\end{split}
\ee
where $\cA_K \equiv \I\(\cK_a \de Z^a -\cK_{\bar a} \de \bZ^{\bar a}\)$
is the K\"ahler connection of the projective special \kahler base $\cM_V$.
A direct computation  of the $(1,0)$ forms \eqref{defPi} then yields
\be
\begin{split}
\Pi^a = & \, \cR^2 \, \de Z^a \, , \\
\tilde{\Pi}_\Lambda = & \, \frac{\I}{2}\, \cR \left( \de \tilde{\zeta}_\Lambda
- F_{\Lambda \Sigma} \de \zeta^\Sigma - F_{\Lambda \Sigma \Xi} \, \zeta^\Sigma \, \de Z^\Xi \right)
\\
&  + \frac{\I}{4r}\, \cR^3
\Big( \Im(F_{\Lambda \Sigma}) \bZ^\Sigma +\frac{\I}{4\cR^2}\,
F_{\Lambda \Sigma \Xi} \,\zeta^\Lambda \zeta^\Xi \Big)
\left( Z^{\rm T} \de \tilde{\zeta}_{\rm T} - F_{\rm T} \de \zeta^{\rm T} \right) \, ,
\\
\tilde{\Pi}_\flat = & \, - \cR \Big[
\frac{r + 2c}{r+c }\,  \de r + c \, \de \cK
\\
& \quad
+ \frac{\I}{4} \left( \de \sigma + \tilde{\zeta}_\Lambda \de \zeta^\Lambda
+\zeta^\Lambda\de\tilde{\zeta}_\Lambda  - F_{\Lambda \Sigma \Xi} \,
\zeta^\Lambda \zeta^\Sigma \de Z^\Xi
- 2 F_{\Lambda \Sigma} \zeta^\Lambda \de \zeta^\Sigma \right)
\\
& \quad
+ \frac{\I}{4r}  \Big( \cR^2 \Im(F_{\Lambda \Sigma} ) \zeta^\Lambda \bZ^\Sigma
+ \frac{\I}{12} \,F_{\Lambda \Sigma \Xi} \, \zeta^\Lambda \zeta^\Sigma \zeta^\Xi \Big)
 \left( Z^{\rm T} \de \tilde{\zeta}_{\rm T} - F_{\rm T} \de \zeta^{\rm T} \right)
\Big]\, .
\end{split}
\ee
where $r\equiv e^{\phi}$.
Taking linear combinations, a basis of (1,0) forms can be chosen as
\be
\begin{split}
\de Z^a\ ,&\qquad
f_a^\Lambda \left( \de \tilde{\zeta}_\Lambda
- F_{\Lambda \Sigma} \de \zeta^\Sigma \right) \ ,\qquad\\
Z^{\Lambda} \de \tilde{\zeta}_{\Lambda} - F_{\Lambda} \de \zeta^{\Lambda}\, ,&\qquad
\frac{r+2c}{r+c}\, \de r + \frac{\I}{4}
 \left( \de\sigma + \tzeta_\Lambda \de
\zeta^\Lambda - \zeta^\Lambda \de \tzeta_\Lambda \right) +  c \, \de \cK \, ,
\end{split}
\ee
where $f_a^\Lambda=e^{\cK/2}(\pa_a Z^\Lambda + \pa_a \cK \, Z^\Lambda)$,
generalizing the one-forms $e^a, E_a, u, v$ introduced in
\cite{Ferrara:1989ik} for $c\neq 0$.
Finally, computing the \kahler form \eqref{om3p} in this basis and raising
the indices, one obtains the one-loop corrected metric on the hypermultiplet
moduli space \cite{Robles-Llana:2006ez,Alexandrov:2007ec},
\bea
ds^2&=&\frac{r+2c}{r^2(r+c)}\,\de r^2
-\frac{1}{r} \(N^{\Lambda\Sigma}-\frac{2(r+c)}{r K} \,Z^\Lambda \bZ^\Sigma\)
\( \de \tzeta_\Lambda - F_{\Lambda\Theta} \de \zeta^\Theta\)
\(\de \tzeta_\Sigma-\bF_{\Sigma\Xi}\de \zeta^\Xi\)
\nonumber \\
&&
+ \frac{r+c}{16 r^2(r+2c)}\(\de\sigma+\tzeta_\Lambda d \zeta^\Lambda
- \zeta^\Lambda \de \tzeta_\Lambda+4c\, \cA_K \)^2
+\frac{4(r+c)}{r}\,\cK_{a{\bar b}}\,\de Z^a \de \bZ^{\bar b}\nn \, .\\
\label{hypmet}
\eea
which identifies $\phi$ as the four-dimensional dilaton.

It is worthwhile noting that the one-loop correction changes the topology of the fibration of the
$\sigma$-circle over the torus coordinatized by $\zeta^\Lambda, \tzeta_\Lambda$,
by a term proportional to  $\cA_K$ \cite{Gunther:1998sc}. Any perturbative
correction to the hypermultiplet metric beyond one-loop would presumably induce extra terms
in the connection on the $\sigma$ circle bundle proportional to a positive power of $r$, and would
therefore conflict with the quantization of its first Chern class. This observation reinforces
the arguments given in~\cite{Robles-Llana:2006ez} ruling out perturbative corrections to
the hypermultiplet metric beyond one-loop.

\section{Linear deformations of $\cO(2)$ \qk spaces}
\label{secdefhkc}

In this section, we study the infinitesimal deformations of $4d$-dimensional
QK spaces $\cM$ with $d+1$ commuting isometries, which preserve the QK
property but may break some or all of the isometries. Our strategy is to apply the general analysis
of linear deformations of $\cO(2)$ HK spaces developed in \cite{Alexandrov:2008ds} to the
Swann bundle $\cS$ of $\cM$, restricting to deformations which preserve the
superconformal invariance property. As explained in the introduction, it is possible
to bypass the Swann bundle and work directly with the twistor space $\cZ_\cM$.
This strategy will be realized in Section \ref{pertqline}.

\subsection{Linear deformations of $\cO(2)$ \hk cones}

As explained in \cite{Alexandrov:2008ds}, deformations of HK spaces $\cS$ are conveniently
described by perturbing the transition functions $\Sij{ij}$ which encode the
holomorphic symplectic structure on the twistor space $\cZ_\cS$,
\be
\Sij{ij}(\nui{i},\mui{j},\zeta)=f_{ij}^{-2} \nui{i}^I\mui{j}_I
-\tHij{ij}(\nui{i},\zeta)-\tHpij{ij}(\nui{i},\mui{j},\zeta) \, ,
\label{genf}
\ee
and working out the perturbations $\nupi{i}^I, \mupi{i}_I$ of the twistor lines,
\be\label{pmult}
\nui{i}^I=\zeta f_{0i}^{-2} \eta^I +\nupi{i}^I \, , \qquad
\mui{i}_I=\muzi{i}_I+\mupi{i}_I
\ee
to first order in the perturbations $\tHpij{ij}$. Here and below, unperturbed quantities
are denoted with $\breve\ $, perturbations with $\hat\ $, and perturbed quantities
with no extra symbol (with the exception of $\eta^I$, $v^I$, $x^I$ which will continue to
denote unperturbed quantities).

As shown in Section 3.1, superconformal invariance restricts the undeformed
transition functions $\tHij{ij}(\nui{i},\zeta)$ to be homogeneous of degree one in
$\nui{i}$, and without any explicit dependence on $\zeta$ except for some factors
of $f_{ij}$. The same reasoning shows  the perturbation $\tHpij{ij}$ should satisfy
the same conditions, namely
\be
\tHpij{ij}(\nui{i},\mui{j},\zeta) =  f_{ij}^{-2} \hHpij{ij}\left( \nui{i}^I,\mui{j}_I + \ci{j}_I \log\left(
f_{ij}^{-2} \nui{i}^\flat\right) \right)\, ,
\ee
where $\hHpij{ij}$ is a homogeneous function of degree one in its first argument\footnote{For
simplicity, we do not consider deformations of the anomalous dimensions.
However, it may be checked that all formulae below continue to hold provided $\ci{i}_I$ denote
the total perturbed anomalous dimensions.}.
Following \cite{Alexandrov:2008ds}, we now trade $\tHpij{ij}$ for
\be
\begin{split}
 \Hpij{ij}(\eta,\muzi{j},\zeta) &\equiv  \zeta^{-1}f_{0j}^2  \,\tHpij{ij}(\zeta f_{0i}^{-2}\eta,\muzi{j},\zeta)\\
 &=\hHpij{ij}\left( \eta^I,\muzi{j}_I + \ci{j}_I \log\left(
f_{0j}^{-2} \zeta \etaf\right) \right)\, .
\end{split}
\ee
We then perform the gauge transformation \eqref{quasG} to obtain
\be
 \Hpij{ij}(\eta,\muzi{j}_T,\zeta)
 =\hHpij{ij}\left( \eta^I,\muzi{j}_{T;I} + \ci{j}_I \log\eta +\ci{0}_I\log\zeta\right)\, .
\ee
Finally, we trade  the argument $\muzi{j}_{I;T}$ for the real multiplet
\be
\rho_I(\zeta) \equiv -\I ( \muzi{0}_{T;I} + \muzi{\infty}_{T;I} )- \I \,\cij{0\infty}_I \log \zeta \, .
\ee
This quantity has the advantage of having non-anomalous $\cO(0)$ transformations,
moreover the reality conditions are also automatically satisfied provided
$\Hpij{i\bi}$ is a real function of $\eta^I$ and $\rho_I$. After these redefinitions,
$\Hpij{ij}$ is now a function of $\eta^I, \rho_I$,  homogeneous of degree one in $\eta^I$,
and with  no explicit dependence on $\zeta$. In addition, it must satisfy the co-cycle
condition \eqref {consist_condH} and is subject to the gauge equivalence \eqref{gaugeHG},
where $\Gi{i}$ is now a function of $\eta^I, \rho_I$ regular in the patch $\hU_i$.

We may now borrow the results from \cite{Alexandrov:2008ds}, Section 5.
In particular, the first order variation of the HK twistor lines is
given by
\bea
\nupi{i}^I &=& \I f_{0i}^{-2}\sum_j\oint_{C_j} \frac{\de\zeta'}{2\pi \I\,\zeta'}\,
\frac{\zeta^3+\zeta'^3}{\zeta' (\zeta'-\zeta)}\, \Hp^{[0j]I}(\zeta') \, ,
\label{nupi}
\\
\mupi{i}_{T;I} &=& \sum_j\oint_{C_j} \frac{\de\zeta'}{2\pi \I\,\zeta'}\,
\frac{\zeta+\zeta'}{2 (\zeta'-\zeta)}\, \Gi{0j}_I(\zeta')
\eea
with
\be
\label{defgi}
\begin{split}
H_I\equiv \p_{\eta^I}H\, ,&
\qquad
H_{IJ}\equiv \p_{\eta^I}\p_{\eta^J} H\, ,
\qquad
{\Hp}_I\equiv \p_{\eta^I}\Hp  \, ,
\qquad
\Hp^I\equiv \p_{\rho_I}\Hp \\
&\Gi{ij}_I \equiv  {\Hpij{ij}}_I+\I\,\Hp^{[ij]J}(\Hij{j0}_{IJ}+\Hij{j\infty}_{IJ})
+\zeta^{-1}f_{0i}^2\,\nupi{i}^J\,\Hij{ij}_{IJ} \, .
\end{split}
\ee

The corresponding deformations of the K\"ahler potential
can be conveniently described by introducing
the deformed Lagrangian
\be
\CL(v,\bv,x,\vrh)=
\sum_j \oint_{C_j} \frac{\de\zeta}{2\pi\I\, \zeta}\(\Hij{0j}(\eta)+\Hpij{0j}(\eta,\rho)\)\, .
\label{defL}
\ee
So defined, it is a function of the complex variables $v^I$. However, after perturbations
$v_I$ are no longer Darboux coordinates. Instead, a system of complex Darboux coordinates
of the deformed HKC, such that $\omega^+_\cS=\de w_I\wedge \de u^I$, is given by
\be
u^I =   v^I+\I\,\p_{\vrh_I} \sum_j \oint_{C_j} \frac{\de\zeta}{2\pi\I}\, \Hpij{0j}  \, ,
\qquad
w_I = \frac{\I}{2} \vrh_I+ \hf \p_{x^I}\CL(u,\bu,x,\vrh) \, ,
\label{holcoor}
\ee
where in the second relation the arguments $v^I$ of the Lagrangian
are replaced by the new complex variables $u^I$
and the derivative is evaluated keeping $u^I, \bu^I$ and $\vrh_I$ fixed. Similarly,
the \kahler potential for the deformed HK metric is given by the Legendre transform
of the deformed Lagrangian \eqref{defL} but written as a function of the new variables
\be
\chi(u,\bu,w,\bar w)=\langl
\CL(u,\bu,x,\vrh)-x^I (w_I+\bw_I)\rangl_{x^I}\, .
\label{chiLeg}
\ee
In particular, the variation of the \hk potential is given by a Penrose-type integral,
\be
\chi_{(1)}(u,\bu,w,\bw)
= \sum_j \oint_{C_j} \frac{\de\zeta}{2\pi\I\, \zeta} \, \Hpij{0j}(\eta,\rho)\, .
\label{defchi}
\ee

We now verify that the perturbed HK manifold is indeed a HKC (as is of course guaranteed
by construction).
Using the quasi-homogeneity property of $\Hp(\eta,\rho)$, and in particular the property
\be
\(u^I \p_{u^I}-\bu^I \p_{\bu^I} + \zeta\p_{\zeta} \)\rho_J=0\, ,
\label{rotrho}
\ee
it is easily checked that $\CL$ satisfies
\be
\label{confsusy}
\begin{split}
 u^I \CL_{u^I}-\bu^I \CL_{\bu^I}=-2\I\, \ci{0}_I  \CL_{\vrh_I}\ \, , &
\\
 x^I \CL_{x^I}+u^I \CL_{u^I}+\bu^I \CL_{\bu^I}=\CL  -2 \ci{0}_I x^I\, . &
\end{split}
\ee
Together with the identities
\bea
&& \(\p_{x^I}+\I\CL_{x^I x^K}\p_{\vrh_K}\)\(\p_{x^J}-\I\CL_{x^J x^L}\p_{\vrh_L}\)\CL
\nonumber \\
&& \qquad\qquad\qquad\qquad\qquad
+ \(\p_{u^I}-\I\CL_{u^I x^K}\p_{\vrh_K}\)\(\p_{\bu^J}+\I\CL_{\bu^J x^L}\p_{\vrh_L}\)\CL=0 \, ,
\label{prop1mod}
\\
&& \(\p_{x^I}+\I\CL_{x^I x^K}\p_{\vrh_K}\)\(\p_{u^J}-\I\CL_{u^J x^L}\p_{\vrh_L}\)\CL
\nonumber \\
&& \qquad\qquad\qquad\qquad\qquad
- \(\p_{x^J}+\I\CL_{x^J x^K}\p_{\vrh_K}\)\(\p_{u^I}-\I\CL_{u^I x^L}\p_{\vrh_L}\)\CL=0 \, ,
\label{prop2mod}
\eea
these equations guarantee that $\CL$ satisfies the constraints of
superconformal invariance. Moreover, from \eqref{su2KV}, one may compute the
homothetic Killing vector and the Killing vectors for the $SU(2)$ isometric action,
\be
\chi^{u^I}=2u^I \, , \qquad \chi^{w_I}=-2\ci{0}_I\, .
\label{homotvec}
\ee
\be
\begin{split}
\delta u^I  = \I \eps_3 u^I + \eps_+ \left( x^I  +\I\CL_{\vrh_I} \right)\, , & \quad
\delta \bu^I = - \I \eps_3 \bv^I + \eps_- \left(x^I-\I\CL_{\vrh_I} \right)\, ,  \quad  \\
\delta w_I = \eps_+ \cL_{u^I} - \I \, \eps_3\, \ci{0}_I\, ,& \quad
\delta \bw_I = \eps_- \cL_{\bu^I} + \I \,  \eps_3\, \ci{0}_I\, ,\quad
 \end{split}
\ee
In particular, the homothetic Killing vector is holomorphic and identical to the
undeformed case. Moreover, one may check that  the one-form obtained
by lowering the index on $k^+$ using the deformed metric reproduces the
Liouville form \eqref{contact1}  on $\cS$ in the patch $i=0$ \cite{deWit:2001dj}.

\subsection{Perturbed contact twistor lines}
\label{pertqline}

In order to extract the deformed \qk metric on $\cM$, one possible strategy is to study
the deformations of the superconformal quotient: this computationally
intensive approach is outlined in Appendix C. However, it turns out to be more economic
and elegant to work directly with the complex contact structure on the
twistor space $\cZ_\cM$,  without reference to the Swann bundle and its twistor space.

For this purpose, let us recast the deformed
symplectomorphism \eqref{genf} into the form of the contact transformations
\eqref{xi:trafo}. Introducing the same coordinates as in \eqref{defxixit},
it is easy to check that the deformed contact transformations are
generated by the following transition functions
\be
\hSij{ij}(\xii{i}^\Lambda,\txii{j}_I) = \txii{j}_\flat
+ \xii{i}^\Lambda \txii{j}_\Lambda
- \hHij{ij}(\xii{i}^\Lambda) - \hHpij{ij}(\xii{i}^\Lambda, \txii{j}_I) \, ,
\ee
where $\txii{j}_I$ should be replaced by $\txii{j}_I+\ci{j}_I \log( \hat f_{ij}^{-2})$.
However, it follows from \eqref{fhflat} that
\be
\hat f_{ij}^{2}\approx 1-\p_{\txii{j}_\flat }\hHpij{ij}\, ,
\label{perthf}
\ee
so that its logarithm is of first order in the perturbation already.
Therefore, to the first order it is consistent to neglect the term $\ci{j}_I \log( \hat f_{ij}^{-2})$
in the argument of $\hHpij{ij}$, and take  $\hHpij{ij}$ to
be an arbitrary function of the undeformed coordinates $\bxii{i}^\Lambda$ and $\btxii{j}_I$.
As a result, one finds the following deformed contact transformations
\be\label{pert:symp}
\begin{split}
\xii{i}^\Lambda = & \, \xii{j}^\Lambda - T_{[ij]}^\Lambda
\; , \qquad
\txii{i}_I =  \txii{j}_I
 - \tilde{T}^{[ij]}_I
\, , \quad
\end{split}
\ee
where, in view of later applications, we abbreviated
\be\label{Tfct}
\begin{split}
T_{[ij]}^\Lambda \equiv & \,
 -\p_{\txii{j}_\Lambda }\hHpij{ij}
+\xii{i}^\Lambda \, \p_{\txii{j}_\flat }\hHpij{ij} \, ,
\\
\tilde{T}^{[ij]}_\Lambda \equiv & \,
 \p_{\xii{i}^\Lambda }
\left(  \hHij{ij} + \hHpij{ij} \right)
- \cij{j}_\Lambda \p_{\txii{j}_\flat} \hHpij{ij}
\, ,
\\
\tilde{T}^{[ij]}_\flat \equiv & \,
 \left(  \hHij{ij}+ \hHpij{ij} \right) \,
- \xii{i}^\Lambda \p_{\xii{i}^\Lambda}
\left(  \hHij{ij} + \hHpij{ij} \right)
+ \cij{ij}_I \xii{i}^I
+ \cij{j}_\Lambda \p_{\txii{j}_\Lambda}  \hHpij{ij} \, .
\end{split}
\ee
As usual, the functions $\hHpij{ij}$ must satisfy the co-cycle condition \eqref{consist_condH}
and are defined up to the gauge equivalence \eqref{realH}. Thus, they define an element
in the \v{C}ech cohomology group\footnote{The twisting by $\cO(2)$ is not apparent
in our formalism, as explained in footnote 14, but follows from the fact that $\hHpij{ij}$
originates from a homogeneous function of degree 1 on $\cS$.}
$H^1(\cZ_\cM, \cO(2))$, realizing Lebrun's
assertion that this group classifies the QK deformations of $\cM$  \cite{MR1327157}.

We now determine the deformed contact twistor lines.
For definiteness we focus on the coordinate $\xii{+}^\Lambda(\varpi,x^\mu)$ around
$\zeta=\zeta_+$, i.e. $\varpi=0$. The pole and  the constant term in the Laurent expansion
\eqref{behap} are readily obtained by contour integrating around $\varpi=0$:
\be
\label{xipp}
\xii{+}^\Lambda(\varpi,x^\mu) =  Y_+^\Lambda \varpi^{-1}
+ A_+^\Lambda + \cO( \varpi) \, .
\ee
where
\be
\label{par_AY}
A_+^\Lambda = \oint_0 \frac{\de\varpi}{2\pi\I\,\varpi}\, \xii{+}^\Lambda\ ,\qquad
Y_+^\Lambda = \oint_0 \frac{\de\varpi}{2\pi\I\,}\, \xii{+}^\Lambda \, .
\ee
On the other hand, the full Laurent series expansion of
$\xii{+}^\Lambda(\varpi,x^\mu)$ at $\varpi=0$
is given by the series
\be\label{xiicont}
\xii{+}^\Lambda(\varpi) =
Y_+^\Lambda  \varpi^{-1}+ \sum_{n=0}^{\infty} \oint_0 \frac{d\varpi'}{2\pi\I\,\varpi'}
\left( \frac{\varpi}{\varpi'}\right)^n \xii{+}^\Lambda(\varpi')\, .
\ee
The contour around $\varpi=0$ may be deformed into a sum of contours $\tilde C_j$ around
the other singularities in the $\varpi$ plane. Using the contact transformations \eqref{Tfct} on
each patch, we obtain
\be
\xii{+}^\Lambda(\varpi) =Y_+^\Lambda  \varpi^{-1}
- \sum_{n=0}^{\infty}  \sum_{j\ne +} \oint_{\tilde C_j}
\frac{d\varpi'}{2\pi\I\varpi'}
\left( \frac{\varpi}{\varpi'}\right)^n
\left[ \xii{j}^\Lambda(\varpi')
- T_{[+j]}^\Lambda(\varpi') \right] \, .
\ee
The first term in the square bracket gives a non-vanishing contribution for  $j=-$ and $n=0,1$ only,
while the second term contributes an infinite Laurent series.
Therefore, we arrive at the following representation
\be
\label{xipm}
\xii{+}^\Lambda(\varpi) =  Y_+^\Lambda  \varpi^{-1}
+ A_-^\Lambda - Y^\Lambda_-  \varpi
+\sum_j \oint_{\tilde C_j} \frac{d\varpi'}{2\pi\I}\, \frac{1}{\varpi'-\varpi}\,
T_{[+j]}^\Lambda(\varpi') \, .
\ee
where now
\be
A_-^\Lambda = -\oint_\infty \frac{\de\varpi}{2\pi\I\,\varpi}\, \xii{-}^\Lambda\, ,
\qquad
Y_-^\Lambda= \oint_\infty \frac{\de\varpi}{2\pi\I\,\varpi^2}\, \xii{-}^\Lambda\, .
\ee
From the reality conditions \eqref{rexixit}, we conclude
that $A_-=(A_+)^*,\ Y^\Lambda_-=(Y^\Lambda_+)^*$.
Comparing the $\cO(\varpi^0)$ terms between \eqref{xipp} and \eqref{xipm}
gives the difference
\be\label{ApAmid}
A_+^\Lambda - A_-^\Lambda = \sum_j \oint_{\tilde C_j} \frac{\de \varpi'}{2 \pi \I \varpi'} \,
T_{[+j]}^\Lambda(\varpi') \, .
\ee
Eliminating $A_-^\Lambda$ in \eqref{xipm} in favor of the real quantity
$A^\Lambda= ( A_+^\Lambda + A_-^\Lambda)/2$ leads to
\be
\label{xiqline}
\xii{i}^\Lambda(\varpi) = A^\Lambda +
\left( \varpi^{-1} Y_+^\Lambda - \varpi Y^\Lambda_-  \right)
+\frac12 \sum_j \oint_{\tilde C_j}\frac{d\varpi'}{2\pi\I\, \varpi'}\,
\frac{\varpi'+\varpi}{\varpi'-\varpi}\, T_{[+j]}^\Lambda(\varpi')
\ee
for $i=+$. As observed below \eqref{eqmuqh}, these equations are in fact valid in
any patch $\cU_i$, since they exhibit the correct discontinuities across the contours
$\tilde C_i$.

In an analogous way one may obtain the deformed conjugate coordinates $\txii{i}_I$. The
Laurent coefficient $\txi_{I,0}^{[+]}$ may be extracted by integrating
\be\label{txi:zero}
\txi_{I,0}^{[+]} =  \,
\oint_0 \frac{\de \varpi'}{2 \pi \I \varpi'} \left( \txii{+}_I - \ci{+}_I \log\varpi' \right)
=  \I B_I^+ - \cij{+}_I \oint_0 \frac{\de \varpi'}{2 \pi \I \varpi'} \,
\log\left(\varpi'\xii{+}^0 \right) \, ,
\ee
where we defined
\be
B_I^+ \equiv -\I \oint_0 \frac{\de \varpi'}{2 \pi \I \varpi'}
\left( \mui{+}_I + \cij{+}_I \log \nui{+}^0  \right)\, .
\ee
On the other hand, the Laurent series expansion around $\varpi=0$ may be obtained
by deforming the contour around  $\varpi=0$  into a sum of contours
around the other singularities in the $\varpi$ plane, and use the symplectomorphism
\eqref{pert:symp} to map $\txii{+}_I$ to $\txii{j}_I$:
\be\label{txi:inf}
\txi_{I}^{[+]}(\varpi) =   \ci{+}_I \log\varpi  - \sum_{n=0}^{\infty}\sum_{j\ne +} \oint_{\tilde C_j}
\frac{\de \varpi'}{2 \pi \I \varpi'}\left( \frac{\varpi}{\varpi'}\right)^n
\left( \txii{j}_I - \cij{+}_I \log\varpi' - \tilde{T}_I^{[+j]} \right) \, ,
\ee
where $\tilde{T}^{[ij]}_I$ are defined in \eqref{Tfct}.
The first two terms in the bracket only contribute when $j=-$.
The cancelation of the logarithmic singularity at $\varpi=\infty$
is ensured by the condition $\ci{+}_I=-\ci{-}_I$, corresponding to the figure-eight contour
prescription discussed in \cite{Alexandrov:2008ds}.
Using \eqref{beham}, we obtain
\be\label{txi:inf2}
\txi_{I}^{[+]}(\varpi) =   \ci{+}_I \log\varpi + \I B^-_I
+ \cij{-}_I \oint_\infty \frac{\de \varpi'}{2 \pi \I \varpi'}\,
\log\frac{ \xii{-}^0}{\varpi'}
+ \sum_j \oint_{\tilde C_j} \frac{\de \varpi'}{2 \pi \I (\varpi'-\varpi)} \, \tilde{T}_I^{[+j]} \, ,
\ee
where
\be
B_I^- \equiv \I \oint_\infty \frac{\de \varpi'}{2 \pi \I \varpi'} \, \left( \mui{-}_I
+ \cij{-}_I \log \nui{-}^0  \right)
\ee
is the complex conjugate of $B_I^+$.
Comparing the $\varpi$-independent terms in \eqref{txi:inf} and \eqref{txi:inf2} establishes the identity
\be
\I \left( B^+_I - B^-_I \right)
=  \cij{+}_I \oint_0 \frac{\de \varpi'}{2 \pi \I \varpi'}\, \log \varpi' \xi^0_{[+]}
- \cij{-}_I \oint_\infty \frac{\de \varpi'}{2 \pi \I \varpi'}\, \log\frac{ \varpi' }{  \xi^0_{[-]}}
+ \sum_j \oint_{\tilde C_j} \frac{\de \varpi'}{2 \pi \I \varpi'} \, \tilde{T}_I^{[+j]} \, .
\ee
Eliminating $B_I^-$ in \eqref{txi:inf2} in favor of $B_I = B_I^+ + B_I^-$  leads to
\be
\begin{split}
\txi_I^{[+]}(\varpi) = &  \frac{\I}{2}\, B_I -
\frac12\, \cij{+}_I  \left[ \oint_0 \frac{\de \varpi'}{2 \pi \I \varpi'}\,
\log\left( \varpi'  \xii{+}^0 \right)
+ \oint_\infty \frac{\de \varpi'}{2 \pi \I \varpi'}\,
\log\left( \xii{-}^0 / \varpi'   \right) \right]
\\
&+\half  \sum_j \oint_{\tilde C_j} \frac{\de \varpi'}{2 \pi \I \varpi'} \,
\frac{\varpi' + \varpi}{\varpi' - \varpi}
\, \tilde{T}_I^{[+j]}+  \ci{+}_I \log\varpi \, .
\end{split}
\ee
Using the fact that $ \xii{+}^0\sim Y^0_+/\varpi'$ and
$ \xii{-}^0\sim Y^0_- \varpi'$ at $\varpi'=0$ and $\infty$, we  finally obtain
\be
\label{txiqline}
\txi_I^{[i]}(\varpi) = \frac{\I}{2}\, B_I
+\half  \sum_j \oint_{\tilde C_j} \frac{\de \varpi'}{2 \pi \I \varpi'} \,
\frac{\varpi' + \varpi}{\varpi' - \varpi}
\, \tilde{T}_I^{[+j]}+  \ci{+}_I \log \left( \varpi \sqrt{\frac{Y^0_-}{Y^0_+}}\right)
\ee
for $i=+$, and in fact also for any $i$. This relation generalizes \eqref{eqmuqh} to the  perturbed case.
Taken together, \eqref{xiqline} and \eqref{txiqline} give the contact twistor lines of the deformed
twistor space in terms of the perturbation $\Hpij{ij}$, which is considered as a function
of the undeformed twistor lines $\bxii{i}^\Lambda,\ \btxii{i}_I$ given in \eqref{xiY} and \eqref{eqmuqh}.
In order to make contact with the construction in Section \ref{sec33}, one should recall that
$Y^\Lambda_+=\cR Z^\Lambda$, where
\be
Z^\Lambda = \oint_0 \frac{\de\varpi}{2\pi\I\,\varpi}\,
\frac{\xii{+}^\Lambda}{\xii{+}^0}\, ,
\qquad
\cR= \oint_0 \frac{\de\varpi}{2\pi\I}\, \xii{+}^0 \, ,
\ee
in such a way that $Z^0=1$.

To obtain the perturbed \qk metric, we should also calculate
the leading Laurent coefficients of the contact potentials
$\Phi\di{\pm}$ given in \eqref{contpot}. It turns out that one can
actually compute the full contact potentials, using the
gluing conditions
\be
e^{-\Phi_{[i]}} - e^{-\Phi_{[j]}}  =  e^{-\phi} \p_{\txii{j}_\flat }\hHpij{ij}\, .
\ee
where $\phi$ is defined by \eqref{coor} in the unperturbed geometry.
These conditions follow from the gluing conditions for  $\tnui{i}^\flat$ in \eqref{transtnu},
using  the results for the transition functions
$\hat f_{ij}^2$ \eqref{perthf} and the unperturbed $\tnui{i}^\flat$ \eqref{nuom}.
Repeating again the same steps as above, one easily arrives at
\be
e^{\Phi_{[i]}} =e^{\phi}\(1+
\hf \sum_j \oint_{\tilde C_j} \frac{\de \varpi'}{2 \pi \I \varpi'} \,\frac{\varpi' + \varpi}{\varpi' - \varpi}
\, \p_{\txii{j}_\flat }\hHpij{0j}(\varpi') \)\, ,
\label{perttnu}
\ee
where $\phi$ is defined in the perturbed case as
\be
\label{phipot}
\phi \equiv \Re{\Phi_{[+]}}(\varpi=0) =
\frac12 \left( \phi_{[+]}^0 + \phi_{[-]}^0 \right)\, .
\ee
Note that this definition coincides with \eqref{coor} in the unperturbed case.
Using \eqref{contpot}, the leading Laurent coefficient of the contact potentials
are given by
\be
\begin{split}
e^{\phi_{[+]}^0} &= \frac12 \, Y^{\Lambda} _+
 \sum_j\oint_{\tilde C_j}\frac{d\varpi}{2\pi\I\varpi^2}\,
 \tilde{T}^{[0j]}_\Lambda
+\half\, \ci{+}_\Lambda \left( A^\Lambda +\frac12\sum_j\oint_{\tilde C_j}
\frac{d\varpi}{2\pi\I\, \varpi}\,T_{[0j]}^\Lambda \right)+ \half\, \ci{+}_\flat \, ,
\\
e^{\phi_{[+]}^0} &= - \frac12\,  Y^{\Lambda}_-
 \sum_j\oint_{\tilde C_j} \frac{d\varpi}{2\pi\I}\,
\tilde{T}^{[0j]}_\Lambda
-\half\, \ci{-}_\Lambda \left( A^\Lambda -\frac12  \sum_j\oint_{\tilde C_j}
\frac{d\varpi}{2\pi\I\, \varpi}\, T_{[0j]}^\Lambda \right)-\half\, \ci{-}_\flat \, .
\end{split}
\label{pertcontpots}
\ee
Inserting in \eqref{phipot}  we therefore obtain
\be
e^\phi=\frac14 \sum_j\oint_{\tilde C_j}\frac{d\varpi}{2\pi\I\, \varpi}
\(\varpi^{-1} Y^{\Lambda}_+ -\varpi Y^{\Lambda}_- \) \tilde{T}^{[0j]}_\Lambda
+\frac14\, \cij{+-}_I  A^I  \, ,
\label{eqchip}
\ee
and the full contact potentials via \eqref{perttnu}. As a useful consistency check,
note that the difference of \eqref{pertcontpots} can be rewritten, after some
considerable work, as
\be
\phi_{[+]}^0-\phi_{[-]}^0 =\sum_j \oint_{\tilde C_j} \frac{\de \varpi}{2 \pi \I \varpi}
\, \p_{\txii{j}_\flat }\hHpij{0j}\, ,
\label{imphi}
\ee
consistently with  \eqref{perttnu}.
Altogether these results allow us to extract the metric
following the procedure outlined at the end of Section \ref{seccontact}.

It is straightforward to relate this contact construction
on $\cZ_\cM$ to the symplectic construction
on  $\cZ_\cS$. For this purpose, one needs to apply the change of variable \eqref{zztpm},
where $\ztpm$ denote the location of the zeros of the perturbed section $\nu^\flat$, to
all contour integrals in the $\varpi$ plane. Under this change of variable,
the integration measure becomes
\be
\frac{\de\varpi}{2 \pi \I \varpi} = \frac{(\ztp-\ztm)}{(\zeta-\ztp)(\zeta-\ztm)}
\frac{\de\zeta}{2\pi \I}\, .
\ee
Its expression to the first order in deformation can be found in \eqref{changevarpi}.
However, in some cases it can be simplified. For example, integrated against a function
which is regular at $\zeta_\pm$, this may be rewritten as
\be
\label{jacopidef}
\frac{\de\varpi}{2 \pi \I \varpi} = \frac{\Rf _\pm\, \de\zeta}{2 \pi \I f^2_{0 \pm} \, \nui{\pm}^\flat}\, ,
\qquad
\frac{1}{\Rf_\pm}\equiv \oint_{C_+} \frac{\de\zeta}{2\pi\I \, f_{0\pm}^2 \nui{\pm}^\flat}\, ,
\ee
where the factor $\rf_\pm$ ensures that the residue at $\zeta=\ztpm$ is equal to one.
Integrated a function with a simple pole at $\zeta_\pm$, \eqref{jacopidef} must be generalized to
\be
\label{jacopidef2}
\frac{\de\varpi}{2 \pi \I \varpi} = \frac{\de\zeta}{2 \pi \I}
\left[ \frac{\Rf _\pm\, }{f^2_{0 \pm} \, \nui{\pm}^\flat}
+ \left( \frac{1}{\zeta_- - \zeta_+} + \frac{s^\flat_\pm}{\Rf_\pm} \right) \right]\, ,
\ee
where $s_\pm^\flat$ is the second coefficient in the Taylor expansion
\be
f^2_{0 \pm} \, \nui{\pm}^\flat = \Rf_\pm (\zeta-\ztpm) + s_\pm^\flat  (\zeta-\ztpm)^2 +\dots \, .
\ee
Note that the correction term in round bracket in \eqref{jacopidef2}
vanishes in the case where $\nu^\flat$ remains a global $\cO(2)$ section.
In this way, we may rewrite the invariant coordinates introduced above
as follows
\be\label{eq:coord}
\begin{split}
Y^\Lambda_\pm \equiv& \Rf_\pm  \oint_{C_\pm} \frac{\de \zeta}{2 \pi \I f_{0\pm}^2} \,
\frac{\nu^\Lambda_{[\pm]}}{(\nui{\pm}^\flat)^2} \, \varpi ^{\pm 1}  \, ,
\\
A_\pm^\Lambda \equiv & \, \pm   \oint_{C_\pm}
\frac{\de\zeta}{2 \pi \I}
\left[ \frac{\Rf _\pm\, }{f^2_{0 \pm} \, \nui{\pm}^\flat}
+ \left( \frac{1}{\zeta_- - \zeta_+} + \frac{s^\flat_\pm}{\Rf_\pm} \right) \right]
\frac{\nui{\pm}^\Lambda}{\nui{\pm}^\flat}\, ,
\\
B^\pm_I \equiv  &\,  \mp  \I  \Rf_\pm \oint_{C_\pm}
\frac{\de \zeta}{2 \pi \I f_{0\pm}^2\,\nui{+}^\flat}
\left( \mu^{[\pm]}_I + \ci{\pm}_I \log \nui{\pm}^0\right)\, .
\end{split}
\ee

From the computation of the deformed \hk potential \eqref{finalchi} in Appendix C,  one may check
that  the relation \eqref{coor}  continues to hold
after perturbation, provided with $\phi$ is defined by \eqref{phipot} and
$\Rf=(\Rf_+ + \Rf_-)/2$. From the general equation \eqref{chinuflat},
this implies that
\be
\Rf = |\vf\hat f_{0+}^2|\,\frac{ 1+z \bz}{|z   |}\,
e^{\Re[\Phi_{[+]} (x^\mu,z) -\phi^0_{[+]} (x^\mu)] }\, .
\ee
We have not attempted to check this relation directly.

In \cite{apsv3}, we shall apply this general framework to the hypermultiplet moduli
space in compactifications of type II string theory on a Calabi-Yau three-fold.

\acknowledgments

We are grateful to A.~Neitzke for discussions and former collaboration on related topics.
The research of S.A. is supported by CNRS and by the contract
ANR-05-BLAN-0029-01. The research of B.P. is supported in part by ANR(CNRS-USAR)
contract no.05-BLAN-0079-01. F.S.\ acknowledges financial support from the ANR grant BLAN06-3-137168.
S.V. thanks the Federation de Recherches ``Interactions Fondamentales'' and LPTHE
at Jussieu for hospitality and financial support.
Part of this work is also supported by the EU-RTN network MRTN-CT-2004-005104
``Constituents, Fundamental Forces and Symmetries of the Universe''.

\appendix

\section{Infinitesimal $SU(2)$ transformations}
\label{ap_SU2}

In this appendix, we study the infinitesimal action of  $SU(2)$
on the local sections introduced in the main text. We parametrize the
Lie algebra of $SU(2)$ by $\eps_\pm=(\eps_\mp)^*$ and $\eps_3=(\eps_3)^*$
such that
\be
\label{alphaeps}
\begin{pmatrix} \alpha & \beta \\ -\bar\beta & \bar\alpha \end{pmatrix}=\begin{pmatrix}
1-\frac{\I}{2} \eps_3 & \eps_+ \\ - \eps_- & 1+\frac{\I}{2} \eps_3
\end{pmatrix} +\cO(\eps^2)\, .
\ee
The infinitesimal action of $SU(2)$ on $\pi^{A'},\bpi_{A'}$ is given in \eqref{su2pi},
\be
\label{su2pi_ap}
\delta \begin{pmatrix} \pi^1 & \pi^2 \\
-\bpi_2 & \bpi_1 \end{pmatrix}=
\begin{pmatrix} \frac{\I}{2} \eps_3 & -\eps_+ \\
\eps_- & -\frac{\I}{2} \eps_3
\end{pmatrix}\cdot
\begin{pmatrix} \pi^1 & \pi^2 \\
-\bpi_2 & \bpi_1 \end{pmatrix}\, ,
\ee
The finite action of $SU(2)$ on $\cO(2n)$ sections was discussed in section 2.4.
At the infinitesimal level, \eqref{su2ze} reduces to
\begin{equation}
\delta \zeta \equiv \zeta'-\zeta =
\epsilon_+ -\I \epsilon_3 \zeta +\epsilon_- \zeta^2 +\cO(\eps^2)\, .
\end{equation}
In the  patch $i=0$,  the $\cO(2n)$ transformation rule \eqref{su2diff} then leads to
\be
\label{infi-su2diff}
\begin{split}
\delta \nu^I_{[0]}(\zeta)
&\equiv \nu^{'I}_{[0]}(\zeta) - \nu^{I}_{[0]}(\zeta) \\
&=\left[\eps_+ \pa_{\zeta} - \I \eps_3 \left( \zeta\pa_{\zeta} - n \right)
+ \eps_- \left( \zeta^2\pa_{\zeta} - 2n \zeta \right) \right]
\nui{0}^I(\zeta)+\cO(\eps^2)\, .
\end{split}
\ee
Thus, the Taylor coefficients of $\nui{0}=\sum_m  \nu_m^I \zeta^m$
around $\zeta=0$ vary under an infinitesimal $SU(2)$ action by
\be
\label{su2tay}
\delta \nu_m^I =\,   (m+1) \nu_{m+1}^I \eps_+ - \I  (m - n) \nu_m^I \eps_3
+  (m-2n-1) \nu_{m-1}^I \eps_- \,  .
\ee
The variation of $\muor_I$ and its Laurent coefficients $\mu_{I,m}$
is obtained by replacing $\nu^I\to \mu_I, n\to 1-n$ in these expressions.

In an arbitrary patch $\cU_i$, the $SU(2)$ action  \eqref{su2diff} is most easily expressed in
terms of $f_{i0}^{-2n}(\zeta) \,\nu_{[i]}^{I}(\zeta)$, which formally transforms
in the same way as $\nuor^{I}(\zeta)$. Similarly, the $SU(2)$ action \eqref{mu-anom}
is most easily stated in terms of $f_{i0}^{-2}(\zeta) \exp( - \mui{i}/c_I^{[i]})$, which also
formally transforms in the same way as $\nuor^{I}(\zeta)$. After the gauge transformation
\eqref{quasG}, the transformation rules of $\mui{i}_I$ are changed to
\be
\delta\mui{i}_{T;I}(\zeta)=\left( \eps_+ -
\I \eps_3 \zeta + \eps_- \zeta^2 \right)\p_\zeta \mui{i}_{T;I}(\zeta)
+\(  \frac{\eps_+}{\zeta} - \I \eps_3  + \eps_- \zeta  \)\ci{0}_I
-\( \frac{\eps_+}{\zeta}- \eps_- \zeta\)\ci{i}_I \, ,
\label{su2defci}
\ee
consistently with the fact that \eqref{truemu}
transforms like a non-anomalous $\cO(0)$ section.

The variation \eqref{su2tay}  applies for the Laurent coefficients of any local
section of $\cO(2n)$. In particular, one may consider a homogeneous function $G(\nu_{\alpha})$
of $\cO(2n_{\alpha})$ multiplets $\nu_{\alpha}$, of homogeneity degree $n$ when each
$\nu_{\alpha}$ are scaled with homogeneity degree $n_{\alpha}$:
\be
\label{homg}
\sum_{\alpha} n_{\alpha}\, \nu_{\alpha} \pa_{\nu_{\alpha}} G = n \,G\, .
\ee
Then, for an arbitrary contour $\Gamma$ (not necessarily surrounding
the origin),  the $SU(2)$ variation of the integrals
\be\label{ionasinv}
G_m \equiv  \oint_\Gamma \frac{\de\zeta}{2\pi\I\, \zeta^{m+1}} \, G(\nu_{\alpha})
\, ,
\ee
is given by
\be
\label{su2G}
\delta G_m = (m+1) G_{m+1} \eps_+ - \I  (m - n) G_m \eps_3
+  (m-1-2n) G_{m-1} \eps_-\, ,
\ee
as one can check from explicit calculation.
In particular, for $n=m=-1$, we recover the remark in
\cite{Ionas:2007gd2}, according to which a contour integral of
a section of $\cO(-2)$ is $SU(2)$-invariant. For $n=-3/2$, the contour
integral of a section of $\cO(-3)$  with $m=-1,-2$ instead produces a
$SU(2)$ doublet. These observations are central to
the superconformal quotient discussed in Section \ref{subsec_QKmap}.

\section{An alternative formulation for hypermultiplet moduli spaces}
\label{ap_hyper}

In this appendix we explain the relation between the formulation of the hypermultiplet
space used in Section 4, and the one introduced  in \cite{Alexandrov:2007ec} using
a different contour prescription, and establish their equivalence up to a local
symplectomorphism.

Aiming for a Lagrangian $\cL^\prime$ whose limit $\vf\to 0$
 is regular, \cite{Alexandrov:2007ec} considered the contour integral
\be
\label{contcmapm}
\CL'(v,\bv,x) = {\rm Im} \, \oint_{C} \frac{\de\zeta}{2 \pi\I\zeta}
\left( \frac{F(\eta^\Lambda)}{\eta^\flat} + 4 \I c \,\etaf \log\etaf \right) \, ,
\ee
where  the contour $C$ appearing in \eqref{contcmap} encircles the poles
$\zeta=0,\ztp$ counter-clockwise direction and the logarithmic brach cuts connect
$0, \zeta_+$ and $\zeta_-, \infty$, respectively. The resulting Lagrangian
\be\label{tenslagr}
\CL'=
\frac{1}{\rf}\,\Im F(\etap)-\xf\Im\frac{F(v)}{(\vf)^2}+x^\Lambda\Im\frac{F_\Lambda(v)}{\vf}
 +4 c \left( \xf - \rf + \xf \log\frac{\xf+\rf}{2} \right)\, ,
\ee
differs from \eqref{cont-cmap}, \eqref{Loneloop} by terms
linear in $x^I$ only and therefore describes the same
metric. In terms of our general discussion,
the contour prescription \eqref{contcmapm} arises from
the transition functions
\be\label{symp-cmapnew}
\begin{split}
&\pHij{0+}= 0\, ,
\qquad
\pHij{0i}=\frac{\I}{2}\,\frac{F(\eta)}{\etaf}\, ,
\\
\pHij{0-}&= \pHij{0\infty}=
\frac{\I}{2}\, \frac{F(\eta)-\bF(\eta)}{\etaf}-4 c\etaf\log\etaf\, ,
\end{split}
\ee
where $i$ labels the patches where $F(\eta)$ is singular.
These transition functions are related to the ones given in \eqref{symp-cmap}, \eqref{1ltrans}
by the gauge transformation generated by
\be
\label{eq:gt}
\begin{split}
\Gi{0} = \frac{\I}{2}\, \frac{F(\eta)}{\etaf}  - 2 c \, \etaf\log(\etaf \zeta)\, &,
\qquad \Gi{\infty}= \frac{\I}{2}\,
\frac{\bF(\eta)}{\etaf}  + 2 c \, \etaf\log(\etaf/ \zeta)\, ,\\
\Gi{+}=\Gi{-}&=\Gi{i}=-2c   \, \etaf\log(\zeta) \, .
\end{split}
\ee
The new non-vanishing quasi-homogeneity coefficients \eqref{def:cij} are
\be
\ci{0}_\flat= \ci{+}_\flat=-2c \, ,
\qquad
\ci{\infty}_\flat= \ci{-}_\flat=2c\, .
\ee
Note that in contrast to the description in Section \ref{sec_firstflat}, the coefficient
$\ci{0}_\flat$ does not vanish, corresponding to the Lagrangian
\eqref{tenslagr} being quasi-homogeneous.

We now discuss the twistor lines arising from this new contour prescription.
The global $\cO(2)$ sections $\eta(\zeta)$ are unchanged
while the gauge transformation \eqref{eq:gt} induces
\be
\pmui{0}_\Lambda(\zeta) = \mui{0}_\Lambda(\zeta)
-\,\frac{\I}{2} \frac{F_\Lambda(\eta)}{\etaf}  \, , \quad
\pmui{0}_\flat(\zeta) = \mui{0}_\flat(\zeta)
+\,\frac{\I}{2}\frac{F(\eta)}{(\etaf)^2} + 2 c\, ( \log( \etaf \zeta) +1)\, .
\ee
Consequently, the coordinates $w_I$ become
\be
\begin{split}
w'_\Lambda = w_\Lambda -\ \frac{\I}{2}\frac{F_\Lambda(v)}{\vf} \, ,
\qquad
w'_\flat = w_\flat +\,\frac{\I}{2}\frac{F(v)}{(\vf)^2} + 2 c\, (\log( \vf)+1) \, .
\end{split}
\ee
The corresponding map between the coordinates $\varrho_I$ and $\varrho_I'$
is readily obtained using $\varrho_I = - \I(w_I - \bw_I)$.
Furthermore, the base coordinates \eqref{intQKmap}-\eqref{coorBqh} are identical,
\be
e^{\phi'} = e^{\phi} \, ,
\quad
Z'^a = Z^a \, ,
\quad
A'^\Lambda = A^\Lambda \, ,
\quad
B'_I = B_I \, ,
\ee
where the last equality follows from \eqref{coorBqh} upon a brief computation.
Taking into account also that only $\cij{+-}_I$, which are equal in the two formulations,
contribute to the general expressions \eqref{eqmuqh},
this, in turn, implies that both contour prescriptions give rise to the same
twistor lines \eqref{1ltwistor}.

\section{Deformed superconformal quotient}
\label{ap_superquot}

In this appendix we generalize the superconformal quotient procedure
of Section \ref{subsec_QKmap} to include deformations. While conceptually straightforward,
this procedure is toilsome compared to the contact geometry approach of
Section \ref{pertqline}. Nevertheless, we include it here for completeness, as it
provides useful consistency checks on our formalism,

\subsection*{Coordinates on the deformed base}
\label{subsec_QKmappert}

As in  the undeformed case described in Section \ref{subsec_QKmap},
$SU(2)$ invariant functions on the Swann bundle $\cS$ can be obtained
by contour-integrating $\cO(-2)$ sections on $\cZ_\cS$. In the presence of
deformations, the global sections   $\eta^I$ in \eqref{intQKmap} must be replaced
by the deformed local sections $\nui{+}^I$, leading to the
definitions\footnote{Note that $d\zeta/f_{0+}^2 = d\zeta\ui{+}$ is
the natural integration measure in the patch $\cU_+$.}
\be
\begin{split}
& \displaystyle
\qquad
\frac{1}{\Rf}\equiv \Re \oint_{C_+}
\frac{\de\zeta}{2\pi\I \, f_{0+}^2}\,\frac{1}{\nui{+}^\flat}\, ,
\qquad
{\cA^\Lambda} \equiv {\Rf}\Re\oint_{C_+}
\frac{\de\zeta}{2\pi\I\, f_{0+}^2}\,\frac{\nui{+}^\Lambda}{(\nui{+}^\flat)^2}\, ,
\\
& \displaystyle
\cZ^a \equiv  {\Rf}\oint_{C_+} \frac{\de\zeta}{2\pi\I\, f_{0+}^2}\,
\frac{\nui{+}^a}{\nui{+}^\flat\nui{+}^0}\, ,
\qquad
\CB_I\equiv 2{\Rf}\Im \oint_{C_+} \frac{\de\zeta}{2\pi\I\, f_{0+}^2}\,
\frac{\mui{+}_{I} + \cij{+}_I \log \nui{+}^0}{\nui{+}^\flat}\, .
\end{split}
\label{intQKmappert}
\ee
To first order in the deformation, this gives
\be
\begin{split}
\Rf &= \brf+ \hf\(\nuppz^\flat-\nupmz^\flat\)
+\frac{\bvf}{\rf}\, \(\nupp^\flat+\nupm^\flat\)  \, ,
\\
\cZ^a &=  \breve Z^a+\frac{\nupp^a-\breve A^a\nupp^\flat}{\ztp\etap^0}
-\frac{\nupp^0-\breve A^0\nupp^\flat}{\ztp\etap^0}\,\breve Z^a\, ,
\\
\cA^\Lambda &= \breve A^\Lambda+\frac{1}{2\rf}\( \nuppz^\Lambda -\nupmz^\Lambda
+\frac{2\bvf}{\rf}\( \nupp^\Lambda+\nupm^\Lambda\)\) \\
&
-\frac{1}{(\rf)^2}\Re\Bigl[\ztp\etap^\Lambda \nuppzz^\flat  +\(2\rf \breve A^\Lambda
- x^\Lambda+2\bv^\Lambda \ztp\) \nuppz^\flat
+2\(2\bvf \breve A^\Lambda-\bv^\Lambda \)\nupp^\flat\Bigr]\, ,
\end{split}
\label{newcR}
\ee
where $\breve\ $ marks the unperturbed quantities defined in
\eqref{intQKmap} and we introduced
\be
\nuppm^I=f_{0\pm}^2\nupi{\pm}^I(\ztpm) \, ,
\qquad
\nuppmz^I=\p_{\zeta}\(f_{0\pm}^2\nupi{\pm}^I\)(\ztpm)\, ,
\qquad
\nuppmzz^I=\p^2_{\zeta}\(f_{0\pm}^2\nupi{\pm}^I\)(\ztpm)\, .
\ee
We omitted the expansion of $\CB_I$ since it will not be needed.
The $SU(2)$ invariant  $\cR$ defined in \eqref{defcR}  may also be
extended to the deformed case as
\be
\cR =\breve\cR\[1+\hf\(\frac{\nupp^0-\breve A^0 \nupp^\flat}{\ztp\etap^0} +\frac{\nupm^0
-\breve A^0 \nupm^\flat}{\ztm\etam^0}\)
-\frac{1}{2\rf}\( \nuppz^\flat-\nupmz^\flat+\frac{2\bvf}{\rf}\( \nupp^\flat+\nupm^\flat\)\)\]\, .
\ee
Using the formulae given at the end of this appendix, one can check explicitly
that the above expressions are indeed $SU(2)$ invariant.

\subsection*{Coordinates on the $\IC^2/\IZ_2$ fiber}

The coordinates $\pi^{A'}$ on the fiber of $\cS$, \eqref{pieq} can be
similarly generalized to the deformed case as follows:
\be
\begin{split}
\label{pertpi}
\pi^1 &= C \int_{C_+}\frac{\de\zeta}{2\pi \I}\,
\frac{\zeta}{f_{0+}^2\nui{+}^\flat\bigl({f_{0+}^2\nui{+}^0}\bigr)^{1/2}}
\\
&=  \breve \pi^1
\[1-\frac{\nupp^0}{2\ztp\etap^0}-\frac{\nuppz^\flat}{\rf}
-\frac{\nupp^\flat}{2\rf}\(\frac{4\bvf}{\rf}+\frac{x^0+2v^0/\ztp}{\ztp\etap^0}\)\]\, ,
\\
\pi^2 &=
- \bC \int_{C_-}\frac{\de\zeta}{2\pi \I}\,
\frac{\zeta}{f_{0-}^2\nui{-}^\flat\bigl({-f_{0-}^2\nui{-}^0}\bigr)^{1/2}}
\\
&= \breve \pi^2
\[1-\frac{\nupm^0}{2\ztm\etam^0}+\frac{\nupmz^\flat}{\rf}
-\frac{\nupm^\flat}{2\rf}\(\frac{4\bvf}{\rf}-\frac{x^0+2v^0/\ztm}{\ztm\etam^0}\)\]
\, .
\end{split}
\ee
The conjugate variables $\bar \pi_{A'}$ can be obtained from \eqref{pertpi} using
\be
\overline{\ztp}=-1/\ztm\, ,
\qquad
\overline{\etap^\Lambda}=\etam^\Lambda\, ,
\qquad
\overline{\nupp^I}=-\nupm^I/\ztm^2\, ,
\qquad
\overline{\nuppz^I}=-\nupmz^I+2\nupm^I/\ztm\, .
\label{conjug}
\ee
In particular, one has very simple relations
\be
\ztp\equiv -\frac{\pi^1}{\bpi_2}=\brztp\(1-\frac{\nupp^\flat}{\rf\ztp}\)\, ,
\qquad
\ztm\equiv \frac{\pi^2}{\bpi_1}=\brztm\(1+\frac{\nupm^\flat}{\rf\ztm}\)\, ,
\label{genzeros}
\ee
whereas the variable $z=\pi^1/\pi^2$ parametrizing the fiber of $\cZ_{\cM}$ is given by
\bea
z&= \breve z &
\[1-\frac{\nupp^0}{2\ztp\etap^0}+\frac{\nupm^0}{2\ztm\etam^0}
-\frac{\nuppz^\flat+\nupmz^\flat}{\rf}
\right.
\nonumber \\
&& \left.
-\frac{\nupp^\flat}{2\rf}\(\frac{4\bvf}{\rf}+\frac{x^0+2v^0/\ztp}{\ztp\etap^0}\)+
\frac{\nupm^\flat}{2\rf}\(\frac{4\bvf}{\rf}-\frac{x^0+2v^0/\ztm}{\ztm\etam^0}\)\]\, .
\label{pertz}
\eea
To first order in the perturbation, the two quantities in  \eqref{genzeros}
provide the zeros $\ztpm$ of the deformed section $\nu^\flat$, and
\eqref{varpidef},\eqref{zztpm} continue to hold. Using
the explicit expressions \eqref{pertpi} for $\pi^{A'}$, one finds
\be
\label{changevarpi}
\frac{\de \varpi}{\varpi}
=\[\frac{\rf }{\zeta\etaf}-\frac{1}{\rf}\(\frac{\nupp^\flat}{(\zeta-\ztp)^2}
+\frac{\nupm^\flat}{(\zeta-\ztm)^2}  \)\]\de\zeta \, .
\ee

\subsection*{Relation to the contact geometric approach}

Using this relation, one may relate the coordinates defined here
to those defined in Section \ref{pertqline}:
\be
\label{relinvpert}
\begin{split}
Y_+^\Lambda & =\cR \, \cZ^\Lambda \(1-3\I\sum_j\int_{\tC_j}\frac{\de \bbz}{2\pi\I \bbz}\,
\p_{\rho_\flat}\Hpij{0j}(\breve\xi(\bbz),\rho(\bbz))\)\, ,\\
A^\Lambda & =\CA^\Lambda+
\I\breve\cR\sum_j\int_{\tC_j}\frac{\de \bbz}{2\pi\I \bbz}\( \bbz^{-1}Z^\Lambda + \bbz\bZ^\Lambda \)
\p_{\rho_\flat}\Hpij{0j}(\breve\xi(\bbz),\rho(\bbz))\, ,
\\
\Rf&=\hf(\Rf_++\Rf_-)\, ,
\qquad B_I=\CB_I\, ,
\end{split}
\ee
where
\be
\label{ap_sections}
\begin{split}
\breve \xi^\Lambda(\bbz) &=\breve A^\Lambda+
\breve\cR\( \bbz^{-1}\breve Z^\Lambda-\bbz\breve\bZ^\Lambda\)\, ,
\\
\rho_I(\bbz)&=  \breve B_I-\I\sum_j\int_{\tC_j}\frac{\de \bbz'}{2\pi\I\, \bbz'}\,
\frac{\bbz'+\bbz}{\bbz'-\bbz}\, \Hij{ij}_I(\breve\xi(\bbz'))
\\
& \qquad \qquad\qquad\qquad
-\I \(\Hij{i0}_I(\breve\xi(\bbz))+\Hij{i\infty}_I(\breve\xi(\bbz))\)
-\I\cij{+-}_I\log \bbz
\end{split}
\ee
parametrize the unperturbed twistor lines and $\bbz$ is related to $\zeta$ through the
(undeformed) relation \eqref{zztpm},
\be
\bbz=-\frac{1}{\breve \bz}\,\frac{\zeta-\brztp}{\zeta-\brztm}\, .
\ee

With these definitions and relations, it is tedious but straightforward
to compute the deformed complex coordinate
$\xi^\Lambda=u^\Lambda/\uf$ on $\cZ$
in terms of the coordinates on $\cM \times \CP$,
\be
\label{pertrelxi}
\xi^\Lambda =A^\Lambda+ z^{-1}Y_+^\Lambda
-zY_-^\Lambda
+\I\sum_j \int_{\tC_j}\frac{\de \bbz}{2\pi\I \bbz}\,
\frac{\bbz+z}{\bbz-z}\[ \p_{\rho_\Lambda}\Hpij{0j}
-\breve \xi^\Lambda(\bbz)\p_{\rho_\flat}\Hpij{0j}\]\, .
\ee
Taking into account the relation
\be
\rho_I=-2\I\btxii{i}_I-\I \(\Hij{i0}_I(\breve\xi)+\Hij{i\infty}_I(\breve\xi)\)\, ,
\label{relrhotxi}
\ee
one verifies that this expression coincides with the result
\eqref{xiqline} from contact geometry,
at $\zeta=0,\ \varpi=z$. A similar derivation of $\txii{i}_I$
in \eqref{txiqline} ought to be possible
but we have not attempted to carry it through.

\subsection*{Deformed \hk potential}
\label{ap_chi}

Having defined an appropriate set of coordinates on the Swann bundle,
we now generalize the representation \eqref{eqchi}  for the \hk potential
to the deformed case, and relate it to the contact potential $\Phi_{[+]}$ of contact
geometry.
For this purpose, we note that the \hk potential \eqref{chiLeg} can be written as
\be
\chi =\(1+\nupk{0}^I\p_{v^I} +\bnupk{0}^I\p_{\bv^I}\)\oint_C \frac{\de\zeta}{2\pi\I\,\zeta}\,
\left[ 1- x^I \(\pa_{\eta^I}+\p_{x^I}\rho_J\p_{\rho_J} \)  \right]\(H(\eta)+\Hp(\eta,\rho)\)\, ,
\ee
where we omitted the summation over patches and used the relation \eqref{holcoor}
between the undeformed and deformed complex coordinates $v^I$ and $u^I\equiv v^I+\nupk{0}^I$.
Following the same steps \eqref{Legtrafo2} which led to the representation \eqref{eqchi},
one finds
\be
\begin{split}
\chi=&\(1+\nupk{0}^I\p_{v^I} +\bnupk{0}^I\p_{\bv^I}\)
\[\rf\breve \cR \oint_\tC\frac{\de \bbz}{2\pi\I \bbz}
\( \bbz^{-1}\breve Z^\Lambda-  \bbz\breve \bZ^\Lambda\)
(H_\Lambda+\HpI{\Lambda})\bigl(\breve \xi(\bbz),\rho(\bbz)\bigr) \right. \\
&\left. +\rf \ci{+-}_I \breve A^I\]
-\oint_\tC\frac{\de\zeta}{2\pi\I\zeta}
\[\(x^I \p_{x^I}-\frac{{\vf}{\zeta^{-1}}
+\bvf\zeta}{\etaf}\,\zeta\p_\zeta\)\rho_J\]{\p_{\rho_J}\Hp}
\, ,
\label{chiprel}
\end{split}
\ee
where $\breve \xi^\Lambda(\bbz)$ and $\rho(\bbz)$ are given in  \eqref{ap_sections}.

The unperturbed quantities $\breve A^\Lambda,\ \breve Z^\Lambda$, {\it etc.}, are not $SU(2)$ invariant.
To arrive at the desired form of the \hk potential, we replace them in the first term
of \eqref{chiprel} by their deformed SU(2) invariant counterparts and collect the remaining terms
which are all of order $\cO(\Hp)$. The first term in \eqref{chiprel} then reads
\be
\label{firstt}
\Rf\, \cR \oint_\tC\frac{\de \bbz}{2\pi\I \bbz}
\( \bbz^{-1}\cZ^\Lambda-  \bbz\bcZ^\Lambda\)(H_\Lambda+\HpI{\Lambda})\(\zxi(\bbz),\rho(\bbz)\)
+\Rf \ci{+-}_I A^I\, ,
\ee
where
\be
\zxi^\Lambda(\bbz)\equiv A^\Lambda+ \bbz^{-1}\, Y_+^\Lambda-\bbz\, Y_-^\Lambda\, .
\ee
The remaining terms are of three types: (i) the second term in \eqref{chiprel},
(ii) the terms coming from the derivatives with respect to $v^I$ and $\bv^I$
in the first term in \eqref{chiprel} and (iii) the terms coming from difference
between deformed and undeformed invariants in \eqref{firstt}.
Altogether they should combine in an invariant expression written as
a contour integral of a $\cO(-2)$ section.
After a long calculation, one obtains
\bea
\chi &=&
\Rf\, \cR \oint_{\tC}\frac{\de \bbz}{2\pi\I \bbz}
\( \bbz^{-1}\cZ^\Lambda-  \bbz\, \bcZ^\Lambda\)(\Hij{0j}_\Lambda
+\HpI{\Lambda}^{[0j]})+\Rf \ci{+-}_I A^I
\nn
\\
&&  -\I\, \Rf\, \cR \oint_{\tC}\frac{\de \bbz}{2\pi\I \bbz}
\( \bbz^{-1} \cZ^\Sigma-\bbz\, \bcZ^\Sigma\)
\(\Hij{j0}_{\Lambda\Sigma}+\Hij{j\infty}_{\Lambda\Sigma}\)
\(\p_{\rho_\Lambda}\Hpij{0j}-\zxi^\Lambda\p_{\rho_\flat}\Hpij{0j}\)
\label{respertchi}
\\
&& +\I\,\Rf\, \cR \oint_{\tC}\frac{\de \bbz'}{2\pi\I \bbz'} \oint_{\tC}\frac{\de \bbz}{2\pi\I \bbz}
\, \frac{\bbz'+\bbz}{\bbz'-\bbz} \( \bbz^{-1} \cZ^\Sigma-\bbz\, \bcZ^\Sigma\)
\Hij{ij}_{\Lambda\Sigma}
\(\p_{\rho_\Lambda}\Hpij{0i}-\zxi^\Lambda(\bbz')\p_{\rho_\flat}\Hpij{0i}\)\, ,
\nn
\eea
where $\Hij{ij}_{\Lambda\Sigma}$ are functions of $\zxi(\bbz)$,
whereas $\Hpij{0i}$ are functions of $\zxi(\bbz)$ and $\rho(\bbz)$ (or, in the last term,
functions of $\bbz'$).

This expression can be further simplified. First, taking into account \eqref{relrhotxi}
and
\be
\Hij{ij}_\flat(\xi)=\hHij{ij}(\xi)-\xi^\Lambda\hHij{ij}_\Lambda(\xi)+\cij{ij}_\Lambda\xi^\Lambda+\cij{ij}_\flat\, ,
\ee
it is easy to check that the first and third terms in \eqref{respertchi}
combine into one with the derivatives of
the transition functions replaced by $\tilde{T}^{[0j]}_\Lambda(\zxi,\txii{j})$ \eqref{Tfct}.
Moreover, the last term in \eqref{respertchi} can be rewritten as
\be
\Rf\cR\sum_j\int_{\tC_j}\frac{\de \bbz}{2\pi\I \bbz}
\( \bbz^{-1} \cZ^\Sigma-\bbz\bcZ^\Sigma\)\Hij{0j}_{\Lambda\Sigma}
\[-T_{[0j]}^\Lambda+
\hf\sum_i \int_{\tC_i}\frac{\de \bbz'}{2\pi\I \bbz'}\,
\frac{\bbz'+\bbz}{\bbz'-\bbz}\,T_{[0i]}^\Lambda\] \, ,
\ee
where for $i=j$ the variable $\bbz'$ lies inside the contour for
$\bbz$ and the first term appears since
in \eqref{respertchi} the situation was opposite. Then the expression in the square brackets
is just $\xii{0}^\Lambda(\bbz)-\zxi^\Lambda(\bbz)$ for $\bbz\in \cU_j$.
Finally,  $Y^\Lambda$ and $\cR\cZ^\Lambda$
differ by a phase factor (see \eqref{relinvpert})
which can be absorbed into a redefinition of the integration variable $\bbz$.
As a result, the \hk potential can be written more compactly as
\be
\chi=\Rf \sum_j\oint_{\tilde C_j}\frac{\de \bbz}{2\pi\I \bbz}
\(\bbz^{-1} Y^{\Lambda}_+ -\bbz Y^{\Lambda}_- \) \tilde{T}^{[0j]}_\Lambda(\xii{0}, \txii{j})
+\Rf \cij{+-}_I  A^I  \, .
\label{finalchi}
\ee
Comparing with the contact potential \eqref{eqchip},
one finds that the relation \eqref{coor} continues to hold in the perturbed case.

\subsection*{Deformed $SU(2)$ transformations}
\label{ap_perttrans}
The $SU(2)$ invariance of the quantities defined in this appendix can be checked
using the following transformation rules, which follow from the general discussion
in Appendix~A:
\be
\begin{split}
& \delta v^I=\I\eps_3 v^I+\eps_+ x^I-\frac{3}{2}\,\eps_-\bnupk{3}^I\, ,
\qquad \delta \bv^I=-\I\eps_3 \bv^I+\eps_- x^I-\frac{3}{2}\,\eps_+\nupk{3}^I\, ,
\label{transv_ap}
\\
& \delta x^I=-2(\eps_- v^I+\eps_+ \bv^I) \, , \qquad\qquad
\delta \vrh_I=\I\(\eps_- \CL_{\bu^I}-\eps_+  \CL_{u^I}\) + \eps_3 c_I\, ,
\\
&\delta w_I = \eps_+ \cL_{u^I} +\frac{\I}{2}\, \eps_3 c_I\, ,\qquad\quad\qquad
\delta \bw_I = \eps_- \cL_{\bu^I} - \frac{\I}{2} \, \eps_3 c_I\, ,
\end{split}
\ee

\be
\begin{split}
\delta\nupk{0}^I&=\I\eps_3 \nupk{0}^I +
i\eps_+\CL_{\vrh_I}+\frac{3}{2}\,\eps_-\bnupk{3}^I \, ,
\\
\delta \nuppm^I&=(\I\eps_3-2\eps_-\ztpm)\nuppm^I-\frac{3}{2}\(\eps_+\ztpm^2
\nupk{3}^I-\eps_-\bnupk{3}^I\)\, ,
\\
\delta \nuppmz^I&=-2\eps_-\nuppm^I-3\eps_+\ztpm \nupk{3}^I\, ,
\\
\delta \nuppmzz^I&=-(i\eps_3-2\eps_-
\ztpm)\nuppmzz^I-2\eps_-\nuppmz^I-3\eps_+ \nupk{3}^I\, ,
\end{split}
\ee
where
\be
\nupk{0}^I=\oint_C \frac{\de\zeta}{2\pi}\, \Hp^I\, ,\qquad
\nupk{3}^I=\oint_C \frac{\de\zeta}{\pi\,\zeta^3}\, \Hp^I\, , \qquad
\bnupk{3}^I=-\oint_C \frac{\de\zeta}{\pi\,\zeta^{-1}}\, \Hp^I
\ee
are Laurent coefficients of the deformation $\nupk{0}^I$ given in \eqref{nupi}
(as usual, we omitted the sum over contours).

The following properties, valid in the absence of perturbations, are also useful:
\be
\delta \ztpm=-\epsp -\epsm\ztpm^2+\I\epst\ztpm
\, ,
\label{transztpm}
\ee
\be
\delta\etapm^\Lambda=-(\eps_+\bar\zeta_\mp+\eps_-\ztpm)\etapm^\Lambda
\, ,
\ee
\be
\delta z = \frac{1+z\bz}{|z|} \sqrt{\frac{\bvf}{\vf}}\, z\,\eps_+ ,\qquad
\delta \bz = \frac{1+z\bz}{|z|} \sqrt{\frac{\vf}{\bvf}}\, \bz\,\eps_- \, ,
\label{transz_ap}
\ee
\be
\delta\rho_I=\(\eps_+ +\eps_- \zeta^2-\I\eps_3\zeta\)\pa_\zeta\rho_I\, .
\label{transrho_ap}
\ee



\providecommand{\href}[2]{#2}\begingroup\raggedright\endgroup

\end{document}